\documentclass[numberedappendix]{emulateapj}
\usepackage[T1]{fontenc}
\usepackage[latin9]{inputenc}
\usepackage{amsbsy}
\usepackage{graphicx}
\usepackage[unicode=true,pdfusetitle,
 bookmarks=true,bookmarksnumbered=false,bookmarksopen=false,
 breaklinks=false,pdfborder={0 0 0},backref=false,colorlinks=false]
 {hyperref}

\makeatletter

\providecommand{\tabularnewline}{\\}

\usepackage{tablefootnote}
\usepackage{amsmath}
\usepackage{times}
\usepackage{url}

\makeatother

\begin{document}

\global\long\def\Mo{M_{\odot}}
\global\long\def\Ro{R_{\odot}}
\global\long\def\Lo{L_{\odot}}
\global\long\def\Mbh{M_{\bullet}}
\global\long\def\Ms{M_{\star}}
\global\long\def\Rs{R_{\star}}
\global\long\def\Ts{T_{\star}}
\global\long\def\vs{v_{\star}}
\global\long\def\s{\sigma_{\star}}
\global\long\def\n{n_{\star}}
\global\long\def\tr{t_{\mathrm{rlx}}}
\global\long\def\Ns{N_{\star}}
\global\long\def\ns{n_{\star}}
\global\long\def\jlc{j_{lc}}

\slugcomment{Draft version 0.0 \today}

\title{Steady state relativistic stellar dynamics around a massive black
hole}

\shorttitle{Steady state relativistic loss-cone}

\author{Ben Bar-Or and Tal Alexander}

\shortauthors{Bar-Or \& Alexander}

\affil{Department of Particle Physics \& Astrophysics, Weizmann Institute
of Science, P.O. Box 26, Rehovot 76100, Israel }
\begin{abstract}
A massive black hole (MBH) consumes stars whose orbits evolve into
the small phase-space volume of unstable orbits, the ``loss-cone'',
which take them directly into the MBH, or close enough to interact
strongly with it. The resulting phenomena: tidal heating and tidal
disruption, binary capture and hyper-velocity star ejection, gravitational
wave (GW) emission by inspiraling compact remnants, or hydrodynamical
interactions with an accretion disk, are of interest as they can produce
observable signatures and thereby reveal the existence of the MBH,
affect its mass and spin evolution, probe strong gravity, and provide
information on stars and gas near the MBH\@. The continuous loss
of stars and the processes that resupply them shape the central stellar
distribution. We investigate relativistic stellar dynamics near the
loss-cone of a non-spinning MBH in steady-state analytically and by
Monte Carlo simulations of the diffusion of the orbital parameters.
These take into account Newtonian mass precession due to enclosed
stellar mass, in-plane precession due to general relativity, dissipation
by GW, uncorrelated two-body relaxation, correlated resonant relaxation
(RR) and adiabatic invariance due to secular precession, using a rigorously
derived description of correlated post-Newtonian dynamics in the diffusion
limit. We argue that general maximal entropy considerations strongly
constrain orbital diffusion in steady-state, irrespective of the relaxation
mechanism. We identify the exact phase-space separatrix between plunges
and inspirals, predict their steady-state rates, and verify they are
robust under a wide range of assumptions. We derive the dependence
of the rates on the mass of the MBH, show that the contribution of
RR is small, and discuss special cases where unquenched RR in restricted
volumes of phase-space may affect the steady-state substantially. 
\end{abstract}


\makeatletter{}

\global\long\def\Mo{M_{\odot}}
\global\long\def\Ro{R_{\odot}}
\global\long\def\Lo{L_{\odot}}
\global\long\def\Mbh{M_{\bullet}}
\global\long\def\Ms{M_{\star}}
\global\long\def\Rs{R_{\star}}
\global\long\def\Ts{T_{\star}}
\global\long\def\vs{v_{\star}}
\global\long\def\s{\sigma_{\star}}
\global\long\def\n{n_{\star}}
\global\long\def\tr{t_{\mathrm{rlx}}}
\global\long\def\Ns{N_{\star}}
\global\long\def\ns{n_{\star}}
\global\long\def\jlc{j_{lc}}

\section{Introduction}

\label{s:intro}

\subsection{Background}

Strong interactions of stars with a massive black hole (MBH) in a
galactic center lead to a variety of extreme phenomena, as well as
provide mass for the growth and evolution of the MBH\@. The small
phase-space volume of orbits whose periapse lies close enough to the
MBH to lead to a strong interaction is called the loss-cone~\citep{fra+76},
since in most cases the interaction destroys the star, either immediately
(for example by a direct plunge through the event horizon or by tidal
disruption outside it~\citep{ree88} or gradually, after the orbit
decays by some dissipation mechanism (for example the emission of
gravitational waves (GW)~\citep{hil+95}, tidal heating of the star
by the MBH~\citep{ale+03a}, or drag against a massive accretion disk~\citep{ost83}). Even when the star survives the encounter, for example
in a tidal scattering event~\citep{ale+01b}, the restricted set of
orbits that allow such near-misses lie very close to the loss-cone
phase-space. Since typically stars do not survive long on loss-cone
orbits, the key questions is how, and at what rate are these orbits
repopulated by new stars. The stellar dynamical study of this question
is known as loss-cone theory. A main re-population channel is by dynamical
relaxation mechanisms, which randomize stable orbits and causes them
to diffuse in phase-space into the loss-cone\footnote{Other possibilities include e.g., in-situ star formation or galaxy
mergers.}. The close interaction event rates in the steady-state of dynamically
relaxed systems are of particular interest, both because these can
be derived from first principles independently of initial conditions,
and because these correspond, statistically, to the cases most likely
to be observed. 

Studies of the loss-cone can be broadly categorized by four criteria:
whether they deal with processes that lead to immediate stellar destruction
(infall) or a gradual one (inspiral); whether they are strictly Newtonian
or include also general relativity (GR), fully or perturbatively;
whether they include only slow non-coherent two-body relaxation (i.e.,
non-resonant relaxation, NR~\citealt{cha44}) or also fast coherent
relaxation (known as resonant relaxation, RR,~\citealt{rau+96}. See
Section~\ref{s:MBHrlx}); and finally by the calculation methods employed:
whether they are analytical, or based on the diffusion approximation
either by direct numerical solutions of the Fokker-Planck (FP) equations
or by Monte-Carlo (MC) methods, or whether they employ direct $N$-body
simulations. 

Early studies focused on the infall rates of tidal disruption events
in the Newtonian approximation, using analytic and FP-based methods~\citep{fra+76,you+77,coh+78,sha+78}. These studies were subsequently
updated and generalized to include some deviations from spherical
symmetry~(\citealt{sye+99,mag+99}; see review by~\citealt{ale12}).
The prospects of detecting low-frequency GW from compact remnants
spiraling into MBHs (extreme mass ratio inspiral (EMRI) events) with
long-baseline space-borne GW detectors~\citep{ama+07} motivated studies
of the inspiral rates for EMRIs in using FP-based methods in the NR-only
limit with perturbative GR~(\citealt{hil+95,sig+97,mir+00,fre01,fre03,iva02,hop+05,hop+06b};
see review by~\citealt{sig03}). The effects of the MBH spin on the
EMRI rates were also considered~\citep{ama+13}. 

A unifying framework relating plunge and inspiral processes was formulated
by~\citet{ale+03b} and used to estimate infall and inspiral event
rates in the Galactic Center in the NR-only limit: infall by direct
plunge and tidal disruption, inspiral by GW emission and tidal heating~\citep{ale+03a}, as well as tidal scattering events~\citep{ale+01b}.
Different attempts to estimate the infall and inspiral rates yielded
a wide, uncertain range of values that spans several orders of magnitudes~\citep{sig03,ale12}.

Fast relaxation by RR can be effective on the small spatial scales
where most EMRIs originate and importantly, where stellar orbits are
observed in the Galactic Center and can therefore provide empirical
constraints~\citep{hop+06a}. This realization motivated a re-evaluation
of relaxation processes and their impact on dynamics very close to
MBHs. An approximate comparison of the relative rates of RR and NR
suggested that the branching ratio between plunges and inspirals depends
strongly on the efficiency of RR, which was then poorly understood~\citep{hop+06a,eil+09}. This added a yet larger uncertainty to EMRI
rate estimates. A key question is the physical origin and characteristics
of the quenching mechanism that perturbs the near-Keplerian symmetry
that generates RR, and causes the orbits to drift in phase-space from
their initial values.

Initial analysis of RR in the relativistic context~\citep{rau+96}
indicated that rapid GR precession on very eccentric orbits likely
plays a key role in quenching RR\@. Importantly, GR quenching can
prevent RR from rapidly pushing all stars into plunge orbits, thereby
allowing slow inspiral to produce detectable periodic GW signals (EMRIs)~\citep{hop+06a}. In these earlier studies the deterministic GR precession
was treated as an effective stochastic perturbation of the Keplerian
orbits. 

First indications that the precession cannot be treated that way,
and that the long-timescale behavior of RR is not well described as
a Markov process (random walk), were uncovered in post-Newtonian small
$N$-body simulations of direct plunge and GW inspiral events~\citep[MAMW11]{mer+11}.
These revealed oscillatory orbital behavior at high eccentricities
that appeared to act as a barrier against further evolution to even
higher eccentricities and subsequent infall or inspiral. MAMW11 dubbed
this dynamical phenomenon the Schwarzschild Barrier (SB), and showed
that the oscillations are well approximated by the simple \emph{ansatz}
of assuming 2-body GR dynamics in the presence of a randomly oriented
fixed force vector~\citep{ale10}, representing the residual force
due to the background stars. While the effect appeared related to
the EMRI-preserving RR quenching predicted by~\citet{hop+06a}, its
magnitude seemed much stronger than anticipated, in that it not only
damped the RR torques, but actually appeared to prevent the orbits
from interacting closely with the MBH at all. The larger-scale relativistic
$N$-body simulations of~\citet[BAS14]{bre+14} confirmed that GR
precession quenches RR roughly on the scale of the SB, and concluded
that the resulting EMRI rates are consistent with those driven by
NR\@. 

The SB phenomenon was subsequently explained rigorously in terms of
the adiabatic invariance (AI) of the GR precession against the coherent
RR torques, when the precession period is shorter than the typical
RR coherence time (the $\boldsymbol{\eta}$-formalism,~\citealt{bar+14};
see review by~\citealt{ale15}). By describing the RR torques due
to the background stars in terms of a correlated noise field, it is
possible to formulate an effective FP description for RR that takes
into account AI, and to derive the corresponding effective diffusion
coefficients (DCs), whose form and behavior depends critically on
the assumed temporal smoothness of the noise model. The continuous
orbital evolution of the stellar background suggests that the physically
correct form of the stochastic torques is that of a smooth (infinitely
differentiable, $C^{\infty}$) noise. The AI is maximal for smooth
noise. In that case its dynamical effect can be described as a faster
than exponential suppression of the diffusion coefficients below some
critical angular momentum limit. The vanishing phase-space density
past this limit grows so slowly ($\sim\log t$), that the limit can
be considered as an effective barrier fixed in time. While this limit
is not a true barrier, nor a reflecting one, it does effectively divide
phase-space into a region where RR can be efficient, and a region
where it cannot. As we show below, the presence of the competing process
of NR substantially limits the significance of AI in the dynamics
of the loss-cone on long timescales (of order of the NR relaxation
time).

This study focuses mainly on the implications of NR and RR around
a MBH for loss-rates. However, these dynamical processes are also
relevant for understanding and modeling other processes around MBHs,
and in particular the Galactic MBH, SgrA$^{\star}$. Although the
inner Galactic Center contains a relatively small and manageable number
of stars by the standards of current Newtonian $N$-body codes, it
is still extremely challenging to simulate it directly both because
of the extreme dynamical range introduced by the high MBH to star
mass ratio, and because of the added complexity of the GR equations
of motion. The impact of MBH spin and RR on orbital tests of GR in
the Galactic Center were studied with post-Newtonian small $N$-body
simulations~\citep{mer+10}. A study of the implications of RR for
the formation mechanisms of the of stars orbiting SgrA$^{\star}$,
either resorted to large Newtonian-only $N$-body simulations~\citep{per+08},
or substantially underestimated the efficiency of AI in quenching
RR by using a MC scheme based on the simple fixed force \emph{ansatz}
with non-differentiable ($C^{0}$) noise, to study the implications
of the SB for stars in the Galactic Center~\citep{ant+13,ant14}.

\makeatletter{}

\global\long\def\Mo{M_{\odot}}
\global\long\def\Ro{R_{\odot}}
\global\long\def\Lo{L_{\odot}}
\global\long\def\Mbh{M_{\bullet}}
\global\long\def\Ms{M_{\star}}
\global\long\def\Rs{R_{\star}}
\global\long\def\Ts{T_{\star}}
\global\long\def\vs{v_{\star}}
\global\long\def\s{\sigma_{\star}}
\global\long\def\n{n_{\star}}
\global\long\def\tr{t_{\mathrm{rlx}}}
\global\long\def\Ns{N_{\star}}
\global\long\def\ns{n_{\star}}
\global\long\def\jlc{j_{lc}}

\subsection{Objectives and overview}

\label{s:outline}

The objectives of this study are to integrate the recent insights
about the role correlated noise plays in determining the properties
of RR and its formulation as an effective diffusion process~\citep{bar+14}
together with the known properties of NR; to derive a rigorous computational
framework for calculating the steady-state phase-space density near
the relativistic loss-cone and the resulting loss-rates, and to use
this framework for a systematic study of the dependence of the results
on the various physical mechanisms involved in the dynamics: mass
(Newtonian) precession, GR precession, GW dissipation, the RR noise
model and coherence time. The ultimate objective is to provide well-defined
estimates of the infall and inspiral rates (including, but not limited
to direct plunges, tidal disruptions and EMRIs) and their scaling
with the properties of the galactic nucleus (MBH mass and stellar
density). These can then inform design decisions about planned surveys
and experiments, and serve as benchmarks for more detailed future
studies.

We focus our study on a simplified galactic nucleus containing a stationary
non-spinning (Schwarzschild) MBH surrounded by a Keplerian, spherically
symmetric (in the time-averaged sense), power-law cusp of single mass
stars (the background cusp). Direct relativistic $N$-body simulations
generate, by construction, the correct dynamics, but are presently
limited by computational costs to unrealistically small $N$, which
generally cannot be scaled up to astrophysically relevant values since
different dynamical mechanisms scale differently with $N$~\citep[e.g.,][]{heg+03}.
Moreover, they allow little freedom to switch on/off the various physical
mechanisms that affect the outcome. It is therefore difficult to disentangle
their contributions and interpret the results. 

Here we follow a different approach. We represent the evolution of
the system in the realistic large-$N$ limit as a superposition of
diffusion processes. This enables us to isolate and study the effect
of the different dynamical mechanisms, and thereby obtain an analytic
description of the system.

We calculate the loss-cone phase-space density and loss-rates by two
complementary methods. We show that the diffusion in phase-space is
well approximated as a separable process: fast diffusion in angular
momentum, superposed on a slow diffusion in energy. We then use this
separation of timescales to derive analytically the steady-state properties
of the system. We also solve the diffusion in energy and angular momentum
phase-space numerically by MC simulations, which are statistics-limited,
but have the advantage of flexibility in introducing additional dynamical
effects and constraints. We cross-validate these two calculation methods,
and also compare the MC results to the $N$-body loss-rates of MAMW11
and BAS14, and reproduce the AI effects of~\citet{bar+14} in the
absence of NR\@.

This paper is organized as follows. In Section~\ref{s:MBHrlx} we
review the dynamical process of relaxation near a MBH in a galactic
nucleus. We present a unified framework for describing both non-coherent
two-body relaxation and coherent resonant relaxation. We discuss the
role of secular processes in the emergence of adiabatic invariance
in the long-term orbital evolution of the system. In Section~\ref{s:analytic}
we describe the structure and properties of phase-space near the loss-cone,
and derive analytic estimates for the steady-state distribution and
loss-rates. We start, in Section~\ref{ss:diff-eq}, by formulating
the diffusion equations which govern the evolution of the system.
In Section~\ref{ss:PSflow} we describe of the diffusion process in
terms of the streamlines of the probability flow, which provide a
powerful visual representation of the dynamics, and guides us in Sections~\ref{ss:ss-flux}--\ref{ss:SSNR}
in solving the steady-state distribution and loss-rates (at this stage,
without RR or GW dissipation). In Section~\ref{s:EMRI-event-rates},
we show that GW dissipation separates the probability flow into two
distinct regions in phase-space: a region where stars can inspiral
into the MBH while emitting GWs, and a region where stars plunge directly
into it. The inspiral event rate is then calculated exactly by locating
the separatrix that demarcates the two regions (Appendix~\ref{a:GWline}).
Finally, in Section~\ref{s:RR-effect} we show that RR has only a
small impact on the steady-state density and loss-rates, and provide
a method to quantify its effect. In Section~\ref{s:MCmodel} we present
our MC procedure for modeling orbital evolution in phase-space (The
MC procedure and the NR DCs that enter it are described in Appendices~\ref{a:MCproc}
and~\ref{a:2bDCs}). In Section~\ref{ss:1DMC} we validate the description
RR and the emergence of AI in angular momentum-only simulations against
the analytic results of~\citet{bar+14}, and show that AI is very
efficiently suppressed by NR on long timescales. We also show that
over short timescales, AI induces the ``Schwarzschild Barrier''
phenomenon seen in angular momentum and energy phase-space~\citep{mer+11},
and demonstrate that this dynamical feature is erased over long timescales.
We compare in Section~\ref{ss:2DMC} the MC code against the small
$N$-body loss-rates of~\citet{mer+11} and~\citet{bre+14}, and show
that the MC and $N$-body give consistent results (The derivation
of steady-state rate estimates from MC simulations is described in
Appendix~\ref{ss:SSrates}). In Section~\ref{s:MCrates} we explore
the robustness of the MC-derived rates under various dynamical approximations
and assumptions. In Section~\ref{ss:GCparms} we estimate the rates
for the prototypical target of low-frequency GW, the Galactic Center.
In Section~\ref{ss:MBHscale} we derive analytically, and confirm
with the MC simulations, the weak scaling of the rates with the MBH
mass. We discuss and summarize our results in Section~\ref{s:discussion}.
In Section~\ref{ss:mainresults} we focus on the role of the principle
of maximum entropy (Appendix~\ref{a:MaxEntropy}) as a guiding principle
in the derivation of the DCs. We argue that RR typically does not
play a major role in the steady-state dynamics of the loss-cone. We
illustrate this analysis by presenting a fine-tuned idealized counter-example
where RR may substantially affect the loss-rates: the interaction
of icy planetesimals with a massive circumnuclear accretion disk.
We conclude by discussing the limitations of our analysis in Section~\ref{ss:caveats}.

\makeatletter{}

\global\long\def\Mo{M_{\odot}}
\global\long\def\Ro{R_{\odot}}
\global\long\def\Lo{L_{\odot}}
\global\long\def\Mbh{M_{\bullet}}
\global\long\def\Ms{M_{\star}}
\global\long\def\Rs{R_{\star}}
\global\long\def\Ts{T_{\star}}
\global\long\def\vs{v_{\star}}
\global\long\def\s{\sigma_{\star}}
\global\long\def\n{n_{\star}}
\global\long\def\tr{t_{\mathrm{rlx}}}
\global\long\def\Ns{N_{\star}}
\global\long\def\ns{n_{\star}}
\global\long\def\jlc{j_{lc}}

\section{Relaxation around a massive black hole}

\label{s:MBHrlx}

We present here an overview of dynamical relaxation around a central
MBH, and in particular of resonant relaxation, using two complementary
approaches. The first, which is closer to the conventional description
of the subject~\citep[e.g.,][]{rau+96,hop+06a}, serves to introduce
the basic terms and ideas, and connect the present work to past studies.
The second presents a unified framework for describing the coherent
and stochastic aspects of relaxation, using adiabatic invariance as
the unifying concept in all forms of relaxation. This connects the
present work to recent results on the representation of RR as an effective
diffusion process, a key tool used here to investigate loss-cone dynamics
in the large-$N$ limit.

\subsection{Two-body relaxation}

To understand the relation between NR and RR, consider a spherical
stellar system composed of stars of mass $\Ms$ orbiting a central
massive object $\Mbh\gg\Ms$, and focus on a test star at radius $r$
from the center, where the local stellar number density is $\ns\left(r\right)$.
The average net force exerted on the test star by the $\mathrm{d}\Ns\left(<b\right)\sim\ns b^{2}\mathrm{d}b$
background stars in a thin small shell around it with radius $b\ll r$
and width $\mathrm{d}b\ll b$, is zero. However, the Poisson fluctuations
in the positions of these $\mathrm{d}\Ns$ discrete masses generate
a residual force of magnitude $\sqrt{\left\langle F^{2}\right\rangle }\sim\sqrt{\mathrm{d}\Ns}G\Ms^{2}/b^{2}$.
This force persists in direction and magnitude until the stars generating
it move substantially. For a random stellar velocity field with dispersion
$\sigma^{2}\sim G\Mbh/r$, this coherence time is $T_{c}^{NR}\sim b/\sigma\ll r/\sigma\sim P$,
where $P$ is the orbital period. Since $T_{c}^{NR}\ll P$, the net
encounter is impulsive---a collision, and so in the case of NR, the
coherence time is the collision time. The change in velocity due to
the residual force over time $T_{c}^{NR}$ is $\delta v\sim\sqrt{\left\langle F^{2}\right\rangle }T_{c}^{NR}/\Ms$.
Over times $t>T_{c}^{NR}$, these impulses add non-coherently, and
the accumulated change in velocity per unit time is $\left\langle \Delta v^{2}\right\rangle _{t}\sim\left(G^{2}\Ms^{2}\ns/\sigma\right)\mathrm{d}b/b$.
Integration over all shells from $b_{\min}$ to $b_{\max}<r$ (assuming
$\ns$ is constant) yields the NR diffusion timescale $T_{NR}\sim\sigma^{2}/\left\langle \Delta v^{2}\right\rangle _{t}\sim\sigma^{3}/\left(G^{2}\Ms^{2}\ns\log\Lambda\right)$
where $\Lambda=b_{\max}/b_{\min}$ is the Coulomb factor. These local
changes in the test star's velocity lead to changes in orbital energy
and angular momentum at a rate $\left\langle \Delta E^{2}\right\rangle /E^{2}\sim\left\langle \Delta J^{2}\right\rangle /J_{c}^{2}\sim1/T_{NR}$,
where $J_{c}=\sqrt{G\Mbh a}$ is the circular angular momentum, and
$a$ the semi-major axis (sma).

\subsection{Resonant Relaxation}

In a nearly-symmetric potential where the background orbits are nearly
fixed on timescales $T_{c}\gg P$, the test star interacts over a
long period of time with the entire phase-averaged background orbits,
and not only instantaneously with the segment of the orbit closest
to it. That is, the interaction is non-local. As in the NR case, the
discrete number of stars on the scale of the test star's orbit, $\Ns\left(<r\right)$,
gives rise to random fluctuations in the force on it, $\sqrt{\left\langle F^{2}\right\rangle }\sim\sqrt{\Ns\left(<r\right)}G\Ms^{2}/r^{2}$,
which persist on timescale $T_{c}$. However, unlike in the NR case,
here $T_{c}\gg P$, and so the orbital energy is adiabatically conserved
(see below) and the residual force affects only the orbital angular
momentum.

\subsection{A unified description of relaxation}

Conventionally, orbital evolution is described by a combination of
two distinct classes of processes: stochastic relaxation processes
(two-body relaxation) and coherent (secular) processes (RR). Here
we present a unified framework (summarized in Table~\ref{tab:A-unified-picture})
for describing and analyzing all the dynamical processes that drive
orbital evolution in terms of short-timescale coherent processes that
effectively contribute as stochastic processes on longer timescales. 

We begin by considering the case where the test star is statistically
indistinguishable from the background stars (this case was the focus
of early works on RR, and is recast here in a more general context). 

It is instructive to consider the various relaxation processes in
the order of their associated coherence times. When two stars interact
impulsively (as in the case where the impact parameter $b$ is much
smaller than the sma $a$ of the orbit around the MBH), the interaction
time is effectively limited to the crossing time of the closest approach.
Since during the interaction, the force on the test star is nearly
constant, the duration of the interaction (the collision timescale)
is the coherence timescale $T_{c}^{NR}\sim b/\sigma$ of two-body
relaxation. As long as only interactions with small enough $b<a$
are considered, so that $T_{c}^{NR}<P(a)$, the Hamiltonian cannot
be orbit-averaged, and therefore all the orbital elements can change
by the interaction. 

The treatment of two-body relaxation is based on the approximation
$T_{c}^{NR}\to0$, that is, that the interaction time is shorter than
any other relevant timescale in the system, and that individual collisions
are uncorrelated. In that limit the process is Markovian~\citep{nel+99,bar+13}
and can therefore be described as diffusion in phase-space. 

Two-body interactions with a large impact parameter $b>a$ (i.e.,
soft encounters) such that $T_{c}^{NR}\sim b/\sigma>P(a)$ are no
longer impulsive. This means they cannot be described as occurring
instantaneously and locally between two point particles, and therefore
can no longer be described by the standard two-body relaxation formalism.
In particular, the interaction is no longer Markovian, since the test
star is repeatedly affected by the same perturbing star. Since it
can be shown (for energy relaxation,~\citealt{bar+13}) that these
soft two-body encounters do not contribute much to the total relaxation,
an approximate cutoff on the maximal impact parameter is introduced
via the Coulomb logarithm term.

Two-body interactions in the extreme soft limit can be described in
terms of an effective diffusion process~\citep{bar+14}. Since $T_{c}>P$,
the Hamiltonian can be double-averaged over both the orbit of the
test star and the orbits of the background stars. The averaged Hamiltonian
is then independent of the mean anomaly, and so the Keplerian energy
(semi-major axis) is adiabatically conserved---proof that the contribution
of soft collisions to energy relaxation is negligible. The double-averaged
Hamiltonian no longer describes point particles, but rather interaction
between Keplerian ellipses (``mass wires''), which mutually torque
each other and exchange angular momentum, but not energy. Here, the
force on a test ellipse by the background ellipses remains constant
as long as the orbital orientations of the background ellipses remain
fixed (i.e., over the coherence time $T_{c}^{RR}$). In analogy to
the case of point-point two-body relaxation, this coherence time can
be considered as the interaction (``collision'') time. The coherence
time is determined by the fastest process that can reshuffle the background
orbital orientations. Since typically the background stars are not
on relativistic orbits, the dominant shuffling process is the retrograde
in-plane drift of the argument of periapse, $\omega$, due the enclosed
stellar mass inside the orbits, on the mass-precession timescale $T_{M}\sim QP/N$~\citep[e.g.,][]{hop+06a}. As long as there are no competing processes
with timescale shorter than $T_{M}$ that could randomize the residual
forces of the orbit--orbit interactions, these will torque the orbits
and change their angular momentum in a coherent ($\propto t$) fashion.
Therefore, the Markovian assumption can used and the diffusion rate
is $\left\langle \tau^{2}\right\rangle _{RR}T_{M}\sim QP$. This regime
of RR is sometimes called ``scalar RR'' since it can change the
magnitude of the angular momentum as well as its direction.

On timescale substantially longer than the precession period, the
Hamiltonian can also be double average over the argument of periapse
and the interaction is then between mass annuli~\citep{koc+15}. In
this case the collision (coherence) time is the self-decoherencing
(or self-quenching) time $T_{c}^{vRR}=T_{sq}\sim J_{c}^{2}/\left\langle \tau^{2}\right\rangle \sim QP/\sqrt{N}$
on which the annuli are re-shuffled by their own mutual torques. The
residual torque $\left\langle \tau^{2}\right\rangle _{RR}$ is now
also averaged over the argument of periapse, which leads some cancellation
of the torques, and therefore $\left\langle \tau^{2}\right\rangle _{vRR}<\left\langle \tau^{2}\right\rangle _{RR}$
and the diffusion rate is $\left\langle \tau^{2}\right\rangle _{vRR}T_{c}^{vRR}\sim QP/\sqrt{N}$.
This regime of RR is sometimes called ``vector RR'' since it can
change only the direction of the angular momentum, but not its magnitude.
In general, as we consider longer coherence timescales, it becomes
possible to average over yet more degrees of freedom (phases), and
the averaging results in an effective potential that is yet more symmetric.
This in turn reduces the magnitude of the residual forces. It is found
that the countervailing effects of smaller torques but longer coherence
times lead to faster relaxation $T_{r}\propto v^{2}/\left(\left\langle F^{2}\right\rangle T_{c}\right)$,
see Table~\ref{tab:A-unified-picture}. 

We now turn to the case where the timescales of the test-star and
the background are different, which was not treated rigorously until
recently~\citep{bar+14}, following the discovery in $N$-body simulations~\citep{mer+11} of an abrupt transition in phase-space to a different
dynamical regime where orbital evolution is governed by deterministic
rather than stochastic processes. In this case the precession of the
test star is due to the combined prograde GR in-plane precession with
period $T_{GR}$ and the retrograde Newtonian precession due to the
stellar mass enclosed inside the orbit. Since the GR precession rate
$T_{GR}^{-1}$ diverges as $1/j^{2}$, where $j=\sqrt{1-e^{2}}$ and
$e$ is the eccentricity, eccentric stars with $j$ smaller than some
critical value $j_{0}$ will precess much faster than the background,
i.e., $T_{GR}<T_{c}$. In this case, the Hamiltonian can be averaged
over both the mean anomaly of the background and of test star, as
well as over $\omega$ of the test-star. As a result, the test star's
$j$ is adiabatically conserved. The transition between unconstrained
RR-driven diffusion at $j\gg j_{0}$ to strict adiabatic invariance
at $j<j_{0}$ was calculated for several models of the effective background
``noise'' (residual torques) in terms of effective diffusion coefficients
by~\citet{bar+14}, who showed that for a smoothly varying background
the transition at $j_{0}$ is extremely sharp, with the $j$-diffusion
coefficients suppressed exponentially with the argument $(T_{c}/T_{GR})^{2}\propto(j_{0}/j)^{4}$.

\begin{table*}
\protect\caption{\label{tab:A-unified-picture}A unified framework for relaxation processes
(see text)}

\centering{}{\tiny{}}\begin{tabular}{ccccccc}
\hline 
Process  & Effective  & Averaged quantity & Conserved  & Coherence time & Residual force magnitude & Relaxation time\tabularnewline
 & particles &  & quantities & $T_{c}$ & $\sqrt{\left\langle F^{2}\right\rangle }$ & $v^{2}/\left(T_{c}F^{2}\right)$\tabularnewline
\hline 
NR & Points & None & None & $T_{c}^{NR}<P$ & $\sim\sqrt{\Ns}G\Ms\sqrt{Q}/a^{2}$~\footnotemark[1]\footnotetext{Integrated over all impact parameters} & $\sim Q^{2}P/\left(N\log Q\right)$\tabularnewline
RR & Ellipses & Mean anomaly & $E$ & $P<T_{c}^{RR}\sim T_{M}<T_{p}$ & $\sim\sqrt{\Ns}G\Ms/a^{2}$ & $\sim QP$\tabularnewline
vRR & Annuli & Argument of periapse & $E,J$ & $T_{p}<T_{c}^{vRR}\sim T_{sq}$ & $\sim\sqrt{\Ns}G\Ms/a^{2}$ & $\sim QP/\sqrt{N}$\tabularnewline
\end{tabular}{\tiny \par}
\end{table*}

\makeatletter{}

\global\long\def\Mo{M_{\odot}}
\global\long\def\Ro{R_{\odot}}
\global\long\def\Lo{L_{\odot}}
\global\long\def\Mbh{M_{\bullet}}
\global\long\def\Ms{M_{\star}}
\global\long\def\Rs{R_{\star}}
\global\long\def\Ts{T_{\star}}
\global\long\def\vs{v_{\star}}
\global\long\def\s{\sigma_{\star}}
\global\long\def\n{n_{\star}}
\global\long\def\tr{t_{\mathrm{rlx}}}
\global\long\def\Ns{N_{\star}}
\global\long\def\ns{n_{\star}}
\global\long\def\jlc{j_{lc}}

\section{The phase space of the loss-cone}

\label{s:analytic}

We describe the dynamics of the loss-cone in the Keplerian approximation,
where the gravitational potential of the stars is assumed to be negligible
relative to that of the central MBH\@. In that limit, the Keplerian
orbital energy is $E=G\Mbh/2a$, where the stellar dynamical convention
$E=-E_{\mathrm{true}}>0$ for bound orbits is adopted. The orbital
angular momentum is parameterized by $j=J/J_{c}(a)$. We marginalize
the dynamics over the orbital angles and consider the evolution in
the $(a,j)$ phase-space. We assumes a stationary non-spinning MBH
of mass $\Mbh$ that is surrounded by an isotropic power-law cusp
of stars of mass $\Ms$ each with number density profile $\ns=n_{0}(r/r_{0})^{-\alpha}$.
The mass ratio is denoted $Q=\Mbh/\Ms$. The cusp is assumed to extend
between $a_{\min}$ and $a_{\max}$. The inner boundary at $a_{\min}$
is an absorbing boundary, set at the innermost stable circular orbit.
The outer boundary at $a_{\max}$ is the interface to the galaxy around
the central cusp, which provides an effectively infinite reservoir
of stars to replace those that are lost into the MBH, or evaporate
back to the galaxy. The Newtonian gravitational dynamics are described
in terms of DCs, whose functional form and normalization are calculated
assuming the background cusp and an isotropic distribution of angular
momentum. The DCs, together with additional non-Newtonian processes
such as an absorbing boundary at the last stable orbit (LSO) and GW
dissipation of energy and angular momentum, are then used to generate
the dynamics of test particles, and derive their steady-state loss-rates
and phase-space density. 

\begin{figure}
\noindent \centering{}\includegraphics[width=1\columnwidth]{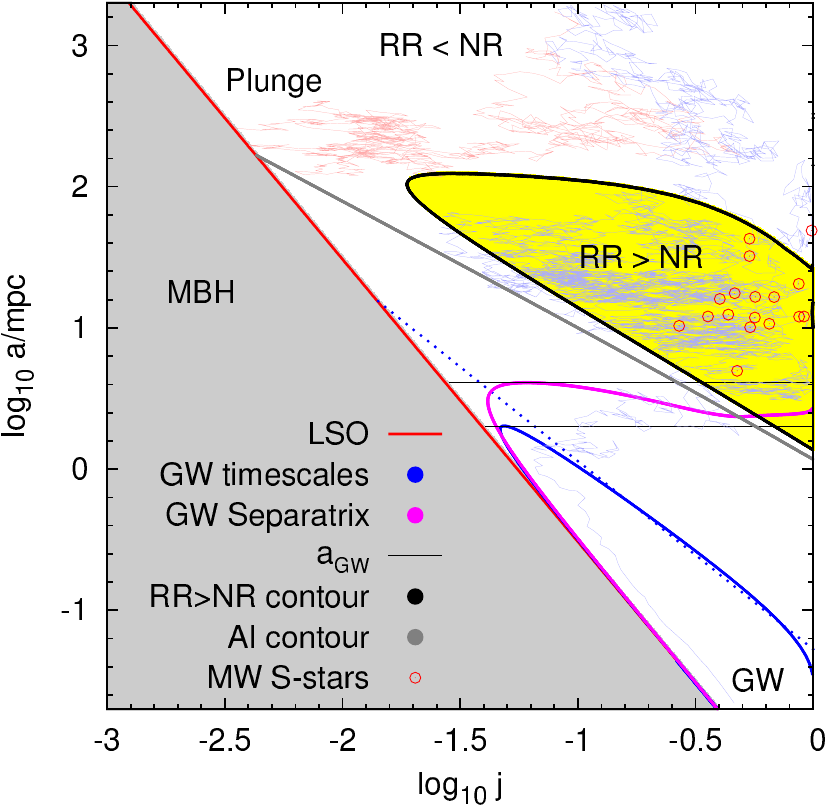}\protect\caption{\label{f:EJ}The $(a,j)$ phase-space of the loss-with the various
critical lines and regions, for a model of the Milky Way model with
$Q=4\times10^{6}$, mass-precession coherence time and a Gaussian
noise model (Section~\ref{ss:GCparms}). Orbits in the gray area below
the LSO line (red) are unstable and promptly plunge into the MBH event
horizon (plunge track example in light red line). Where RR diffusion
is faster than NR diffusion (yellow region), RR dominates the dynamics.
The S-stars observed near the MBH of the MW (red circles)~\citep{gil+09_short}
lie in the RR dominated region. AI suppresses RR torquing below the
AI line (gray). Inside the phase-space region delimited by the GW
line (blue), GW dissipation is faster than NR $J$-scattering and
orbits spiral into the MBH by the emission of GW (inspiral track example
in light blue line). The critical sma for EMRIs, $a_{GW}$ (thin black
line), corresponds to the maximum of the GW curve; below it stars
become EMRIs before they cross the LSO\@. The approximate power-law
GW line with the often assumed simplification $\protect\jlc\to0$
(dotted blue line), substantially over-estimates $a_{GW}$. The exact
separatrix streamline (magenta) provides a more accurate estimate
of $a_{GW}$ than either of the timescale-based GW lines. }
\end{figure}

Stars reach the MBH by crossing the LSO loss-line in $(a,j)$ phase-space
(Figure~\ref{f:EJ}) at $\jlc(a)=4/\sqrt{a/r_{g}}$, where $r_{g}=G\Mbh/c^{2}$
(This value of $J_{lc}=J_{c}j_{lc}$ is exact for a zero energy orbit).
In addition, it is useful to define in the statistical sense the locus
of ``no-return'' for GW inspiral (EMRI). Conventionally, this is
defined by a comparison of timescales as the locus where the time
to spiral into the MBH by the emission of GWs, $t_{GW}(a,j)$, is
shorter than the time needed to scatter across the LSO line by NR,
$(j-j_{lc})^{2}T_{J}(a,j)$, where $T_{J}$ is the $J$-diffusion
timescale (Appendix~\ref{a:GWline}). Note that the GW timescale line
shown in Figure~\ref{f:EJ} is calculated both using the common simplification
$t_{GW}=j^{2}T_{J}$, and with the more accurate form $t_{GW}=(j-j_{lc})^{2}T_{J}$,
which yields an arc-like shape that peaks well below the point where
the approximate power-law GW line intersects the LSO line. The maximal
sma along these lines, $a_{GW}$, is then interpreted as the critical
sma for EMRIs, below which phase-space trajectories cannot (statistically)
avoid crossing the GW line and becoming EMRIs. The EMRI rate scales
as $\propto a_{GW}$ (Eq.~\eqref{eq:Ri_SS}). In Section~\ref{s:EMRI-event-rates}
and Appendix~\ref{a:GWline} we formulate a more rigorous criterion
for the GW line by identifying the exact separatrix between phase-space
streamlines that plunge directly into the MBH, and those that inspiral
into it (see Figure~\ref{f:flow2Dall}). This results in an intermediate
value of $a_{GW}$. These three estimates of the GW line are plotted
in Figure~\ref{f:EJ} for reference, and correspond to different EMRI
rate predictions. It should be emphasized that the GW line does not
enter the MC procedure directly, but is an emergent property. Here
we use the separatrix method for analytic rate estimates, which accurately
reproduce the MC results (Figure~\ref{fig:MC-Rp-MBH}).

We derive here analytic estimates for the steady-state distribution
and the flux of stars through the loss-cone, quantify the contribution
of RR to the loss-rates, and validate our estimates by MC simulations.

\subsection{Diffusion equations}

\label{ss:diff-eq}

On long-enough timescales, where relaxation can be described as a
diffusion process~\citep{bar+13,bar+14}, the evolution of the probability
density function, $n\left(E,J,t\right)$, in ($E$, $J$) phase-space
can be describe by an FP equation
\begin{eqnarray}
\frac{\partial n\left(E,J,t\right)}{\partial t} & = & -\frac{\partial S_{E}\left(E,J,t\right)}{\partial E}-\frac{\partial S_{J}\left(E,J,t\right)}{\partial J}\,,\label{eq:FP_EQ}
\end{eqnarray}
where the probability current densities in the $E$ and $J$ ``directions''
are
\begin{eqnarray}
S_{E}\left(E,J,t\right) & = & D_{E}n\left(E,J,t\right)-\frac{1}{2}\frac{\partial}{\partial E}\left[D_{EE}n\left(E,J,t\right)\right]\nonumber \\
 &  & -\frac{1}{2}\frac{\partial}{\partial J}\left[D_{EJ}n\left(E,J,t\right)\right]\,,\label{eq:S_E}
\end{eqnarray}
and
\begin{eqnarray}
S_{J}\left(E,J,t\right) & = & D_{J}n\left(E,J,t\right)-\frac{1}{2}\frac{\partial}{\partial J}\left[D_{JJ}n\left(E,J,t\right)\right]\nonumber \\
 &  & -\frac{1}{2}\frac{\partial}{\partial E}\left[D_{EJ}n\left(E,J,t\right)\right]\,,\label{eq:S_J}
\end{eqnarray}
and where $D_{EE}$, $D_{E}$, $D_{JJ}$, $D_{J}$ and $D_{EJ}$ are
the DCs that describe the combined effect of NR and RR\@.

Integrating over $J$ from $J_{lc}$ to $J_{c}$ we obtain the total
probability current density gradient (loss-rate) per unit energy
\begin{eqnarray}
\frac{\partial n\left(E,t\right)}{\partial t} & = & -\int_{J_{lc}}^{J_{c}}\frac{\partial S_{E}}{\partial E}dJ-S_{J}\left(E,J_{c}\right)+S_{J}\left(E,J_{lc}\right)\,,\nonumber \\
\end{eqnarray}
where $n\left(E\right)=\int_{J_{lc}}^{J_{c}}n\left(E,J\right)dJ$
is energy probability density.

It is convenient to transform these expressions to $(E,j)$ since
for $j=J/J_{c}$, the current density in the $j$-direction is zero
at the boundary ($j=1)$. Generally, under the coordinate change $\mathbf{x}\to\mathbf{x}^{\prime}$
(here $(E,J)\to(E,j)$), the probability density currents, $S_{i}\to S_{i}^{\prime}$,
transform as 
\begin{equation}
S_{i}^{\prime}\left(\mathbf{x}^{\prime}\right)=\left|\frac{\partial\mathbf{x}}{\partial\mathbf{x}^{\prime}}\right|\frac{\partial x_{i}^{\prime}}{\partial x_{k}}S_{k}\left(\mathbf{x}\right)\,.
\end{equation}
Thus
\begin{equation}
S_{j}\left(E,j\right)=S_{J}\left(E,J\right)-j\frac{\partial J_{c}}{\partial E}S_{E}\left(E,J\right)\,,
\end{equation}
and since $S_{j}\left(E,j=1\right)=0$, we have
\begin{eqnarray}
\frac{\partial n\left(E,t\right)}{\partial t} & = & -\int_{J_{lc}}^{J_{c}}\frac{\partial S_{E}}{\partial E}dJ-\frac{\partial J_{c}}{\partial E}S_{E}\left(E,J_{c}\right)+S_{J}\left(E,J_{lc}\right)\,,\nonumber \\
\end{eqnarray}
and
\begin{equation}
\frac{\partial n\left(E,t\right)}{\partial t}=-\frac{\partial}{\partial E}\int_{J_{lc}}^{J_{c}}S_{E}dJ+S_{J}\left(E,J_{lc}\right)\,.
\end{equation}
Using Eq.~\eqref{eq:S_E} and the fact that $D_{EJ}\left(E,J_{c}\right)=D_{EE}\left(E,J_{c}\right)\partial J_{c}/\partial E$
(Appendix~\ref{a:2bDCs}), allow us to obtain the $J$-averaged FP
equation for the energy probability density~$n\left(E\right)$,
\begin{eqnarray}
\frac{\partial n(E)}{\partial t} & = & \frac{1}{2}\frac{\partial^{2}}{\partial E^{2}}\left[\bar{D}_{EE}n\left(E\right)\right]-\frac{\partial}{\partial E}\left[\bar{D}_{E}n\left(E\right)\right]\nonumber \\
 &  & +S_{J}\left(E,J_{lc}\right)\,,\label{eq:SJ(Jlc)}
\end{eqnarray}
in the presence of a loss term $S_{J}\left(E,J_{lc}\right)$, resulting
from the flux of stars through the loss-cone, per unit energy\footnote{This generalizes the simpler situation where stars are only destroyed
once they reach some high energy threshold, where the loss is expressed
instead by a boundary condition~\citep[cf][]{bah+76}.}, with the $J$-averaged diffusion coefficients 
\begin{equation}
\bar{D}_{E}\left(E\right)=n^{-1}(E)\int_{J_{lc}}^{J_{c}}D_{E}\left(E,J\right)n\left(E,J\right)dJ\,,
\end{equation}

\begin{equation}
\bar{D}_{EE}\left(E\right)=n^{-1}(E)\int_{J_{lc}}^{J_{c}}D_{EE}\left(E,J\right)n\left(E,J\right)dJ\,.
\end{equation}

\subsection{Probability flow in phase space}

\label{ss:PSflow}

The effects of the various physical mechanisms are more clearly apparent
in the flow patterns in phase-space. Since the physical flow is stochastic,
it is more useful to describe it in terms of the flow of the probability
density. In steady-state, the FP equation (Eq.~\eqref{eq:FP_EQ})
can be written as a continuity equation of a compressible flow
\begin{equation}
\frac{\partial}{\partial E}\left[n\left(E,J\right)v_{E}\right]+\frac{\partial}{\partial J}\left[n\left(E,J\right)v_{J}\right]=0\,,\label{eq:sl-cont}
\end{equation}
with effective velocities $v_{E}=S_{E}/n\left(E,J\right)$, and $v_{J}=S_{J}/n\left(E,J\right)$.
The two-dimensional flow in phase-space $\mathbf{v}=(v_{J},v_{E})$
can be visualized by the streamlines\footnote{The streamlines are immutable under coordinates transformation and
therefore do not depend on the specific choice of coordinate system.}, $v_{J}/dJ=v_{E}/dE$, which are derived below from the steady-state
solution of the FP equation~\eqref{eq:EJ_ss}.

\begin{figure}
\includegraphics[width=1\columnwidth]{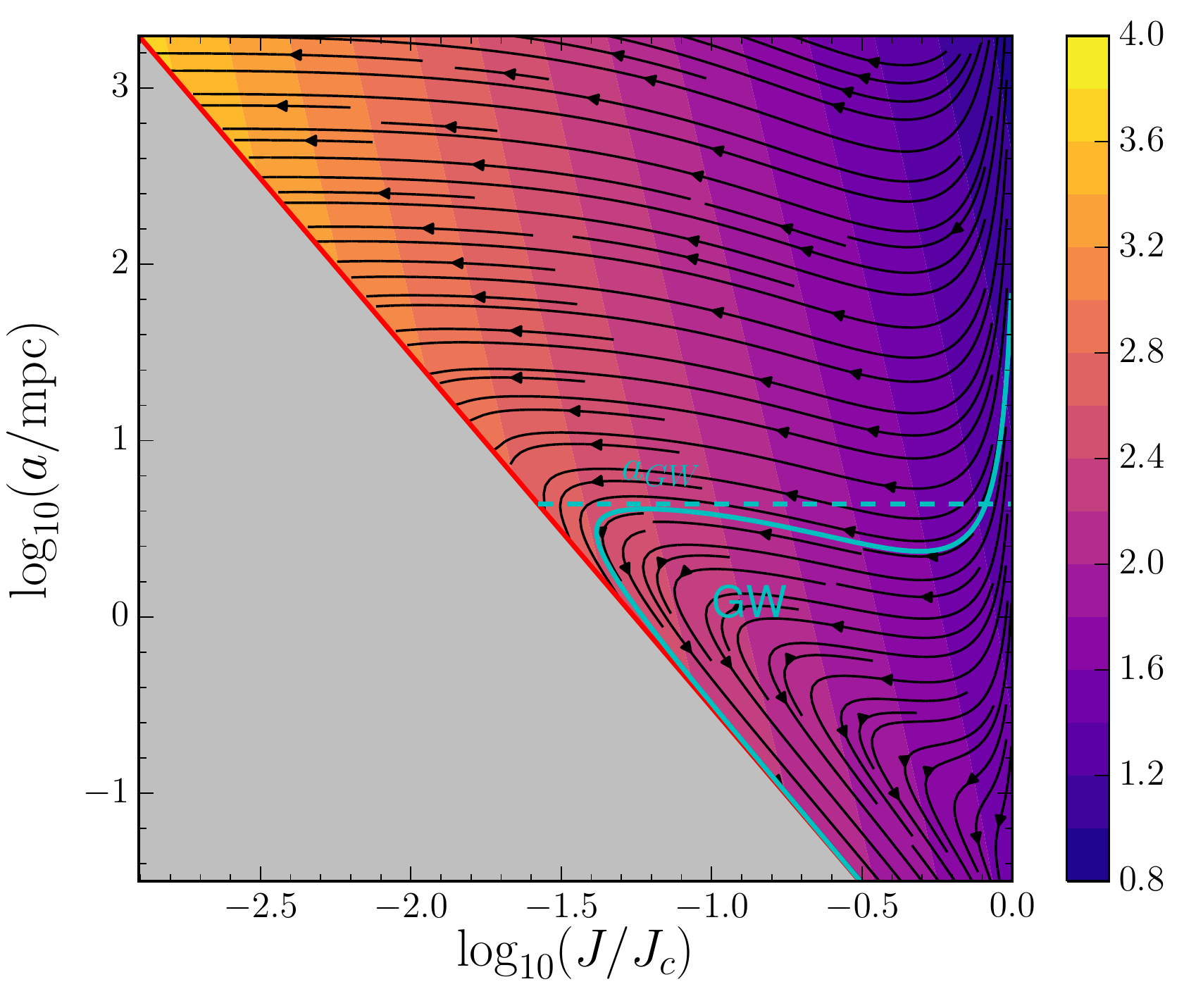}\protect\caption{\label{f:flow2DonlyNR}Streamlines of the phase-space flow $\bar{\mathbf{v}}=(\dot{j}/j,\dot{a}/a)$.
All dynamical effects apart from resonant relaxation, are included
(i.e., two-body relaxation, GW emission, GR precession and mass precession).
The DCs are calculated for a Milky Way-like model (isotropic cusp
$f\left(E\right)\propto E^{1/4}$ with a MBH of $4\times10^{6}M_{\odot}$,
a mass-ratio of $M_{\bullet}/M_{\star}=5\times10^{5}$, and total
stellar mass $M_{\star}\left(r_{h}\right)=2M_{\bullet}$ where $r_{h}=2\mathrm{pc}$).
In addition, the probability current densities are calculated assuming
a DF $f\left(E,J\right)\propto E^{1/4}(2J/J_{c}^{2})\log\left(J/J_{lc}\right)/\log\left(J_{c}/J_{lc}\right)$,
which is the steady-state solution in the presence of a loss-cone.
The color map describes the magnitude of the DCs. The solid cyan line
is the GW separatrix. The dashed cyan line indicates the critical
sma for EMRIs. }
\end{figure}

\begin{figure}
\includegraphics[width=1\columnwidth]{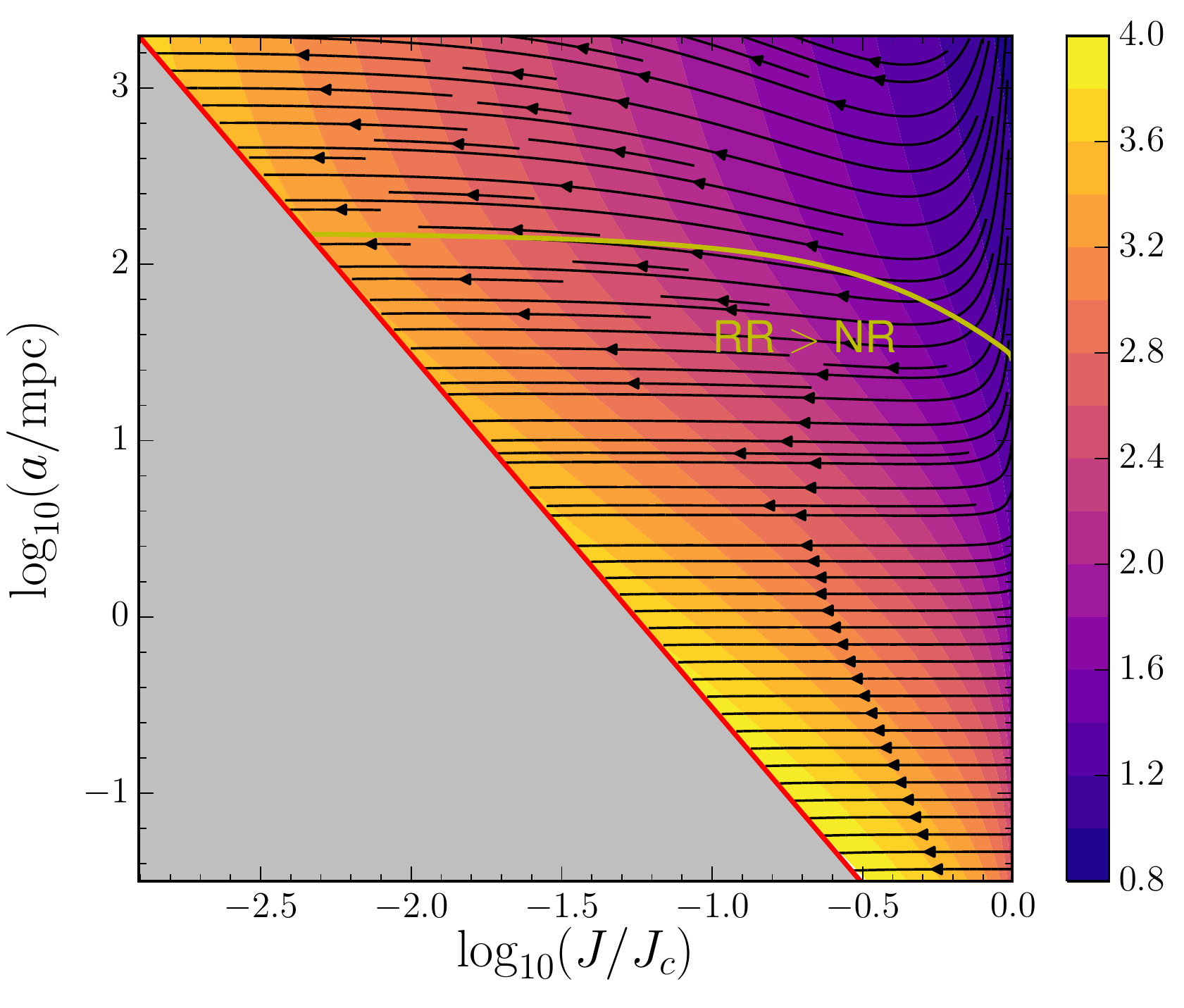}\protect\caption{\label{f:flow2DnoGR}Streamlines of the phase-space flow $\bar{\mathbf{v}}=(\dot{j}/j,\dot{a}/a)$.
All dynamical effects apart from GR precession are included (same
cusp model as in Figure~\ref{f:flow2DonlyNR}). In this case RR dominates
the dynamics all the way to the loss-cone. Note that this scenario
leads to strong depletion of the cusp near the MBH (see for example
Figure~\ref{f:2D} Bottom). This means that in that region our assumption
of a power-law steady-state cusp does not hold. The solid black line
marks the region where RR is effective (stronger than NR). The color
map describes the strength of the DCs.}
\end{figure}

Using the streamlines, we show in Figures~\ref{f:flow2DonlyNR}--\ref{f:flow2DnoGR}
the effects of the various physical mechanisms. The probability current
densities are determined by the distribution function (DF) of the
background stars through the DCs and by DF of the test stars. We begin
by assuming that a relaxed cusp will be approximately isotropic (i.e.,
$f\left(E,J\right)\propto E^{1/4}$)~\citep[BW76]{bah+76}. The existence
of a loss-cone introduces a logarithmic correction, so that the DF
is of the form $f\left(E,J\right)\propto E^{1/4}\log\left(J/J_{lc}\right)/\log\left(J_{c}/J_{lc}\right)$
(see Eq.~\eqref{eq:EJ_ss}, Figures~\ref{fig:dNx},~\ref{fig:dNj}
and~\citealt{hop+05}). Since the DCs are not strongly affected by
this small anisoptropy, it is convenient to assume that the DF of
the background is isotropic. Therefore, in the calculation of the
streamlines we use a DF of the form $f\left(E,J\right)\propto E^{1/4}$
for the background and a DF of the form $f\left(E,J\right)\propto E^{1/4}\log\left(J/J_{lc}\right)/\log\left(J_{c}/J_{lc}\right)$
for the test stars.

\begin{figure}
\includegraphics[width=1\columnwidth]{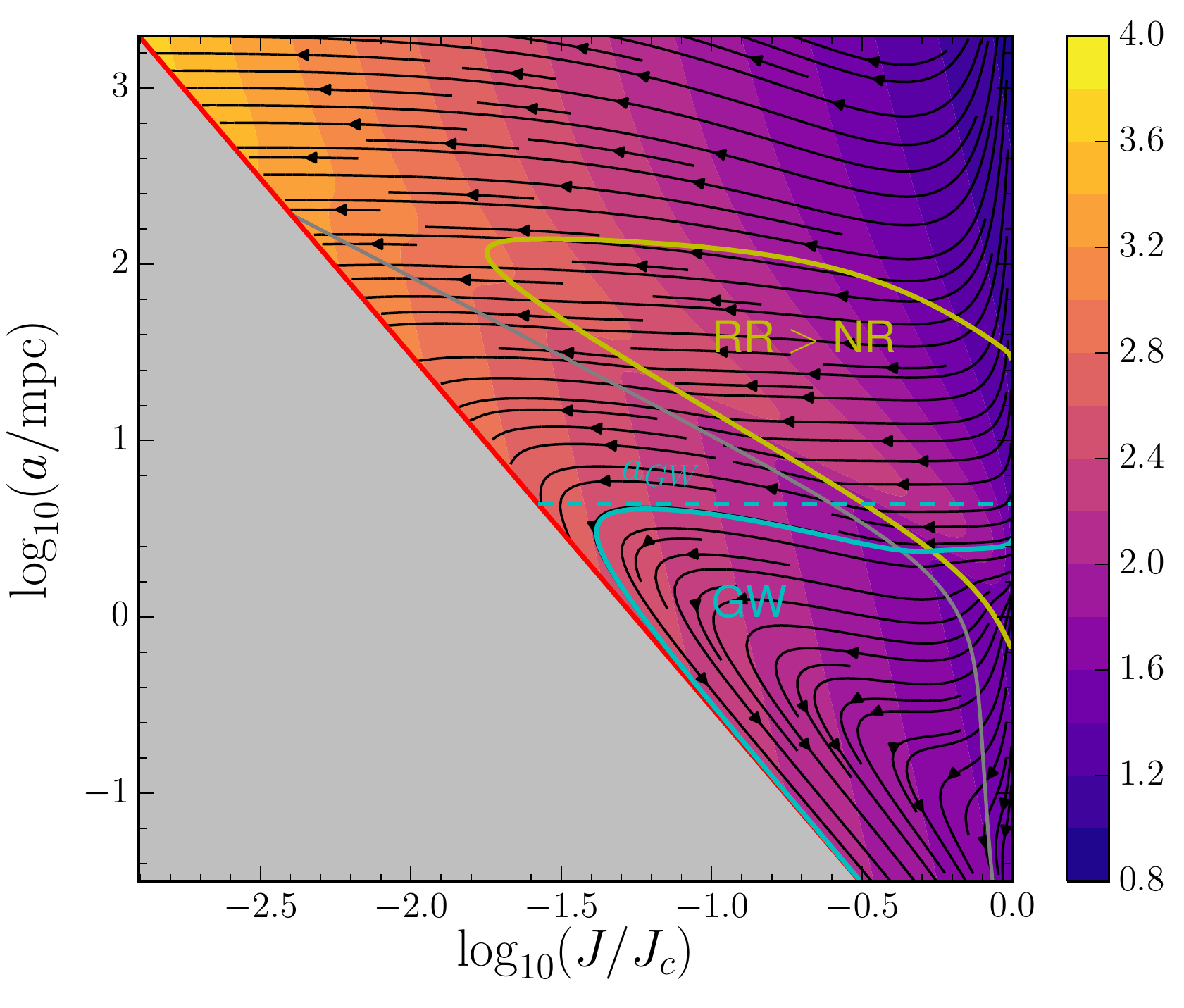}\protect\caption{\label{f:flow2Dall}Streamlines of the phase-space flow $\bar{\mathbf{v}}=(\dot{j}/j,\dot{a}/a)$.
All dynamical effects included (same cusp model as in Figure~\ref{f:flow2DonlyNR}).
The solid black line marks the region where RR is effective (stronger
than NR). The gray line is the locus beyond which RR is totally ineffective
due to adiabatic invariance. The color map describes the strength
of the DCs. The solid cyan line is the GW separatrix. The dashed cyan
line indicates the critical sma for EMRIs. }
\end{figure}

The flow at a point in phase-space is considered $j$-dominated when
the streamlines are approximately horizontal (i.e., $\left|\dot{j}\right|/j\gg\left|\dot{a}\right|/a$)
and $a$-dominated when the streamlines are approximately vertical
(i.e., $\left|\dot{j}\right|/j\ll\left|\dot{a}\right|/a$). As shown
in Figures~\ref{f:flow2DonlyNR}--\ref{f:flow2Dall}, the flow is
$j$-dominated, apart for two restricted regions in phase-space ($J\to J_{c}$
and the GW-dominated region). Therefore, the full flow field can be
separated into two one-dimensional flows, a fact that will be used
below to simplify the analytic treatment. Since RR drives stars only
in the $j$-direction, this separation is enhanced in the phase-space
region where $D_{jj}^{RR}(a,j)>D_{jj}^{NR}(a,j)$, and RR governs
the dynamics (see Figures~\ref{f:flow2Dall}--\ref{f:flow2DnoGR}).

As shown in Figure~\ref{f:flow2DnoGR}, in the absence of GR precession
(i.e., mass precession-only) interior to some sma, RR dominates the
dynamics all the way to the loss-cone. In this hypothetical case,
the loss rate could be high enough to actually empty the cusp close
to the MBH\@. This would then invalidate the assumption of a single
power-low cusp. However, in realty, GR precession does play a crucial
role in determining the dynamics and the steady-state. In fact, due
to GR precession, RR is totally quenched by the adiabatic invariance
(AI) of the angular momentum. This happens for stars with angular
momenta below the locus where the precession frequency $\omega_{p}(a,j)=|\omega_{GR}+\omega_{M}|$
equals the coherence frequency $2\pi/T_{c}$, where $\omega_{GR}$
and $\omega_{M}$ are the GR and mass precession frequencies (see
AI curve in Figure~\ref{f:flow2Dall}). Only above the AI line can
RR be effective. Figure~\ref{f:flow2Dall} shows a closed contour,
somewhat above the AI line, where RR is faster than NR and therefore
dominates the dynamics. As we show in Section~\ref{s:RR-effect},
the fact the RR is not effective near the loss-lines means that RR
does \emph{not} play an important role in setting the steady-state
and the loss-rates.

\subsection{Steady state distribution and loss-cone flux}

\label{ss:ss-flux}

\begin{figure}
\centering{}\includegraphics[width=1\columnwidth]{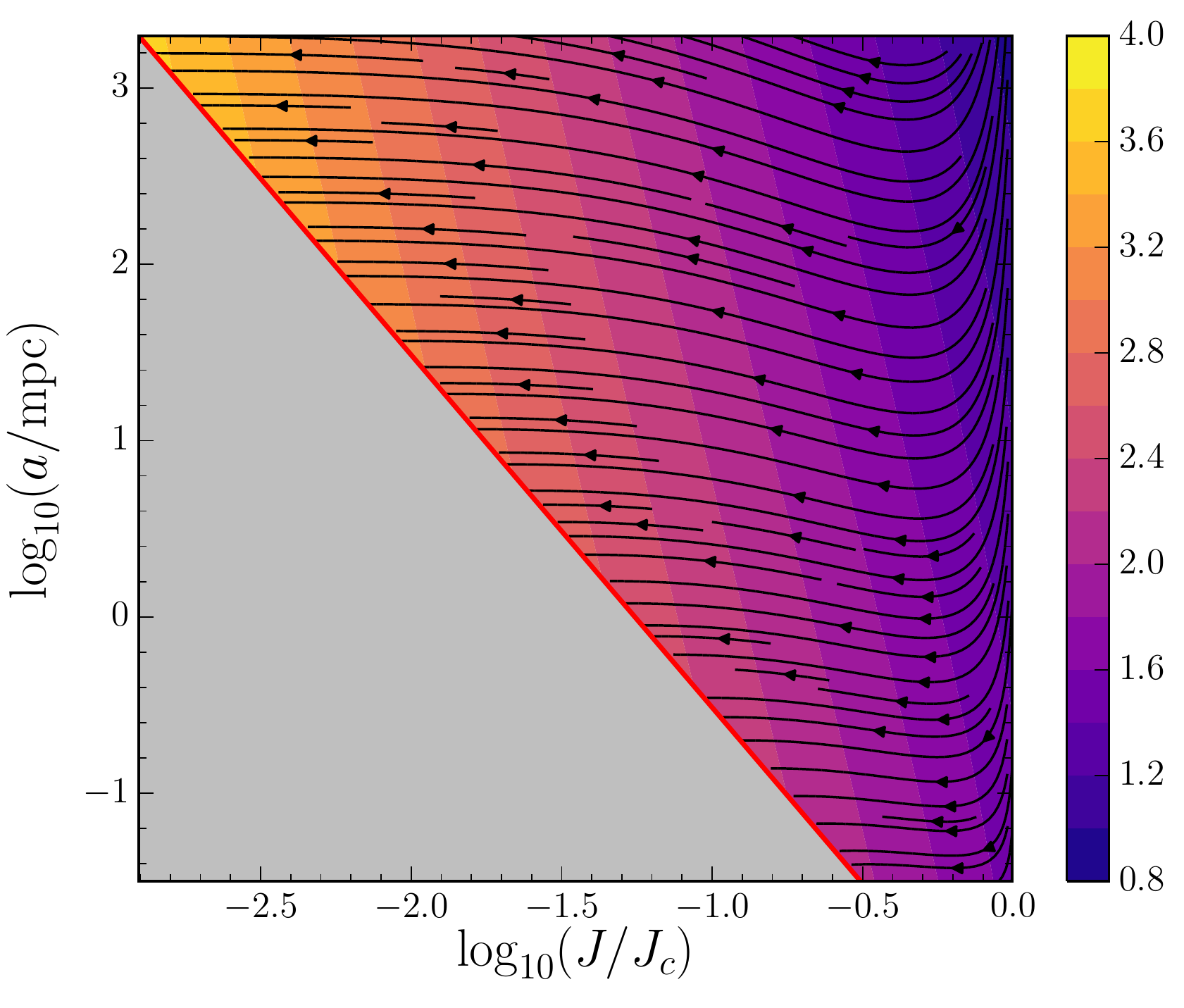}\protect\caption{\label{fig:Streamlines-NR-only}Streamlines of the phase-space flow
$\bar{\mathbf{v}}=(\dot{j}/j,\dot{a}/a)$. Resonant relaxation and
GW emissions are not included (same cusp model as in Figure~\ref{f:flow2DonlyNR}).
The color map describes the strength of the DCs.}
\end{figure}

We assume that the system relaxes in $J$ much faster than it relaxes
in $E$. This assumption can be justified by noting that $J\in[J_{lc},J_{c}(E)]$
is bound, whereas $E$ is unbound. We therefore assume that the relaxation
process is separable: on short timescales stars exchange only angular
momentum but not energy and reach their steady-state $J$-distribution
at fixed $E$, and only on a much longer timescale do they reach global
steady-state in $E$. This assumption of local equilibrium (i.e.,
in each energy bin the $J$-distribution is relaxed) is further supported
by the pattern of the probability current densities (Eqs.~\eqref{eq:S_E},~\eqref{eq:S_J}) described by the streamlines shown in Figure~\ref{fig:Streamlines-NR-only}
for the NR-only case. The inclusion of RR only strengthens this separability.
This demonstrates that the motion in the $E$-direction (a-direction)
occurs only at $j\to1$, whereas it is almost entirely in the $j$
direction at $j<1$. The validity of this assumption is verified by
the excellent match between our analytic predictions and the result
of MC simulations, which do not assume separability a-priori, as shown
in Figures~\ref{fig:dNx}--\ref{fig:RP_a}. In this section we use
this separability assumption and the fluctuation-dissipation relation
(Appendix~\ref{a:MaxEntropy}) to derive the steady-sate state (E,J)
distribution and the flux loss-cone flux. 

In the limit where there is no energy exchange between stars, the
FP equation can written as~\citep{bar+14}
\begin{eqnarray}
\frac{\partial n(E,J,t)}{\partial t} & = & \frac{1}{2}\frac{\partial}{\partial J}\left\{ JD_{JJ}(E,J)\frac{\partial}{\partial J}\left[\frac{1}{J}n(E,J,t)\right]\right\} \nonumber \\
 & = & -\frac{\partial S_{J}(E,J,t)}{\partial J}\,,\label{e:J_FP}
\end{eqnarray}
which generally follows from the maximum entropy principle and in
fact provides a necessary test for the validity of the DCs (Appendix~\ref{a:MaxEntropy}). In the absence of a loss-cone (i.e., $J_{lc}\to0$),
the steady-state probability current density $S_{J}\left(E,J,t\right)$
is zero, and the local equilibrium distribution is isotropic 
\begin{equation}
n\left(E,J\right)=2J/J_{c}^{2}(E)n(E)\,.
\end{equation}
For a finite $J_{lc}$, it follows from the separability assumption,
that the probability current density is non-zero and independent of
$J$. Therefore, from Eq.~\eqref{e:J_FP} we obtain 
\begin{equation}
JD_{JJ}(E,J)\frac{\partial}{\partial J}\left[\frac{1}{J}n(E,J,t)\right]=-2S_{J}\left(E\right)\,.
\end{equation}
By integrating over $J$ and using the normalization $n\left(E\right)=\int_{J_{lc}}^{J_{c}}n(E,J)dJ$,
we obtain
\begin{equation}
S_{J}\left(E\right)=-n\left(E\right)\left/\int_{j_{lc}}^{1}\frac{1-j^{2}}{d_{jj}(E,j)}\frac{dj}{j}\right.\,,\label{eq:S_J_djj}
\end{equation}
and
\begin{eqnarray}
n\left(E,J\right) & = & \frac{2J}{J_{c}^{2}}S_{J}\left(E\right)\int_{j_{lc}}^{j}\frac{1}{j^{\prime}d_{jj}(E,j^{\prime})}dj^{\prime},\label{eq:local_eq}
\end{eqnarray}
were $d_{jj}\left(E,j\right)=D_{JJ}\left(E,J\right)/J_{c}^{2}$.

Once the system achieves local equilibrium in $J$ at any $E$ (Eq.~\eqref{eq:local_eq}), the subsequent steady-state in $E$ is obtained
by solving Eq.~\eqref{eq:SJ(Jlc)} for $n(E)$, given $S_{J}(E)$,
\begin{equation}
\frac{dR_{P}}{dE}=-S_{J}\left(E\right)=\frac{\partial}{\partial E}\left\{ \frac{1}{2}\frac{\partial}{\partial E}\left[\bar{D}_{EE}n\left(E\right)\right]-\bar{D}_{E}n\left(E\right)\right\} \,,\label{eq:steady_state_FP}
\end{equation}
where $R_{p}\left(E\right)=-\int_{J_{lc}}^{J_{c}}S_{E}\left(E,J\right)dJ$,
is the cumulative number of stars lost through the loss-cone per unit
time. Note that in steady-state, the probability current density in
the $J$ direction equals the probability current density gradient
in $E$ (from continuity considerations: the density carried by the
$J$-current at fixed $E$ and lost through $j_{lc}$ is balanced
by the $E$-gradient of the total $E$-current).

\subsection{Steady state distribution for two body relaxation}

\label{ss:SSNR}

We now show that in the case of two-body relaxation, the solution
of the energy FP equation (Eq.~\eqref{eq:steady_state_FP}) with non-zero
flux can be approximated analytically to derive the steady-state density
distribution and the plunge rate. Since the plunge rate is small compared
to the relaxation rate, the energy distribution asymptotes to the
zero-flux (i.e., no plunges)~\citet{bah+76} power-law solution. 

For two-body relaxation, $d_{jj}^{NR}\left(E,j\right)$ asymptotes
to a finite value $d_{NR}^{0}\left(E\right)=d_{jj}^{NR}\left(E,j=0\right)$
as $j\to0$\footnote{Since $d_{jj}^{NR}=2jd_{j}^{NR}$ for $j\to0$~(\citealp[e.g.,][]{sha+78};
Appendix~\ref{a:2bDCs}) and $2jd_{j}^{NR}=\partial_{j}jd_{jj}^{NR}$
(Appendix~\ref{a:MaxEntropy}).}. Since most of the contribution to the current density, $S_{J}\left(E\right)$,
reflects the value of $d_{jj}$ at small $j$ (Eq.~\eqref{eq:S_J_djj}),
it can approximated by $d_{jj}^{0}(E)$, so that 
\begin{eqnarray}
S_{J}\left(E\right) & \approx & -n\left(E\right)\frac{2d_{NR}^{0}(E)}{\log\left(J_{c}^{2}/J_{lc}^{2}\right)-1+J_{lc}^{2}/J_{c}^{2}}\nonumber \\
 & \approx & -n\left(E\right)\frac{d_{NR}^{0}(E)}{\log\left(J_{c}/J_{lc}\right)}\,,\label{eq:SJ_nE}
\end{eqnarray}
and
\begin{eqnarray}
n\left(E,J\right) & \approx & n\left(E\right)\frac{2J}{J_{c}^{2}}\frac{\log\left(J^{2}/J_{lc}^{2}\right)}{\log\left(J_{c}^{2}/J_{lc}^{2}\right)-1+j_{lc}^{2}}\nonumber \\
 & \approx & n\left(E\right)\frac{2J}{J_{c}^{2}}\frac{\log\left(J^{2}/J_{lc}^{2}\right)}{\log\left(J_{c}^{2}/J_{lc}^{2}\right)}\,.\label{eq:nEJ_nE}
\end{eqnarray}
Since $d_{jj}^{NR}$ scales as $\bar{D}_{EE}/E^{2}=T_{E}^{-1}$ (Appendix~\ref{a:2bDCs}), it is convenient to represent $S_{J}(E)$ explicitly
in terms of the energy relaxation time, $T_{E}$, as $S_{J}(E)=-n\left(E\right)\chi(E)/T_{E}(E)$,
where 
\begin{eqnarray}
\chi(E) & = & d_{NR}^{0}(E)/\log\left(J_{c}/J_{lc}\right)=2d_{NR}^{0}(E)/\log\left(E_{lc}/E\right)\,,\nonumber \\
\end{eqnarray}
expresses the logarithmic suppression of the flux due to the decreasing
size of the loss-cone away from the MBH, and where $E_{lc}\equiv G\Mbh/32r_{g}$
corresponds to the limit $J_{c}=J_{lc}$ for $J_{lc}=4r_{g}c$ and
for Keplerian energy. Note that this is not the true innermost stable
circular orbit, but rather a formal extrapolation of the approximations
adopted here, used for normalization only. Here we are interested
in stars with $E\gg E_{lc}$ where $1/\log\left(E_{lc}/E\right)$
is small. The cumulative plunge rate $R_{P}=-\int S_{J}dE$ can then
be approximated as
\begin{eqnarray}
R_{p} & \approx & \frac{1}{2}N\left(E\right)\frac{\chi\left(E\right)}{T_{E}}+\frac{3}{4}\int_{E}^{E_{\max}}\frac{\chi\left(E^{\prime}\right)}{T_{E}\left(E^{\prime}\right)}N\left(E^{\prime}\right)\frac{dE^{\prime}}{E^{\prime}}\,,\nonumber \\
\end{eqnarray}
where $N\left(E\right)$ is the number of stars with energy larger
than $E$.

For an infinite isotropic cusp $n\left(E\right)\propto E^{p-5/2}$
where $0<p<1/2$ (to ensure the DCs are finite), the $J$-averaged
DCs are~\citep{bar+13}
\begin{eqnarray}
\bar{D}_{E} & = & -\frac{4(1-4p)\log\Lambda}{1+p}\frac{N(E)}{Q^{2}P(E)}E\propto E^{p+1},\\
\bar{D}_{EE} & = & \frac{16\left(3-2p\right)\log\Lambda}{(1+p)(1-2p)}\frac{N(E)}{Q^{2}P(E)}E^{2}\propto E^{p+2}\,,
\end{eqnarray}
and the plunge rate is
\begin{equation}
R_{p}\approx\frac{3-2p}{3-4p}N\left(E\right)\frac{\chi\left(E\right)}{T_{E}}\,.
\end{equation}
Since $\chi$ is nearly constant in $E$ for $E\ll E_{lc}$, it can
be approximated in that limit by evaluating it at $E_{\min}=GM_{\bullet}/2a_{\max}$.
In that case, Eq.~\eqref{eq:steady_state_FP} has an analytical solution,
\begin{equation}
\frac{\left(4p-3\right)\left(1-4p\right)}{4\left(3-2p\right)}=\chi\,,
\end{equation}
which connects the current to the power-law exponent of the cusp.
The physical branch of the solution is\footnote{This generalizes the analytic BW76 solution ($p=1/4$,$\chi=0$),
which applies for a power-law DF in steady-state with a constant $E$-current
(which then must be zero). Here, the ``leakage'' of stars through
the loss-cone at all $E$ ($\chi>0$) implies a non-constant current,
which allows flatter cusp solutions with $p<1/4$.} 
\begin{eqnarray}
p & = & \frac{1}{4}\left(-\sqrt{\chi^{2}+8\chi+1}-\chi+2\right)\approx\frac{1}{4}\left(1-5\chi\right)\,.
\end{eqnarray}
Since for $p=1/4$, $d_{jj}^{0}\approx1/\left(8T_{E}\right)$ (Appendix~\ref{a:2bDCs}), it follows that $\chi\approx1/\left[8\log\left(1/j_{lc}\right)\right]\ll1$,
and so the~\citet{bah+76} solution $p=1/4$ (i.e., $\chi\to0$) is
a reasonable approximation in the $E\to E_{\min}$ limit, where $j_{lc}\ll1$.
Thus in steady-state the energy (or sma) distribution is $n\left(E\right)\propto E^{-9/4}$
(or $n\left(a\right)\propto a^{1/4}$). Using Eq.~\eqref{eq:nEJ_nE}
we obtain the ($E,J$) steady-state distribution 
\begin{eqnarray}
n\left(E,J\right) & \approx & \frac{5}{4}N\left(E_{\min}\right)\frac{2J}{J_{c}^{2}}\frac{\log\left(J/J_{lc}\right)}{\log\left(J_{c}/J_{lc}\right)}\left(E/E_{\min}\right)^{-9/4}\,,\nonumber \\
\label{eq:EJ_ss}
\end{eqnarray}
and a steady-state plunge rate 
\begin{eqnarray}
R_{p}\left(a\right) & \approx & \frac{5}{32}\frac{1}{\log\left(1/j_{lc}\left(a\right)\right)}\frac{N\left(a\right)}{T_{E}\left(a\right)}\,,\label{eq:Rp_SS}
\end{eqnarray}
where the energy relaxation time is~\citep{bar+13}
\begin{equation}
T_{E}\left(a\right)=\frac{1}{64}Q^{2}\frac{P\left(a\right)}{N\left(a\right)\log Q}\,.
\end{equation}

\begin{figure}
\includegraphics[width=1\columnwidth]{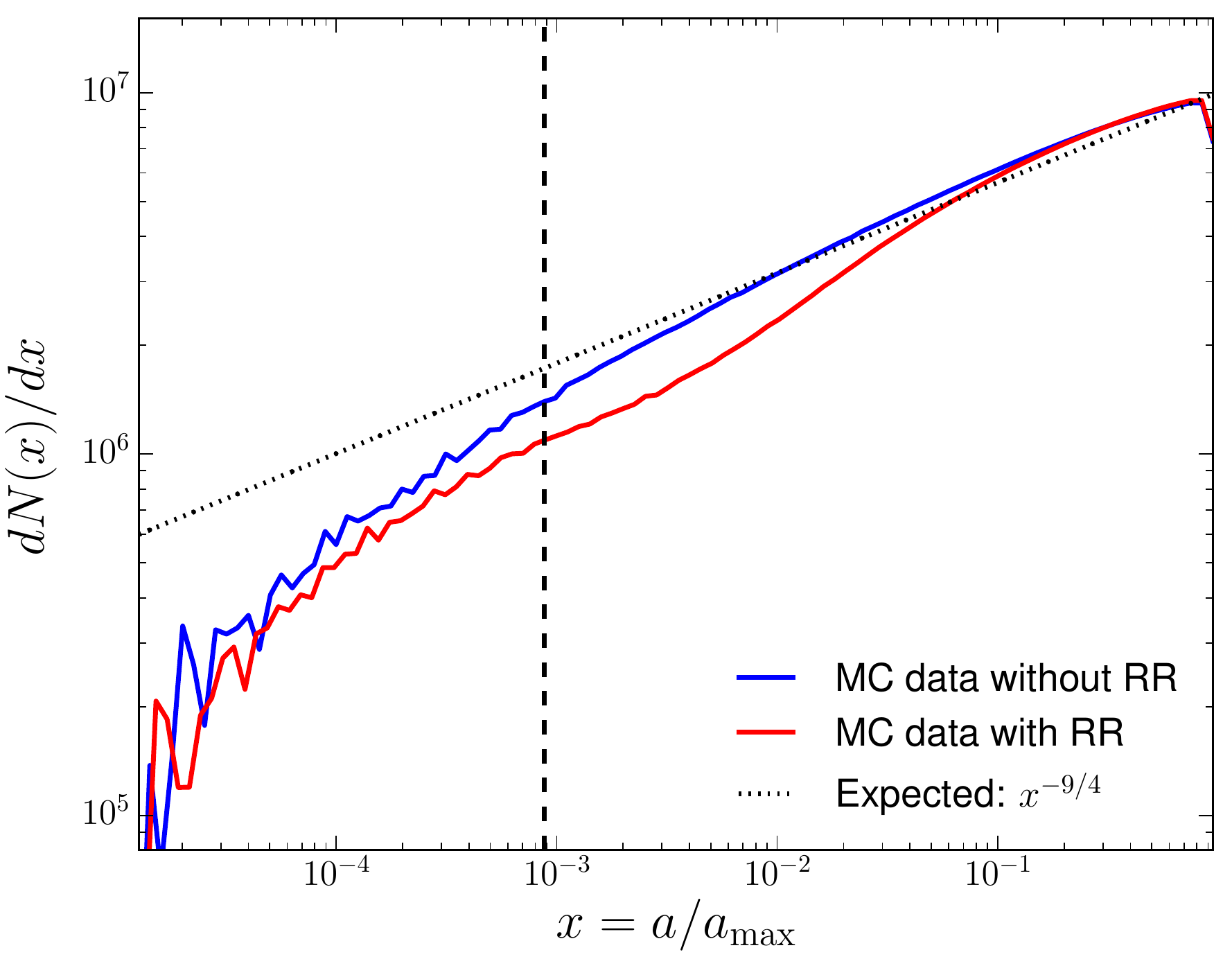}\protect\caption{\label{fig:dNx}The energy distribution as function of energy shows
a good agreement between the analytic BW76 cusp solution and MC results.
The critical sma, $a_{GW}(E_{GW})$ as defined by the sparatrix (see
Section~\ref{s:EMRI-event-rates}) is shown by a dashed line.}
\end{figure}

\begin{figure}
\includegraphics[width=1\columnwidth]{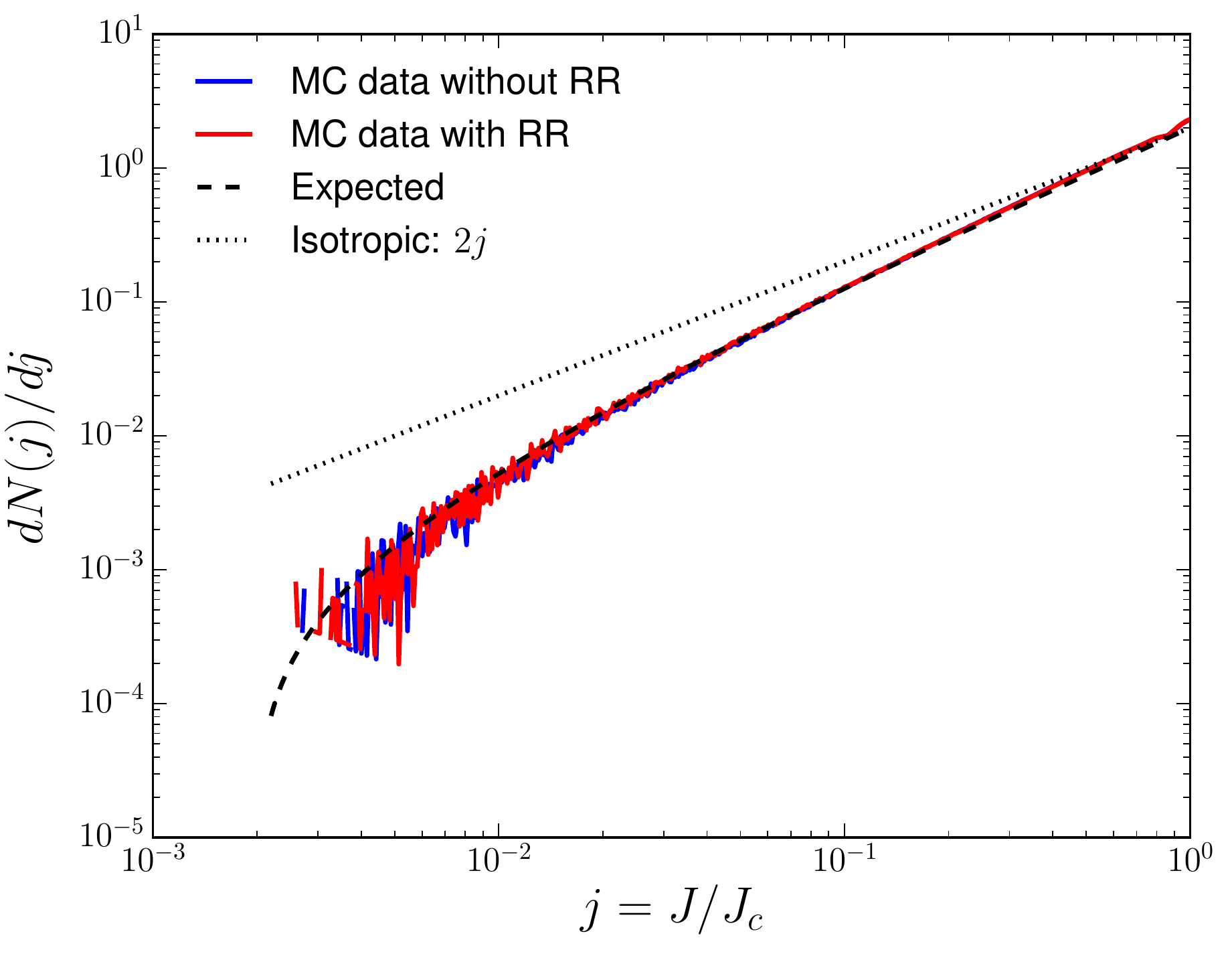}\protect\caption{\label{fig:dNj}The angular momentum distribution in a bin centered
around $a=0.4a_{\max}$. The MC data shows the expected logarithmically-suppressed
distribution (Eq.~\eqref{eq:EJ_ss}) compared with the isotropic one.}
\end{figure}

\begin{figure}
\includegraphics[width=1\columnwidth]{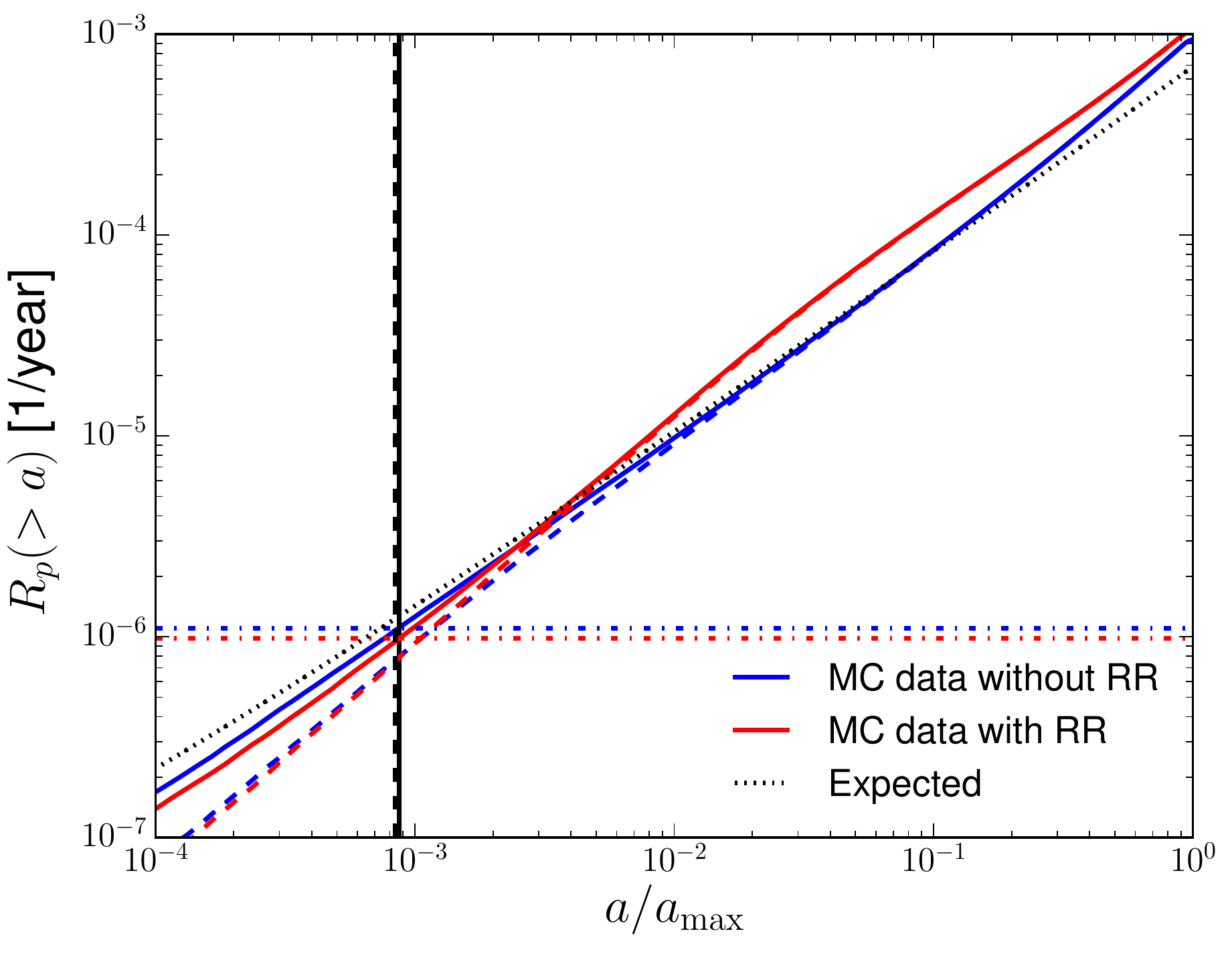}\protect\caption{The plunge rate as a function of semi-major axis. The MC data (without
RR) agree with the analytic expression (Eq.~\eqref{eq:RpMW}). The
MC results demonstrate that the contribution of RR to the plunge rate
is small. The critical sma, $a_{GW}$ (vertical dashed line) as defined
by the separatrix (see~\ref{s:EMRI-event-rates}) agrees very well
with the sma (vertical solid line) where the inspiral rate, $R_{i}^{tot}$
(dash-doted lines), equals the plunge rate $R_{p}^{no\,GW}$ in the
absence of GW emission.\label{fig:RP_a}}
\end{figure}

In the limit $E\to E_{lc}$, $\chi(E)$ can no longer be approximated
as fixed nor as small, and the power law solution breaks down. We
now argue that this power-law approximation is valid for almost the
entire range of MBH masses and their host galactic nuclei. Define
$a_{BW}$ as the minimal sma where $\chi$ and $\partial\log\chi/\partial\log E$
are still small enough for this approximation to hold. Since $\partial\log\chi/\partial\log E\propto\chi\propto1/\log\left(1/j_{lc}\right)$
where $j_{lc}\propto\sqrt{r_{g}/a}$, it follows that $a_{BW}\propto M_{\bullet}$.
We are interested to resolve the dynamics close to the MBH in order
to obtain a reliable estimate for the EMRI event rate. As we show
in Section~\ref{s:EMRI-event-rates}, the rate is determined by the
dynamics near the critical GW sma, $a_{GW}$. Thus it is sufficient
do show that $a_{BW}<a_{GW}$, since then power-law approximation
is valid over the entire range of relevant radii. The MC results shown
in Figures~\ref{fig:dNx} and~\ref{fig:RP_a} for the case of $M_{\bullet}=4\times10^{6}M_{\odot}$,
demonstrate that our analytic estimates (Eqs.~\eqref{eq:EJ_ss},~\eqref{eq:Rp_SS})
for the steady-state energy distribution and plunge rate reproduce
the simulated results at least down to $a_{GW}$. This holds also
for more massive MBHs. The $M/\sigma$ relation and the scalings $a_{BW}\propto\Mbh$
and $a_{GW}\propto\left(\log Q\right)^{-4/5}r_{h}$ (Eq.~\eqref{eq:aGW_approx})
 imply $a_{BW}/a_{GW}\propto\left(\log Q\right)^{4/5}M_{\bullet}^{2/\beta}$
and therefore $a_{BW}\lesssim a_{GW}$ up to $M_{\bullet}\sim10^{9}M_{\odot}$.


\subsection{EMRI event rates}

\label{s:EMRI-event-rates}

So far we ignored the contribution of GW emission to the dynamics.
Compact objects can withstand the tidal field of the MBH\@. When
on eccentric orbits, their orbital decay by the emission of GWs can
be faster than the diffusion of angular momentum due to the stochastic
perturbations of the stellar background. In that case, they inspiral
gradually all the way down to the innermost stable circular orbit
(ISCO) as EMRIs~\citep{pet+63,pet64,gai+06}, instead of plunging
directly into the MBH with $J<J_{lc}$ (Figure~\ref{f:EJ}). The GW
signature of plunges and inspirals is very different. The low mass
of the compact objects generates weak signals, well below the noise.
Plunges result in short, very hard to detect broad spectrum GW flares.
In contrast, EMRIs are of special interest since their quasi-periodic
signal can be integrated and detected against the noise if the waveform
is approximately known. 

In the absence of GWs (Figure~\ref{fig:Streamlines-NR-only}), the
streamlines are approximately constant in $a$. In contrast, GW emission
diverts the streamlines to tracks that are almost parallel to the
loss-cone (i.e., nearly constant $J$) in the phase-space region where
GW dominates the dynamics (Figure~\ref{f:flow2DonlyNR}). The outermost
inspiraling streamline separates phase-space into two distinct regions.
Above this separatrix all streamlines are plunges, while below it
all streamlines are inspirals (Figure~\ref{f:flow2Dall}). The continuity
equation (Eq.~\eqref{eq:sl-cont}) implies that the probability current
in steady-state is constant along a streamline bundle. Since the streamlines
in the GW-dominated region below the separatrix originate in phase-space
regions where the density is much higher, the small depletion due
to EMRI losses is not expected to affect the density at the origin
of the streamlines. The EMRI rate can therefore be estimated by identifying
the terminal point ($a_{p},j_{lc}$) of the plunge streamline (without
GW) corresponding to the separatrix. This is the effective critical
sma for EMRIs, $a_{GW}$. The EMRI rate is then obtained by integrating
the differential plunge rate in the absence of GW emission from $a_{isco}$
to $a_{GW}$, 

\begin{equation}
R_{i}^{tot}=-\int_{E_{GW}}^{E_{isco}}S_{J}\left(E\right)dE=R_{p}\left(a_{GW}\right)\,.
\end{equation}

\begin{figure}
\begin{centering}
\includegraphics[width=1\columnwidth]{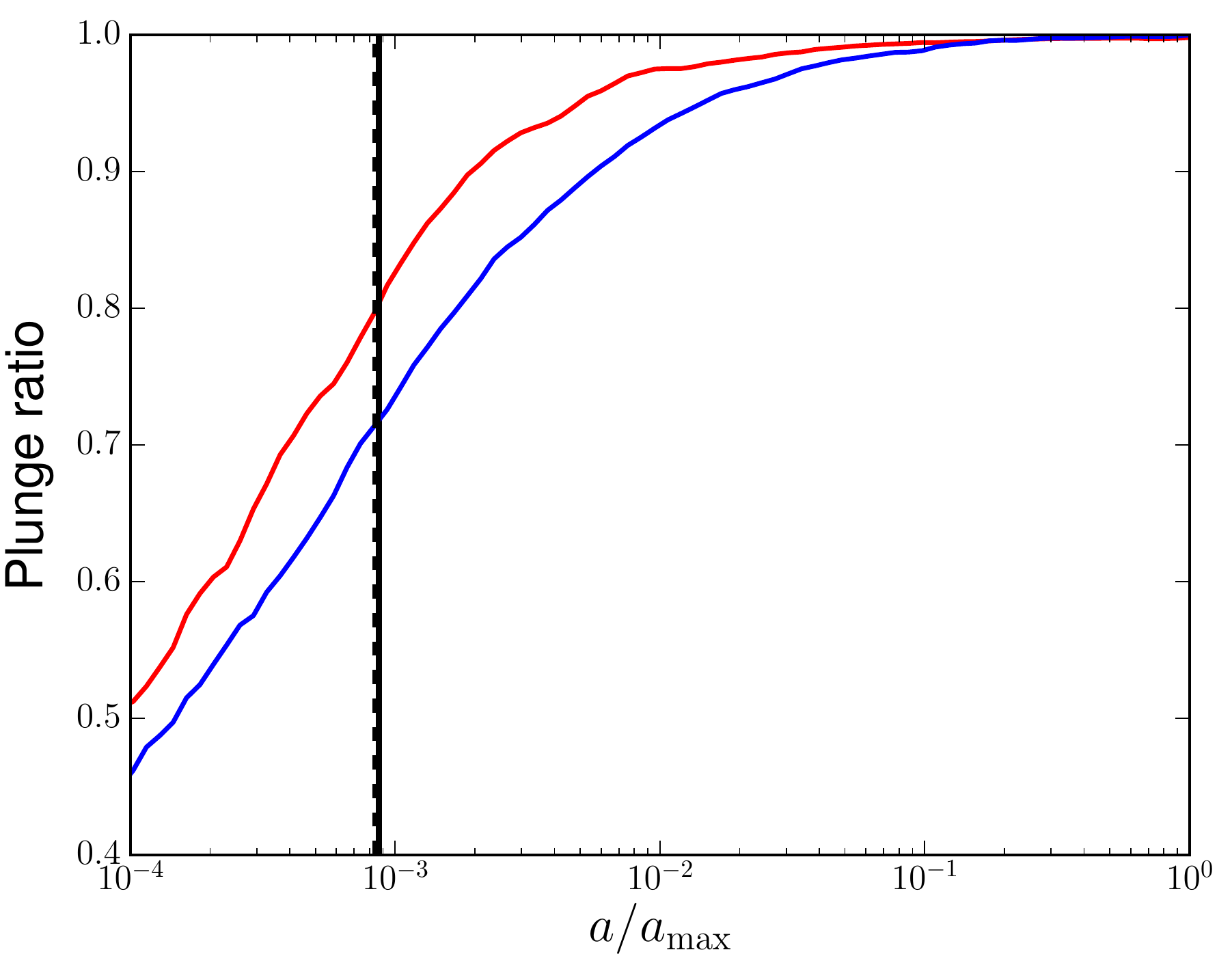}
\par\end{centering}

\protect\caption{The plunge ratio $R_{p}^{GW}/R_{p}^{no\,GW}$, where $R_{p}^{GW}$
and $R_{p}^{no\,GW}$ are the cumulative plunge rates obtained with
and without GW dissipation. The critical sma, $a_{GW}$ (vertical
dashed line) as defined by the separatrix (see~\ref{s:EMRI-event-rates})
agrees very well with the sma (vertical solid line) where the inspiral
rate, $R_{i}^{tot}$, equals the plunge rate $R_{p}^{no\,GW}$ in
the absence of GW emission. \label{fig:Plunge-ratio}}
\end{figure}

Thus, for a BW76 cusp the EMRI rate is
\begin{eqnarray}
R_{i}^{tot} & = & \frac{5}{32}\frac{1}{\log\left(1/j_{lc}\left(a_{GW}\right)\right)}\frac{N\left(a_{GW}\right)}{T_{E}\left(a_{GW}\right)}\nonumber \\
 & = & \frac{a_{GW}}{a_{\max}}\frac{\log\left(1/j_{lc}\left(a_{\max}\right)\right)}{\log\left(1/j_{lc}\left(a_{GW}\right)\right)}R_{p}^{no\,GW}\left(a_{\max}\right)\,,\label{eq:Ri_SS}
\end{eqnarray}
which is approximately linear in $a_{GW}$. The value of $a_{GW}$
is determined by solving the streamline equation $dE/dJ=S_{E}/S_{J}$
with the boundary condition that the streamline trajectory reaches
$J_{lc}$ at $E_{\max}$. The probability current densities are estimated
by assuming that $S_{J}$ is constant in $J$ and $S_{E}$ results
only from GW dissipation. This means that in the absence of GWs, that
streamline is constant in $E$ and $a_{GW}$ can be estimated by taking
the value of $a$ at $J\gg J_{lc}$. The exact value of $a_{GW}$
depends on the GW emission approximation used and is calculated in
Appendix~\ref{a:GWline}. As shown in Figure~\ref{fig:RP_a}, this
definition of $a_{GW}$ is indeed a good approximation to the sma
where the plunge rate (without GW) is equal to the inspiral rate and
can be used to predict the inspiral rate in the MC simulations (Figure~\ref{fig:MC-Rp-MBH}). As expected, inside $a_{GW}$ the plunge rate
with GW decreases relative to the plunge rate without GW (See Figure~\ref{fig:Plunge-ratio}). 

\subsection{Effect of Resonant Relaxation\label{s:RR-effect}}

Due to the long coherence time, resonant relaxation is a much more
effective process than two-body relaxation. However, in regions of
phase-space where in-plane GR precession is faster than the coherence
time, $j$ becomes an adiabatic invariant and the RR process is quenched.
RR it is therefore limited to a small region of phase-space (see Figure~\ref{f:flow2Dall}). The locus where in-plane precession quenches
RR by AI (Section~\ref{ss:1DMC}) defines the outer envelope of the
region where RR may be efficient relative to NR\@. The region where
RR dominates the dynamics even on long timescales is where by the
ratio of the 2nd order DCs\footnote{The transformation of the DC from $J$ to $j=J/J_{c}$ is $D_{jj,NR}=D_{JJ,NR}/J_{c}^{2}+(j^{2}/4)D_{EE}/E^{2}+jD_{EJ}/J_{c}E$
(Appendix~\ref{a:2bDCs}).} exceeds unity, i.e., $D_{jj,RR}/D_{jj,NR}>1$ (See Figure~\ref{f:flow2Dall}).
The typical phase-space configuration is shown in Figure~\ref{f:EJ}:
the region where RR dominates is detached from the loss-lines; NR
is required for the stars to evolve towards them, and therefore the
slow NR timescale remains the bottleneck for the loss-rates which
are mostly unaffected by RR (see Figure~\ref{fig:MC-Rp-MBH}). This
can be shown formally by re-estimating the probability current density
in the presence of RR\@.

For a smooth noise, the RR diffusion coefficient can be written as~\citep{bar+14}
\begin{equation}
D_{JJ}^{RR}=2T_{\mathrm{coh}}\nu_{J}^{2}e^{-4\pi j_{0}^{4}/j^{4}},
\end{equation}
where $\nu_{J}=\sqrt{\left\langle \tau_{J}^{2}\right\rangle }$ is
the residual torque in the $\mathbf{J}$ direction (see Appendix~\ref{a:RRtorque})
\begin{equation}
\nu_{J}\approx0.28\sqrt{1-j}\sqrt{N\left(a/2\right)}J_{c}\nu_{R}/Q,
\end{equation}
and $j_{0}\left(a\right)$ is the AI locus where the GR precession
frequency $\nu_{GR}\left(a,j\right)$ equals to the coherence time
\begin{equation}
j_{0}=\sqrt{2\pi T_{c}\nu_{GR}\left(a,j=1\right)}.
\end{equation}
Here we adopt 
\begin{equation}
T_{c}=\sqrt{\pi/2}\nu_{p}^{-1}\left(2a,\sqrt{1/2}\right)
\end{equation}
where $\nu_{p}$ is the combined mass and GR precession.

\begin{figure}
\begin{centering}
\includegraphics[width=1\columnwidth]{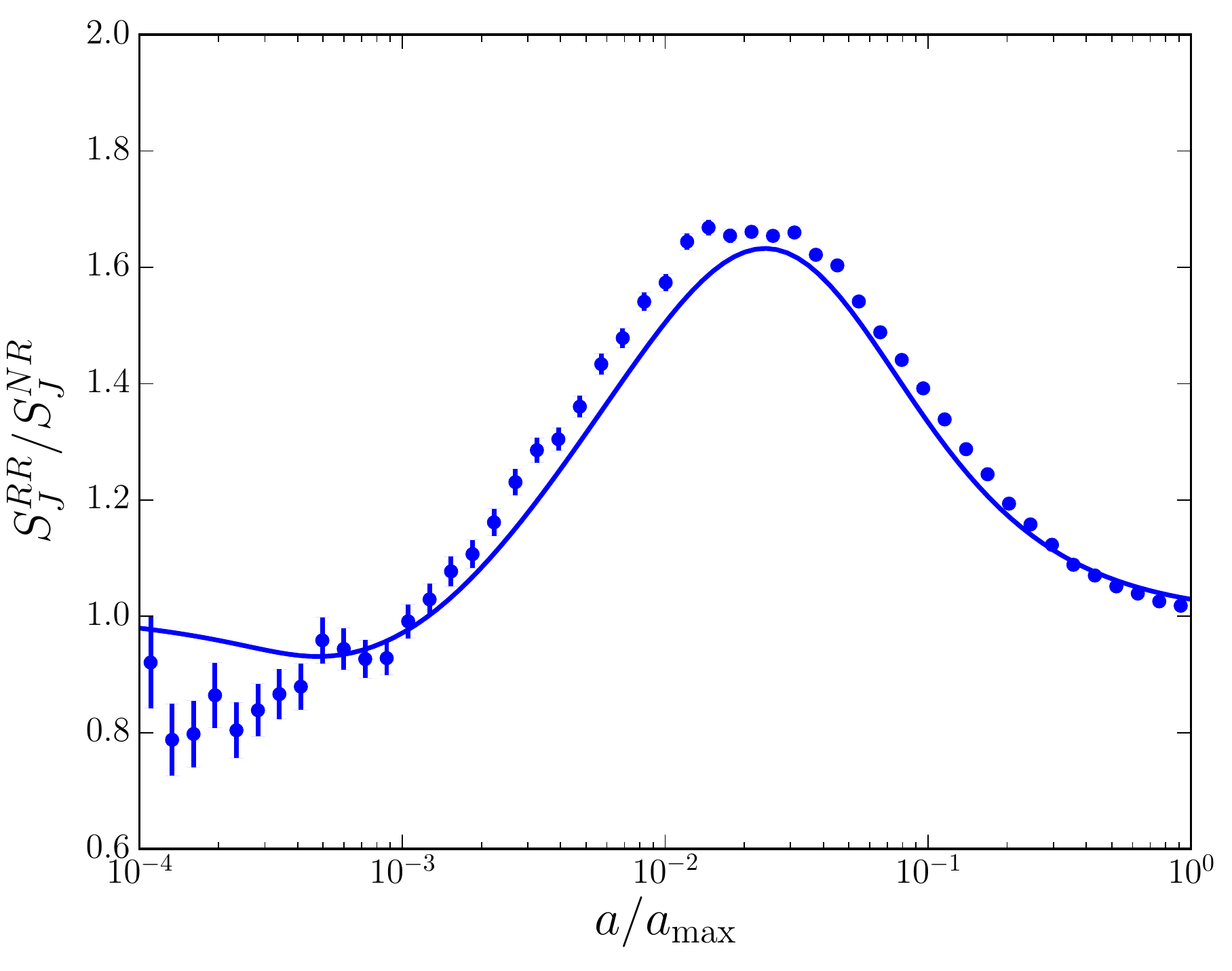}
\par\end{centering}

\protect\caption{\label{fig:FjRR/FjNR}The ratio between the probability current density
in $J$ with and without RR, as function of $a$ (Eq.~\eqref{eq:SJRR_SJNR}),
which expresses the differential plunge rate. The limited increase
in the ratio over its asymptotic value of $1$ in the $a\to0$ and
$a\to\infty$ limits expresses the fact that the contribution of RR
to the plunge rate is small.}
\end{figure}

The combined diffusion coefficients are
\begin{eqnarray}
D_{J} & = & D_{J}^{NR}+D_{J}^{RR}\,,\\
D_{JJ} & = & D_{JJ}^{NR}+D_{JJ}^{RR}\,,
\end{eqnarray}
and using Eq.~\eqref{eq:S_J_djj}, the flux is given by
\begin{equation}
S_{J}\left(E\right)=-n\left(E\right)/\int_{j_{lc}}^{1}\frac{1-j^{2}}{D_{JJ}^{NR}/J_{c}^{2}+D_{JJ}^{RR}/J_{c}^{2}}\frac{dj}{j}\,.
\end{equation}
Since $D_{RR}$ is rises up to some maximal value before sharply drops
as it approaches $j\to j_{0}$, we can approximate the RR DC as
\begin{equation}
d_{jj}^{RR}\approx\begin{cases}
d_{RR}^{0}\left(1-j\right)e^{-4\pi j_{0}^{4}} & j\ge j_{m}\\
0 & j<j_{m}
\end{cases}\,,
\end{equation}
where $d_{RR}^{0}=2T_{\mathrm{coh}}\nu_{J}^{2}\left(j=0\right)$ and
the maximum of $d_{RR}\left(E,j\right)$, occurs at $j_{m}$, given
by $j_{m}^{5}/j_{0}^{5}=16\pi\left(1-j_{m}\right)/j_{0}\approx16\pi\left(1-j_{0}\right)/j_{0}$.
The differential flux is therefore given by
\begin{eqnarray}
\frac{S_{J}^{RR}\left(E\right)}{S_{J}^{NR}\left(E\right)} & = & \left[\int_{j_{lc}}^{j_{m}}\frac{1-j^{2}}{d_{jj}^{NR}}\frac{dj}{j}+\int_{j_{m}}^{1}\frac{1-j^{2}}{d_{jj}^{NR}+d_{jj}^{RR}}\frac{dj}{j}\right]^{-1}\nonumber \\
 & \approx & \left[1-\frac{\chi_{RR}}{1+\chi_{RR}}\frac{\log\left(16\pi\left(1-j_{0}\right)j_{0}^{4}\right)}{5\log\left(j_{lc}\right)}\right]^{-1}\,,\label{eq:SJRR_SJNR}
\end{eqnarray}
where $\chi_{RR}=d_{RR}^{0}e^{-4\pi j_{0}^{4}}/d_{NR}^{0}$. As shown
in Figure~\ref{fig:FjRR/FjNR}, this analytic approximation reproduces
the MC results. 

The small effect of RR on the loss-rates can be estimated by integrating
Eq.~\eqref{eq:SJRR_SJNR} over the relevant region 
\begin{equation}
R_{p}^{RR}\left(E\right)=-\int_{E}^{E_{\max}}S_{J}^{RR}\left(E^{\prime}\right)dE^{\prime}\,.\label{eq:RpRR}
\end{equation}

\makeatletter{}

\global\long\def\Mo{M_{\odot}}
\global\long\def\Ro{R_{\odot}}
\global\long\def\Lo{L_{\odot}}
\global\long\def\Mbh{M_{\bullet}}
\global\long\def\Ms{M_{\star}}
\global\long\def\Rs{R_{\star}}
\global\long\def\Ts{T_{\star}}
\global\long\def\vs{v_{\star}}
\global\long\def\s{\sigma_{\star}}
\global\long\def\n{n_{\star}}
\global\long\def\tr{t_{\mathrm{rlx}}}
\global\long\def\Ns{N_{\star}}
\global\long\def\ns{n_{\star}}
\global\long\def\jlc{j_{lc}}

\section{Monte Carlo models}

\label{s:MCmodel}

We complement and validate our analytic study of the relativistic
loss-cone by numerically evolving the FP equation in both $E$ and
$J$ using a MC procedure (described in Appendix~\ref{a:MCproc}).
Unlike the analytic treatment, this procedure does not assume that
the evolution in $J$ can be decoupled from that in $E$. The advantages
of the MC method over direct $N$-body simulations are the high degree
of flexibility it offers for isolating and studying the different
mechanisms that affect the dynamics of the loss-cone, the ease of
including additional physical effects and of modifying the initial
and boundary conditions, and importantly, its scalability to systems
with a realistically large number of stars. In Section~\ref{s:MCrates}
below we employ the MC procedure to calculate the phase-space density
and rates of relaxed galactic nuclei, and in particular, that of a
Milky Way-like nucleus with a $4\times10^{6}\,\Mo$ MBH and $N\sim{\cal O}(10^{7})$
stars on its radius of influence, $r_{h}\sim2\,\mathrm{pc}$. Such
nuclei are considered archetypal for future space-borne missions to
detect low-frequency gravitational waves from inspiraling compact
objects. 

We begin here by validating the MC procedure. We study the dynamics
in the restricted case where $E$ remains fixed, which allows a direct
test of the impact of adiabatic invariance on the long-term dynamics
(Section~\ref{ss:1DMC}). We then compare rate results from our MC
procedure in both $E$ and $J$ with the currently available results
from direct post-Newtonian $N$-body simulations of small-$N$ ($N<100$)
systems (Section~\ref{ss:2DMC}).

\subsection{$j$-only Monte Carlo simulations}

\label{ss:1DMC}

The maximal entropy limit (Appendix~\ref{a:MaxEntropy}) provides
a basic test for the physical validity of the DCs and of the MC procedure
for evolving the Fokker-Planck equation. The probability density of
a closed system with zero net angular momentum must asymptote to the
maximal entropy solution $\mathrm{d}N/\mathrm{d}j=2j$. Experimentation
shows that this is a sensitive test of both the functional form of
the DCs and the details of the MC procedure, in particular the implementation
of the boundary conditions. We verify the maximal entropy limit in
Section~\ref{sss:1DNR}. In the absence of NR (for example on timescales
$\ll T_{NR}$), a relativistic system that is subject to RR with a
smooth background noise should display adiabatic invariance (AI) in
the form of a sharp drop in the phase-space density below some small
value of $j$ where the GR precession period falls below the coherence
time~\citep{bar+14}. RR with non-smooth background noise is not expected
to display such an AI barrier (the ``Schwarzschild barrier'',~\citealt{mer+11}).
We demonstrate that our MC procedure reproduces this behavior in Section~\ref{sss:1DRR}. Finally, we study the realistic case where NR smears
the RR-generated AI in Section~\ref{sss:1DNRRR}, and also show how
this smearing appears in the unrestricted case where both $a$ and
$j$ evolve.

Since the $j\to0$ limit is of special interest, it is efficient to
use logarithmic bins to collect statistics on the phase-space density.
In that case, it is more useful to represent the density as $\mathrm{dN}/\mathrm{d}\log j$,\footnote{The decreasing size of the bins in linear space leads to a misleading
graphical representation of $\mathrm{d}N/\mathrm{d}j$ when the statistics
are low, as is the typical case at low-$j$, since the normalized
bin density $\Delta N/\Delta j$ diverges for $\Delta N\ge1$ as $\Delta j\to0$.}where the maximal entropy solution is $\mathrm{d}N/\mathrm{d}\log j=2j^{2}$
(or $\mathrm{d}N/\mathrm{d}\log_{10}j=2\log(10)j^{2}$).

\subsubsection{NR only}

\label{sss:1DNR}

\begin{figure}
\begin{centering}
\includegraphics[width=0.95\columnwidth]{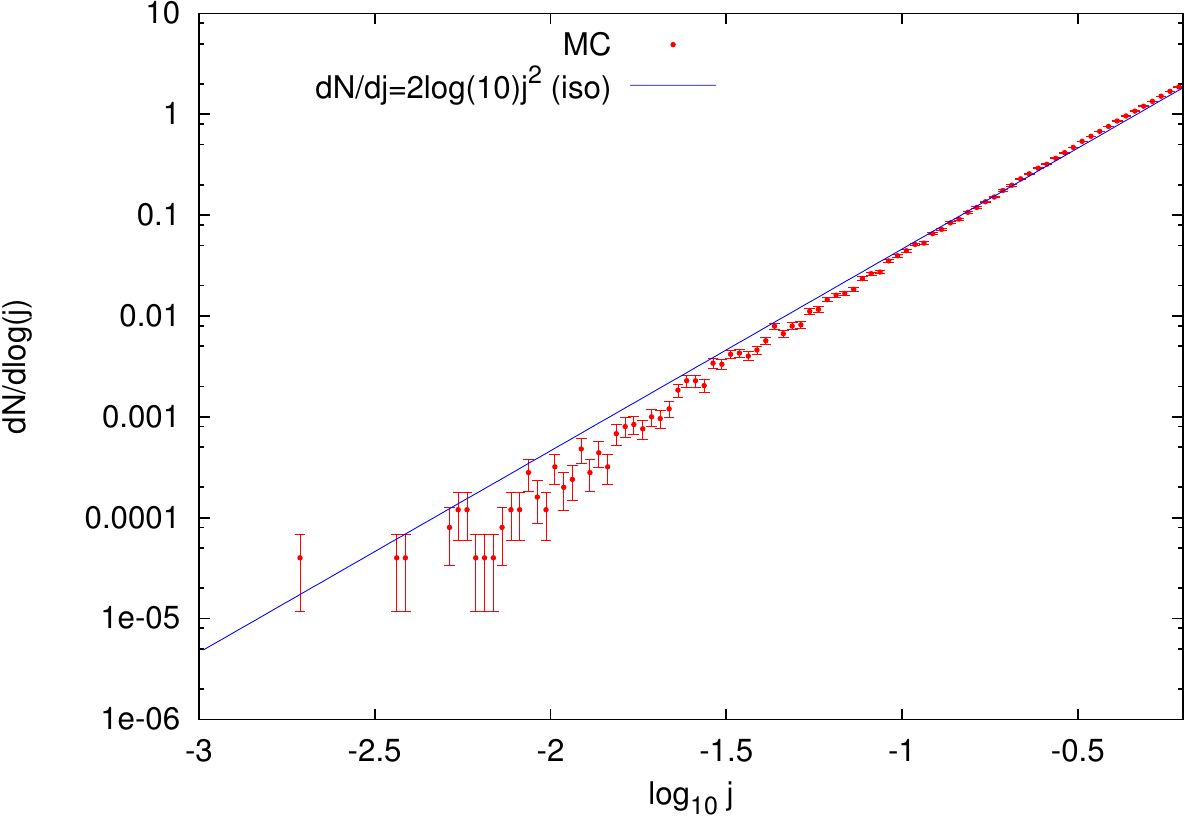}
\par\end{centering}

\protect\caption{\label{f:ME1Dj} The convergence of a MC $j$-only NR diffusion simulation
to the maximal entropy solution $\mathrm{d}N/\mathrm{d}\log_{10}j=2\ln10j^{2}$
}
\end{figure}

Figure~\ref{f:ME1Dj} shows the $j$-PDF at $a=a_{\max}/4$ for an
$\alpha=7/4$ cusp with $a_{\max}=10^{4}a_{\min}$, after time $t=100T_{E}$
($T_{E}=[E^{2}/\Delta(E^{2})]_{a_{\max}}$). Near-complete convergence
is already reached at $T\gtrsim T_{E},$ (Section~\ref{sss:1DNRRR}).
The convergence to the expected maximal entropy solution is apparent,
although a bias toward a somewhat steeper slope for $j\lesssim0.1$
is observed. In addition to the $\alpha=7/4$ case shown in Figure~\ref{f:ME1Dj}, simulations with other values of $\alpha$ or of $a_{\max}/a_{\min}$
confirm that the maximal entropy solution holds generally.

\subsubsection{RR only}

\label{sss:1DRR}

\begin{figure}
\begin{centering}
\begin{tabular}{c}
\includegraphics[width=0.95\columnwidth]{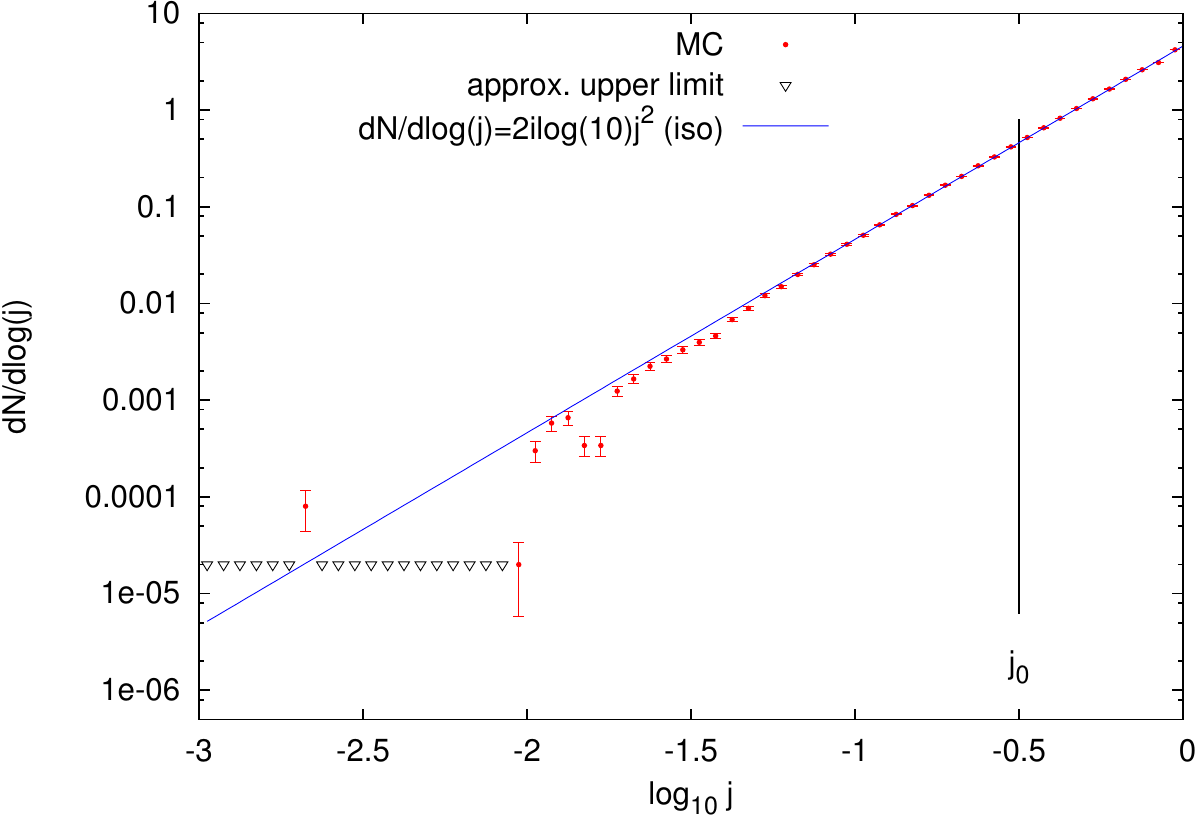}\tabularnewline
\includegraphics[width=0.95\columnwidth]{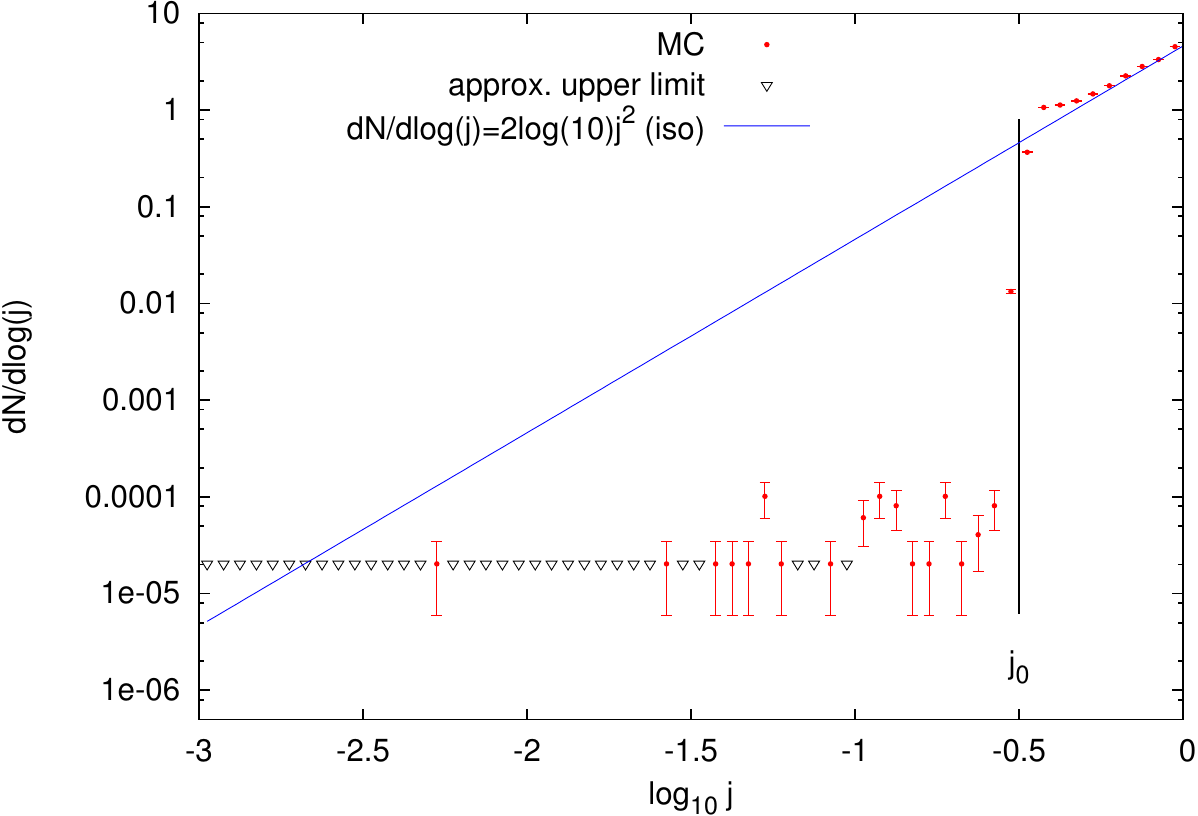}\tabularnewline
\end{tabular}
\par\end{centering}

\protect\caption{\label{f:AI1Dj} Adiabatic invariance and the convergence of $j$-RR
to the maximal entropy solution $\mathrm{d}N/\mathrm{d}\log_{10}j=2\ln10j^{2}$
for two different background noise models: non-smooth noise $(C^{0})$
with an exponential ACF (top) and smooth noise ($C^{\infty}$) with
a Gaussian ACF (bottom).  The predicted characteristic position of
the AI front at $j_{0}$ is marked by the vertical line. An approximate
upper limit on the density (1 count per bin) was estimated for empty
bins (triangles). }
\end{figure}

Figure~\ref{f:AI1Dj} shows the $j$-PDF for the RR-only case. The
MC code reproduces the AI barrier for smooth noise at $j_{0}=\sqrt{T_{c}\omega_{GR}/2\pi}$~\citep{bar+14}, while non-smooth noise asymptotes as expected to
the maximum entropy solution---a demonstration that this limit does
not depend on the nature of the relaxation process. 

\subsubsection{NR+RR}

\label{sss:1DNRRR}

\begin{figure*}
\begin{centering}
\begin{tabular}{cc}
\includegraphics[width=0.475\textwidth]{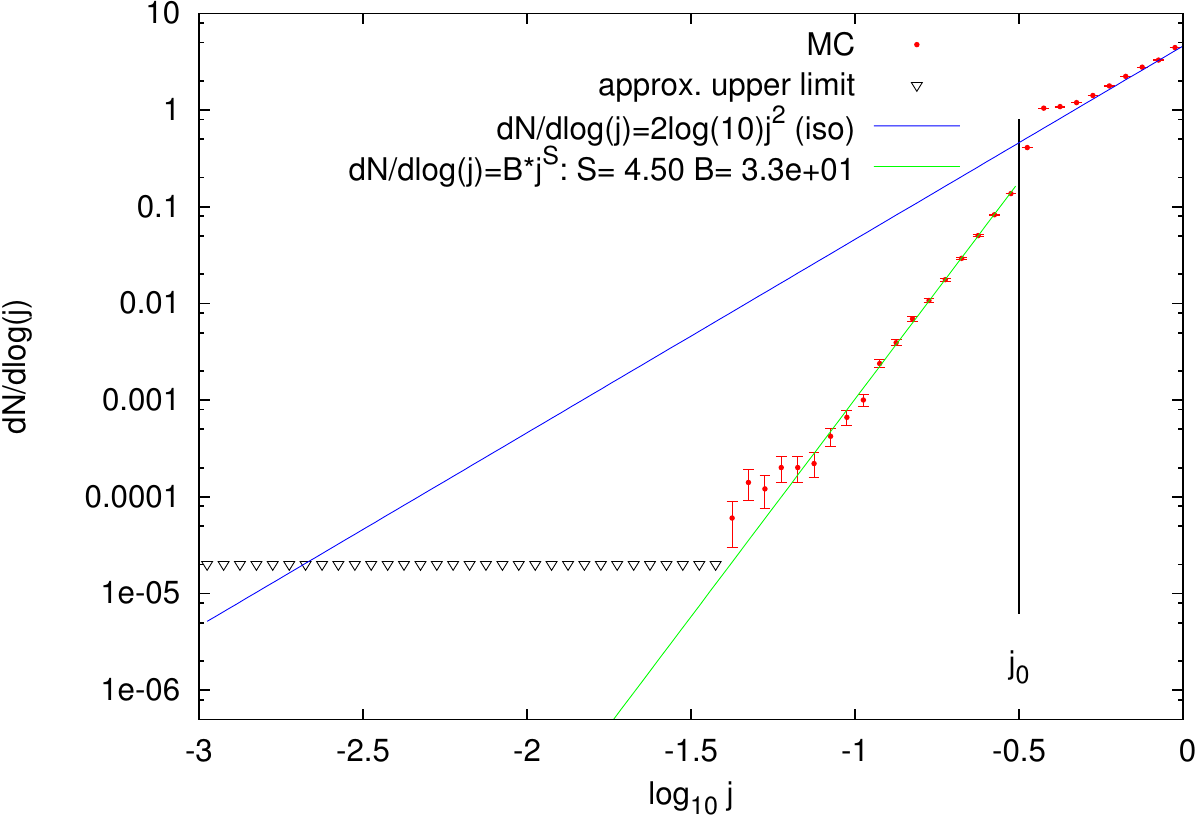} & \includegraphics[width=0.475\textwidth]{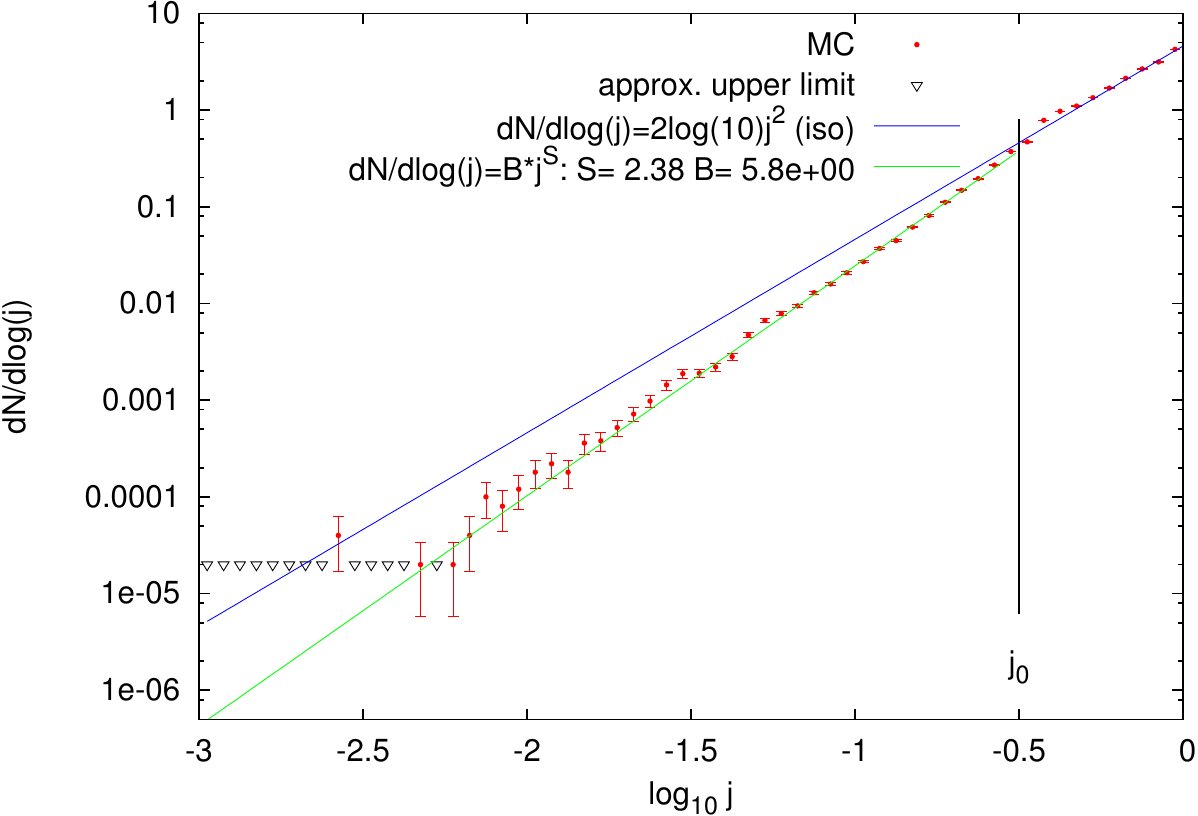}\tabularnewline
\includegraphics[width=0.475\textwidth]{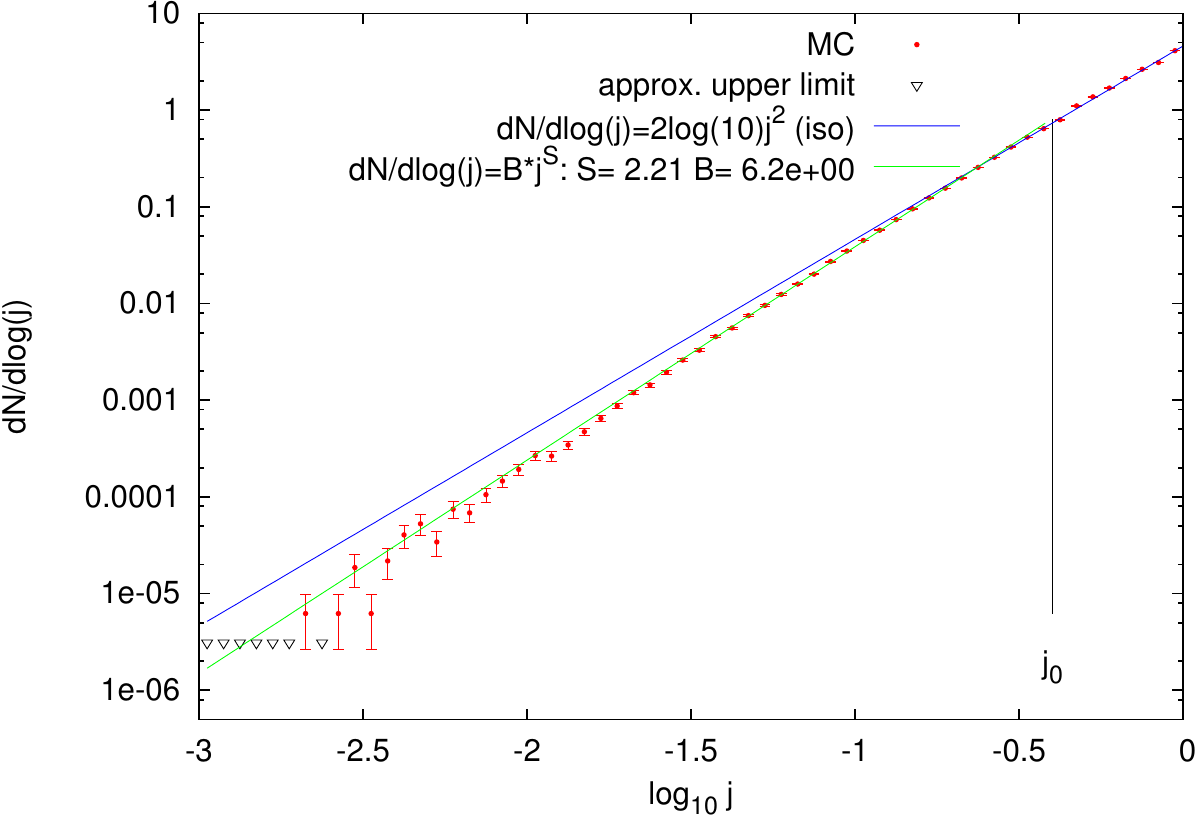} & \includegraphics[width=0.475\textwidth]{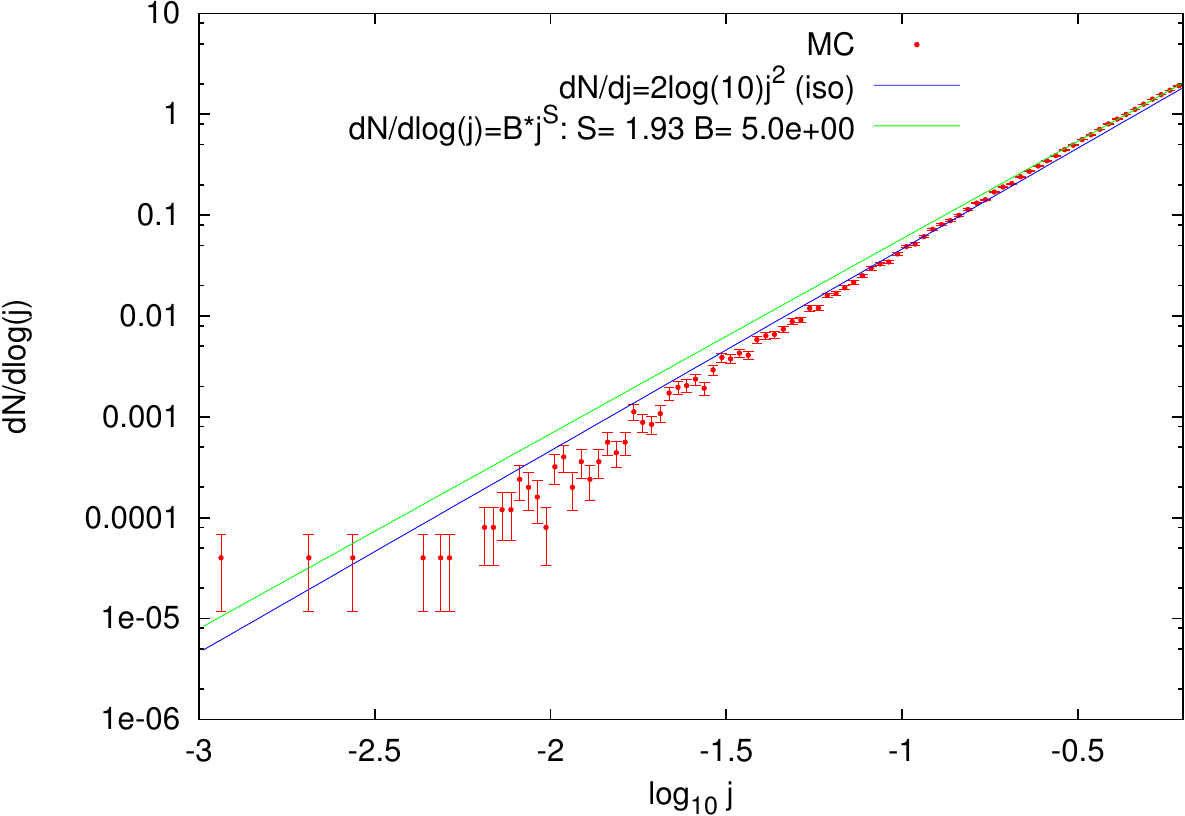}\tabularnewline
\end{tabular}
\par\end{centering}

\protect\caption{\label{f:AI1DjRRNR} The suppression of adiabatic invariance by NR
and the convergence of $j$-only MC simulations with NR and RR with
a Gaussian ACF noise to the maximal entropy solution $\mathrm{d}N/\mathrm{d}\log_{10}j=2\ln10j^{2}$.
 All MC runs lasted a fixed time $T_{\mathrm{sim}}=100T_{E}$, were
$T_{E}$ is the energy diffusion timescale at $a_{\max}$, but the
NR relaxation time was artificially extended to $T_{E}^{\prime}=X_{D}T_{E}$
($X_{D}\ge1$) so that $T_{\mathrm{sim}}=(100/X_{D})T_{E}^{\prime}$
(the larger $X_{D}$, the less significant is NR). The predicted characteristic
location of the AI barrier at $j_{0}$ is marked by the vertical line.
The best fit power-law to the $j$-PDF slope is also shown (green
line). An approximate upper limit on the density (1 count per bin)
was estimated for empty bins (triangles). Top left: $T_{\mathrm{sim}}=0.1T_{E}^{\prime}$.
Top right: $T_{\mathrm{sim}}=T_{E}^{\prime}$. Bottom left: $T_{\mathrm{sim}}=100T_{E}^{\prime}=100T_{E}$.
Bottom right: NR-only MC simulation for comparison. }
\end{figure*}

The presence of NR erases the AI cutoff in the $j$-PDF on timescales
approaching or exceeding the NR timescale (quantified by the energy
diffusion timescale $T_{E}$). This is demonstrated in Figure~\ref{f:AI1DjRRNR},
which shows a sequence of $j$-only MC simulations that include both
NR and RR\@. All the simulation runs had a fixed duration $T_{\mathrm{sim}}=100T_{E}$,
which kept the number of binned points, and hence the statistical
sampling fluctuations, fixed (the MC values are sampled every $\Delta t=1\,T_{E}$).
However, the effective NR timescale was artificially extended to $T_{E}^{\prime}=X_{D}T_{E}$
($X_{D}\ge1$), so that $T_{\mathrm{sim}}=(100/X_{D})T_{E}^{\prime}$,
and $X_{D}$ was varied from $X_{D}=10^{3}$ ($T_{\mathrm{sim}}=0.1T_{E}^{\prime}$)
down to $X_{D}=1$ ($T_{\mathrm{sim}}=100T_{E}^{\prime}$): the larger
$X_{D}$, the less significant is NR over $T_{\mathrm{sim}}$. Figure~\ref{f:AI1DjRRNR} shows results for $T_{\mathrm{sim}}=0.1T_{E}^{\prime}$,
$T_{\mathrm{sim}}=T_{E}^{\prime}$, and $T_{\mathrm{sim}}=100T_{E}^{\prime}$.
The AI cutoff is substantially smeared already when $T_{\mathrm{sim}}=0.1T_{E}^{\prime}$
(compare Figure~\ref{f:AI1DjRRNR} top left with Figure~\ref{f:AI1Dj}
right), and the AI remains only as a moderate steepening of the slope
below $j_{0}$ for $T_{\mathrm{sim}}=T_{E}^{\prime}$. For $T_{\mathrm{sim}}=100T_{E}$,
the $j$-PDF is almost indistinguishable from the case of NR-only.
\texttt{}

\begin{figure*}
\centering{}\includegraphics[width=0.33\textwidth]{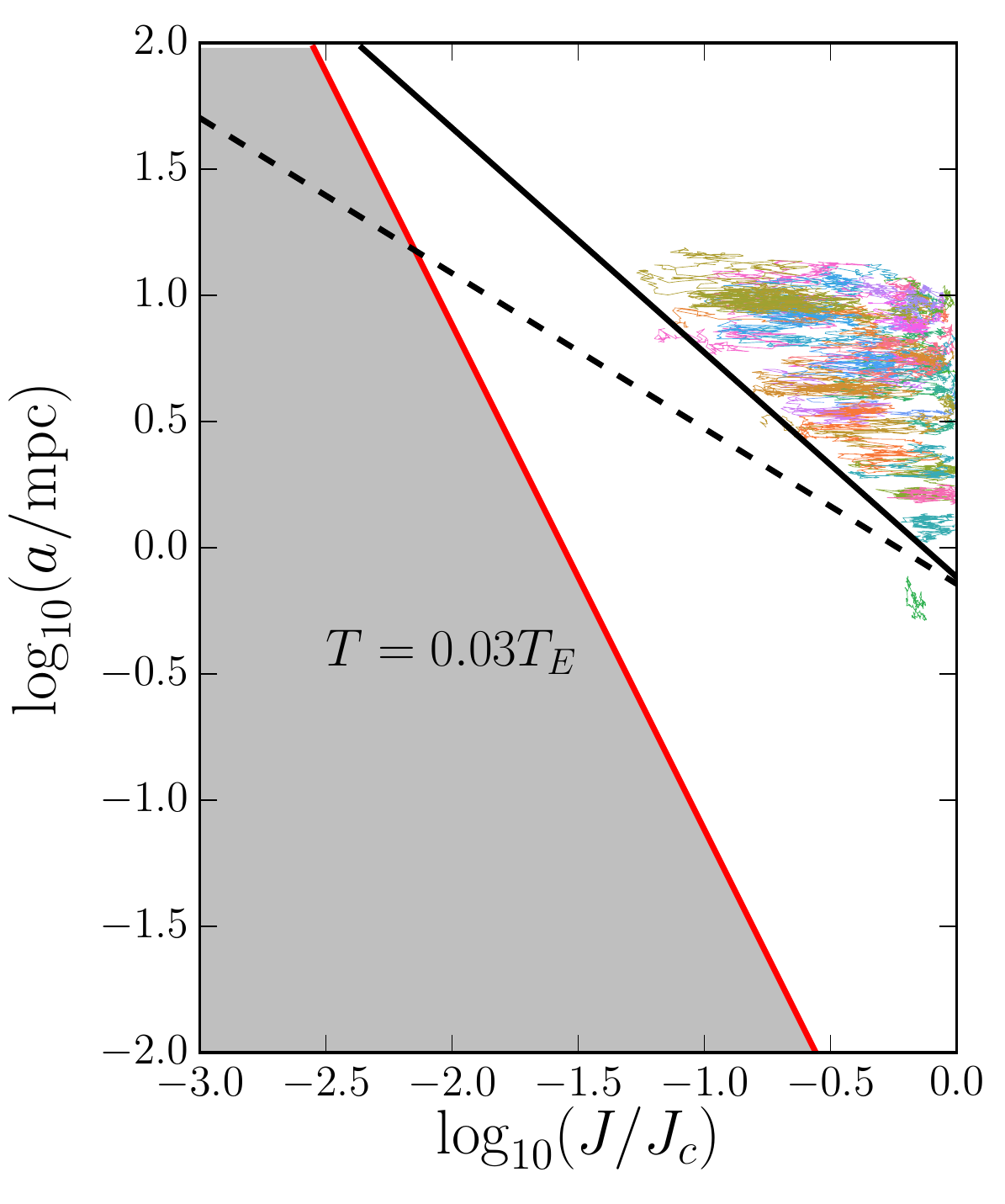}\includegraphics[width=0.33\textwidth]{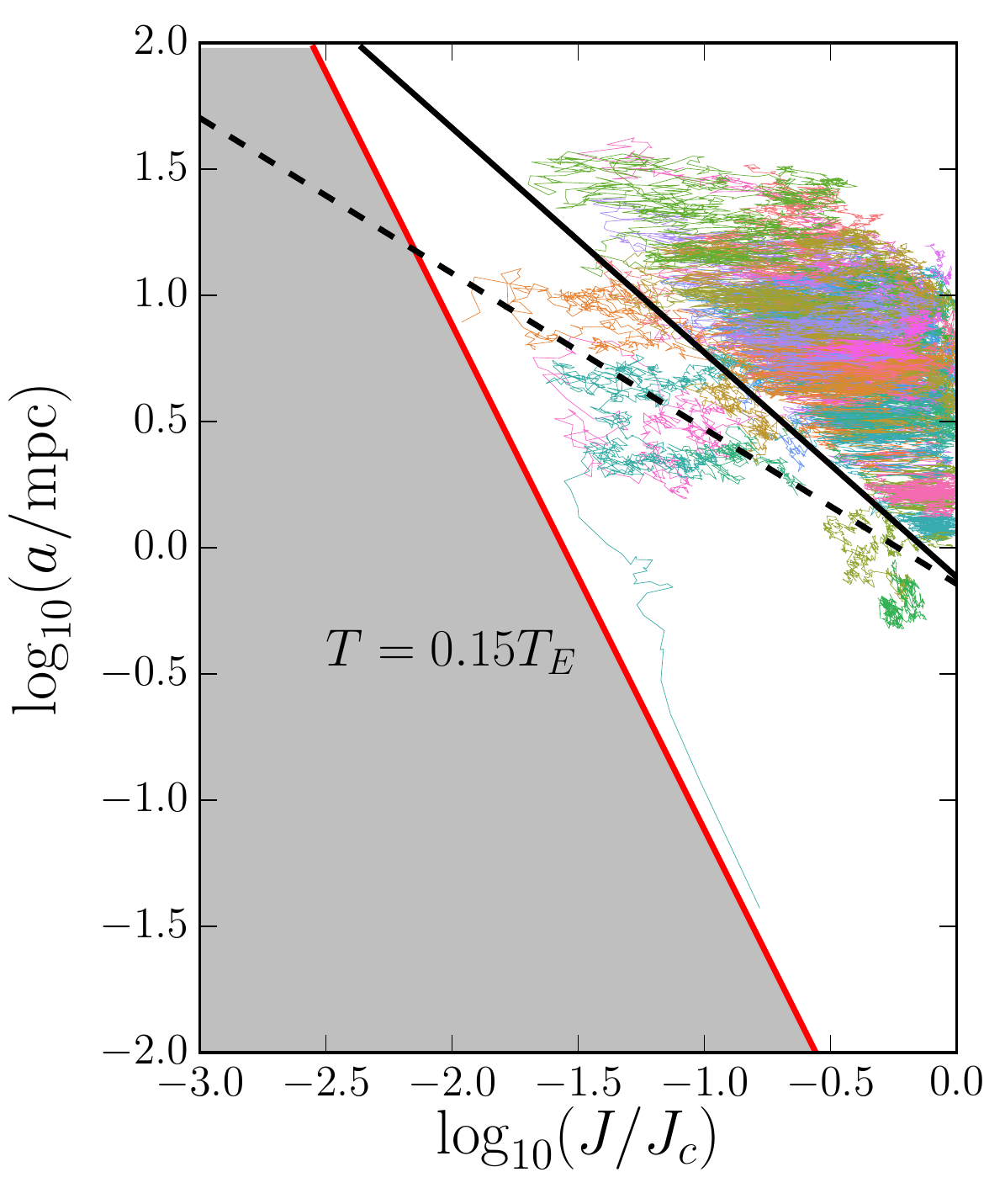}\includegraphics[width=0.33\textwidth]{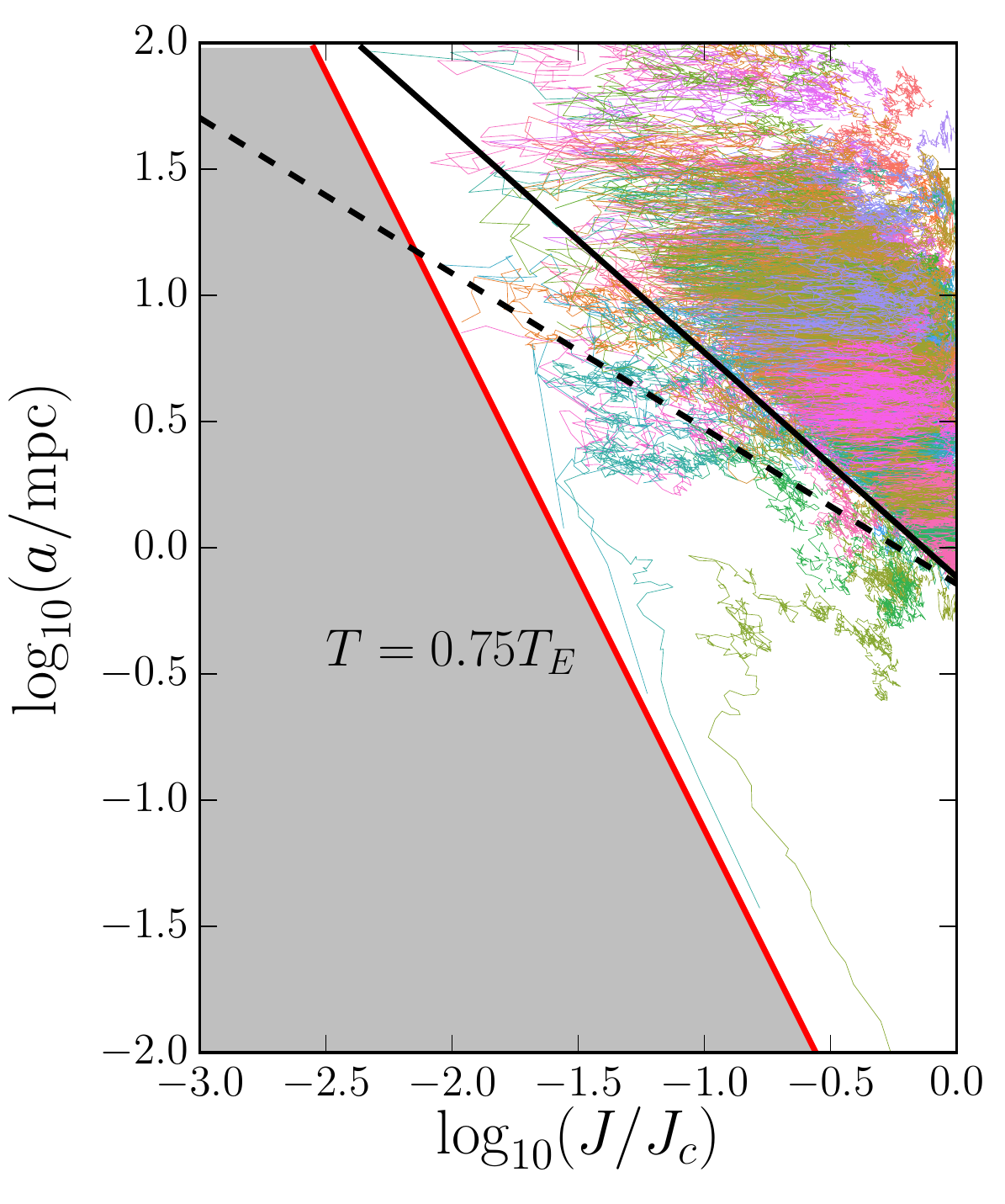}\protect\caption{\label{fig:2dSB}MC snapshots of the trajectories of individual stars
at different times, corresponding to increasing fractions of the energy
relaxation timescales $T_{E}$ at $a_{\max}$. On short timescales,
$t\ll T_{E}$, stars do not cross the AI line (solid line), while
on longer timescales, $t\to T_{E}$, NR progressively drives stellar
diffusion across the AI line to the entire available phase-space.
The MC simulation assume a MBH of $10^{6}M_{\odot}$ and a cusp of
50 stars of $50M_{\odot}$ each with initial conditions drawn from
an isotropic cusp with $\alpha=7/4$ and $a_{\max}=10\,\mathrm{mpc}$.
We also plot for comparison the Schwarzschild barrier for this cusp
model (dashed line), as suggested by MAMW11. }
\end{figure*}

This trend is evident also in the general case where both $a$ and
$j$ are free to evolve, as shown in Figure~\ref{fig:2dSB}. On timescales
of order of the RR relaxation time, but much shorter than the NR timescale,
the stellar trajectories are bound by the AI line. However, on longer
timescales, NR drives stellar diffusion across the AI line and beyond.
The existence of a \emph{persistent} Schwarzschild Barrier with a
locus $a_{SB}(j)\propto j^{-2/(5-\alpha)}$, as suggested by~MAMW11~(Eq.
35 there), is neither supported by our analysis nor observed in our
MC simulations. 

\subsection{Comparison with $N$-body simulations}

\label{ss:2DMC}

\begin{figure}
\centering{}\begin{tabular}{c}
\includegraphics[width=0.95\columnwidth]{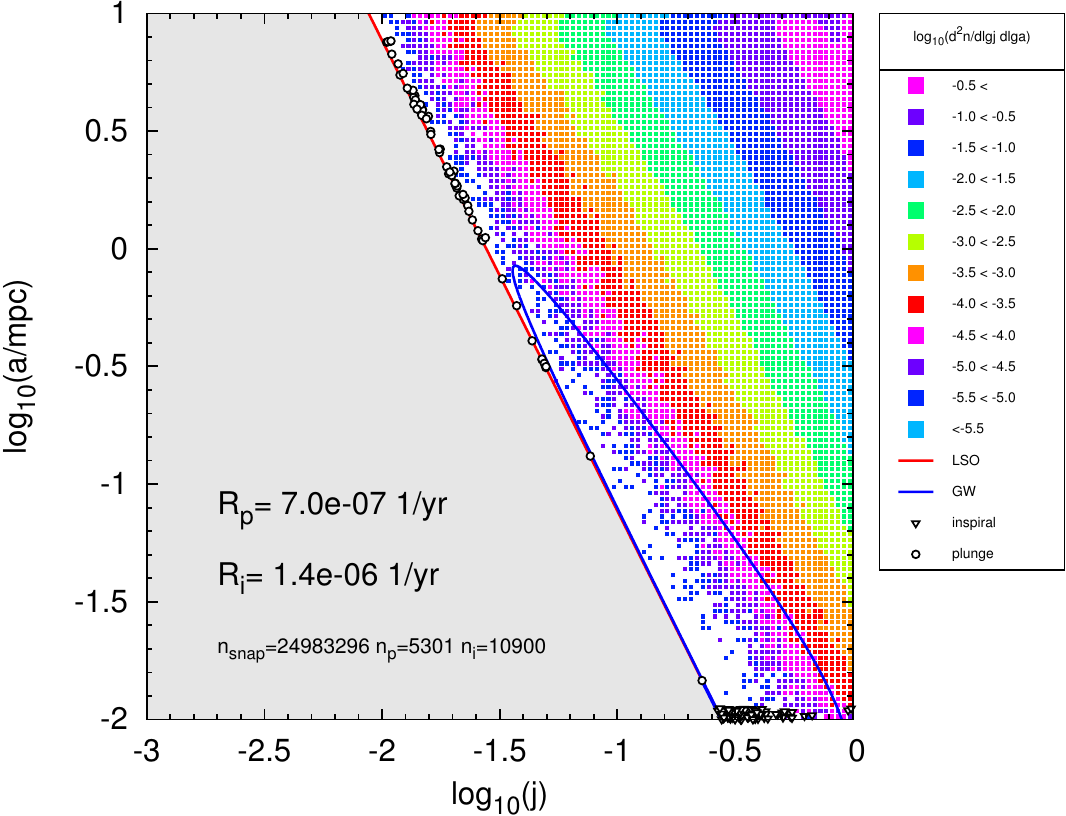}\tabularnewline
\includegraphics[width=0.95\columnwidth]{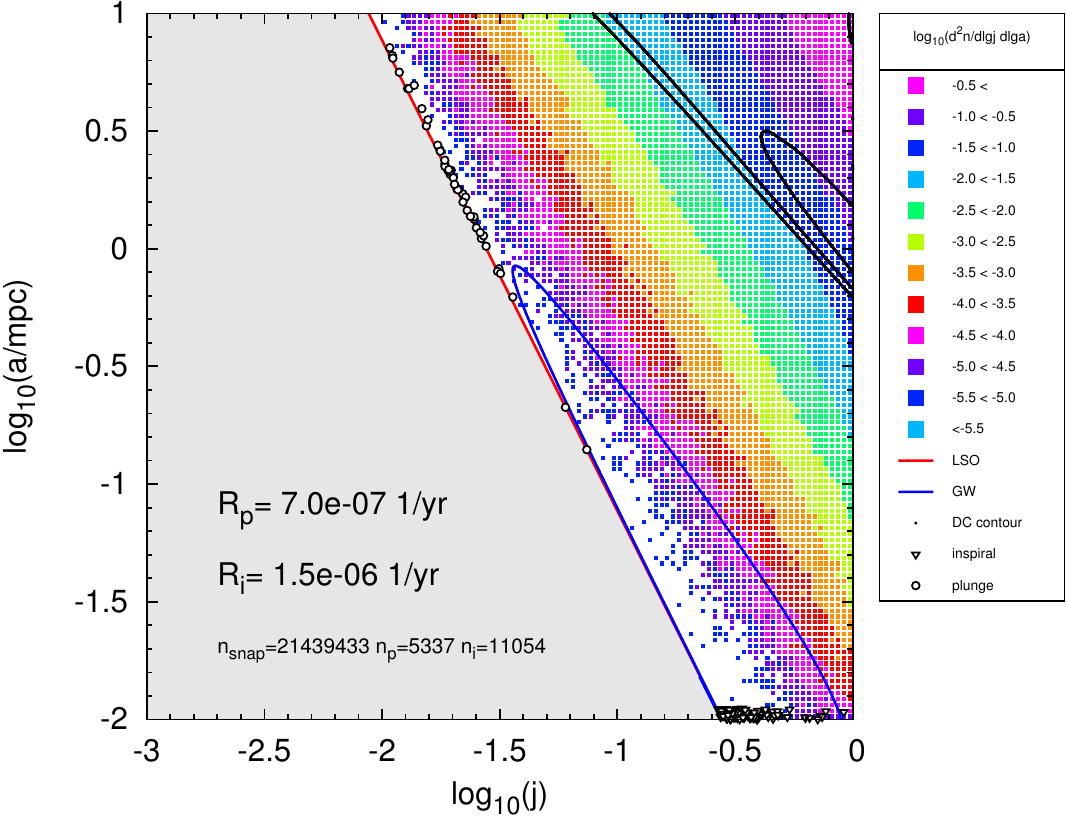}\tabularnewline
\includegraphics[width=0.95\columnwidth]{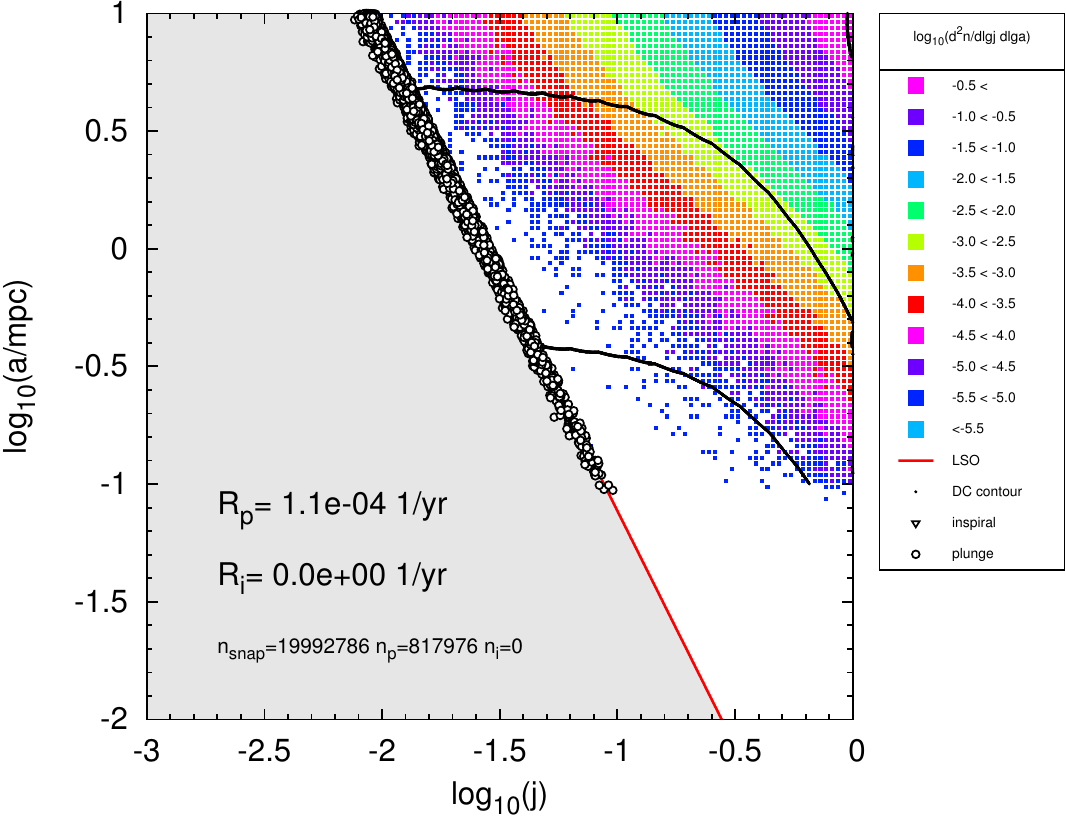}\tabularnewline
\end{tabular}\protect\caption{\label{f:2D}The 2D loss-cone phase-space density in a MAMW11-like
cusp model with mass-precession coherence time and Gaussian noise,
calculated by high phase-space resolution MC simulations. Top: Only
NR\@. Middle: Full model with all the dynamical processes. Bottom:
Same as full model, with GR precession switched off. The black contours
denote the loci where RR dominates NR, $D_{2}^{RR}/D_{2}^{NR}=1,10,100$.
To avoid clutter, only 1\% of the plunge and inspiral terminal track
points (circles and triangles) are displayed. }
\end{figure}

Matching results of MC simulations to results from direct $N$-body
simulations is not straightforward. The MC procedure enforces boundary
conditions at $a_{\max}$ and assumes an approximate steady-state
background, whereas the $N$-body simulations of MAMW11 and BAS14
provide $a_{\max}$ only as initial conditions (the cluster subsequently
expands), and allow the stellar number to fall with time as stars
are lost into the MBH\@. In addition, the MAMW11 and BAS14 models
all have an initial $\ns(r)\propto r^{-2}$ cusp, which is away from
the BW76 steady-state configuration of a single mass population, $\ns(r)\propto r^{-7/4}$.
Thus, these $N$-body simulations always remain out of steady-state
due to relaxation, expansion and stellar loss. The loss-rates of the
MC (when its parameters are matched to initial state of the $N$-body
simulations) should therefore be compared to the \emph{initial} loss-rates
of the $N$-body simulations. To further reduce this incompatibility,
we modified our MC procedure to reproduce the initial conditions of
the $N$-body simulations by introducing the test stars into the interior
of the cusp according to an $\ns\propto r^{-2}$ probability density.

Figure~\ref{f:2D} shows the phase-space density derived from MC simulations
for the MAMW11 cusp model (see details below), which is used here
for comparison with $N$-body results. Results are shown for the case
of Newtonian dynamics without RR, and for the case where all dynamical
effects are switched on (Newtonian dynamics, GR, NR and RR). Also
shown are the endpoints in phase-space of a representative fraction
of the plunge and inspiral events. Figure~\ref{f:2D} and Table~\ref{t:MAMW11}
demonstrate that GR precession plays a critical role in making EMRIs
possible; in its absence, RR remains unquenched at all low values
of $j\ge\jlc$ and therefore stars are rapidly driven to plunging
orbits before they can reach the EMRI line and inspiral by the emission
of GW\@. However, when GR precession is included, the region where
RR dominates over NR is restricted to regions that are far from the
loss-lines (black contours, Figure~\ref{f:2D} top right). This creates
a bottleneck in the flow from $a_{\max}$ and $j\sim{\cal O}(1)$
to the loss-lines, where the orbital evolution is driven by slow NR,
and therefore the effect of RR on the loss-rates in steady-state is
not large. This near-independence of the steady-state loss-rates from
RR is analyzed in Section~\ref{s:analytic}. The MC results also show
that mass precession cannot play a similar role, since it becomes
efficient only for $j\to1$ and $a\to r_{h}$ orbits.

Computational costs limited the MAMW11 simulations to a non-realistic
cusp of only $\Ns=50$ heavy objects of mass $\Ms=50\,\Mo$ each around
a MBH of mass $\Mbh=10^{6}\,\Mo$, extending between $a_{\min}=10^{-3}$
mpc to $a_{\max}=10$ mpc\footnote{These constraints lead to atypical dynamical properties in this model.
(1) Because $\Ns$ is so small that $\sqrt{\Ns}\sim{\cal O}(\Ns),$
the difference between the mass precession coherence time and the
self-quenching coherence time is small, and the two are hard to discriminate.
(2) Since the NR relaxation time is $T_{NR}\sim Q^{2}P/\Ns\log Q$,
while the mass-precession RR timescale is $T_{RR}\sim QP$, $T_{RR}/T_{NR}\ll1$
everywhere in the cusp, that is, RR is atypically efficient. }.  The stars were initially set on stable orbits, isotropic in orientation
and eccentricity. GR was introduced to the equations of motion perturbatively,
up to post-Newtonian (PN) order PN2.5. Orders PN1 and PN2 contribute
only to the in-plane (Schwarzschild) periapse precession, while order
PN2.5 contributes only to dissipative GW emission. By selectively
switching on or off the various PN terms, the $N$-body simulations
tested the cases of Newtonian gravity (all PN terms switched off),
no GR precession (only the PN2.5 term switched on), or full perturbative
GR (all PN terms switched on). Table~\ref{t:MAMW11} compares the
loss-rates for the corresponding MC and $N$-body simulations, as
well as for an artificial model that can only be realized by the MC
method, a Newtonian case where RR is switched off. The MAMW11 loss-rates
were reported as function of time, so it is possible to extrapolate
to $t\to0$ and obtain lower or upper limits on the rates. The BAS14
rates where reported only in the average.

Table~\ref{t:MAMW11} shows that the MC loss-rates for the full GR
models are quite similar, irrespective of the RR noise model, and
that they are also similar to the rates predicted in the artificial
case where RR is switched off. The MC loss-rates are somewhat higher
than those derived from the $N$-body simulations. The MC model with
smooth (Gaussian) noise provides a better fit to the $N$-body results,
while the coherence models are virtually indistinguishable, with only
a marginal preference for mass precession. 

Overall the MC loss-rates are in agreement with the $N$-body simulations;
the systematic trends can be explained by the differences between
the computational techniques and the physical assumptions. Compared
to the MAMW11 $N$-body simulations, the MC plunge rate is $\mbox{3.4 }$
higher and the inspiral rate is $1.7$ higher; compared to the BAS14
$N$-body simulations, the MC plunge rate is $1.5\pm0.3$ higher
and the inspiral rate is $1.8_{-0.2}^{+0.2}$  higher. Since the
incompatibility of the MC and $N$-body treatment of the boundary
condition at $a_{\max}$ results in a lower stellar density in the
$N$-body simulations, the fact that the $N$-body loss-rates are
systematically lower is to be expected. The same systematic trend
is also seen in models where GR is switched off (i.e., no GR precession,
no GW). When only GW is switched on but GR precession is switched
off (i.e., no GR quenching), the MC inspiral rate is zero in agreement
with MAMW11. An additional difference between these studies is the
plunge criterion. MAMW11 used $r<8r_{g}$, BAS14 $r<6r_{g}$ and here
the criterion was based directly on the angular momentum, $J<J_{lc}$
(where $J$ was evaluated in the Keplerian limit). As noted by~\citet{gai+06},
plunging orbits (i.e., parabolic orbits with $J=J_{lc}=4GM/c$) correspond
to Keplerian orbits with periapse $r_{lc}=8GM/c^{2}$, or to relativistic
orbits with periapse $r_{lc}=4GM/c^{2}$. This can explain some of
the systematic differences in the rates, since if $r_{lc}$ \emph{over-estimates}
the true value, stars that should inspiral plunge prematurely, thereby
biasing the rates to too-high plunge rates and too-low inspiral rate.
Conversely, if $r_{lc}$ \emph{under-estimates} the true value, a
too-high inspiral rate will follow. We believe that this explains
the discrepant non-zero inspiral rate that BAS14 find for the case
where GR precession is switched off (i.e., Keplerian dynamics where
$r_{lc}\to8r_{g}$ is the correct limit). Our approximate angular
momentum plunge criterion for parabolic orbit applies generally in
both the Keplerian and relativistic regimes~\citep{gai+06}, unlike
the periapse criteria used by the $N$-body simulations. This may
explain why the MC inspiral rates are somewhat higher than the $N$-body
results, which assume a too-high value of $r_{lc}$ for the relativistic
regime.

\begin{table}
\protect\caption{\label{t:MAMW11}The plunge and inspiral rates in MAMW11-like cusp
models}

\noindent \centering{}\begin{tabular}{llllll}
\hline 
Method$^{1}$ & Processes$^{2}$ & $T_{c}$~$^{3}$ & Noise$^{4}$ & Plunge$^{5}$ & Inspiral$^{5}$\tabularnewline
\hline 
\hline 
MC & All & SQ & E & $0.96$ & $1.8$\tabularnewline
MC & All & SQ & G & $0.71$ & $1.5$\tabularnewline
MC & All & M & E & $0.92$ & $1.8$\tabularnewline
MC & All & M & G & $0.71$ & $1.4$\tabularnewline
\hline 
MC  & No RR & --- & --- & $0.68$ & $1.5$\tabularnewline
MC & No GR & M & G & $110$ & ---\tabularnewline
MC  & No GR prec. & M & G & $110$ & $0$\tabularnewline
MC  & No mass prec. & M & G & $0.71$ & $1.4$\tabularnewline
\hline 
NB1 & \multicolumn{3}{l}{With GR ($t\to0$)} & $\sim0.2$ & $\sim0.9$\tabularnewline
NB1 & \multicolumn{3}{l}{No GR ($t\to0$)} & $>20$ & ---\tabularnewline
NB1 & \multicolumn{3}{l}{No GR prec. ($t\to0$)} & $>20$ & $<1$\tabularnewline
\hline 
NB2 & \multicolumn{3}{l}{With GR} & $0.5\pm0.1$ & $0.8\pm0.1$\tabularnewline
NB2 & \multicolumn{3}{l}{No GR } & $28\pm2$ & ---\tabularnewline
NB2 & \multicolumn{3}{l}{No Gr prec.} & $26\pm2$ & $4.3\pm0.6$\tabularnewline
\hline 
\hline 
\multicolumn{6}{l}{{\scriptsize{}$^{1}$~Method: MC = Monte Carlo, NB = $N$-body (1:
MAMW11, 2: BAS14).}}\tabularnewline
\multicolumn{6}{l}{{\scriptsize{}$^{2}$~Processes: All includes NR, RR, GW~\citep{gai+06},
mass prec., GR prec.}}\tabularnewline
\multicolumn{6}{l}{{\scriptsize{}$^{3}$~Coherence time: M = Mass prec., SQ = Self-quenching.}}\tabularnewline
\multicolumn{6}{l}{{\scriptsize{}$^{4}$~Noise model: G = Gaussian noise, E = Exponential
noise.}}\tabularnewline
\multicolumn{6}{l}{{\scriptsize{}$^{5}$~Event rates in units of $10^{-6}\,\mathrm{yr}^{-1}$.}}\tabularnewline
\hline 
\end{tabular}
\end{table}

\makeatletter{}

\global\long\def\Mo{M_{\odot}}
\global\long\def\Ro{R_{\odot}}
\global\long\def\Lo{L_{\odot}}
\global\long\def\Mbh{M_{\bullet}}
\global\long\def\Ms{M_{\star}}
\global\long\def\Rs{R_{\star}}
\global\long\def\Ts{T_{\star}}
\global\long\def\vs{v_{\star}}
\global\long\def\s{\sigma_{\star}}
\global\long\def\n{n_{\star}}
\global\long\def\tr{t_{\mathrm{rlx}}}
\global\long\def\Ns{N_{\star}}
\global\long\def\ns{n_{\star}}
\global\long\def\jlc{j_{lc}}

\section{Loss rates}

\label{s:MCrates}

\subsection{The Galactic Center test case}

\label{ss:GCparms}

The MBH in the center of the Milky Way and the stars and compact objects
around it are a system of particular relevance, both because it is
uniquely accessible to observations, and can therefore place constraints
on dynamical models and theories~\citet{ale05}, and because planned
space-borne GW detectors with ${\cal O}(10^{6}\,\mathrm{km})$ baseline
will be optimally sensitive to GWs emitted by a mass orbiting a $10^{6}-10^{7}\,\Mo$
MBH near the last stable orbit~\citep{ama+07}. Therefore, although
it remains an open question whether the Galactic center (GC) is surrounded
by a high density relaxed cusp of stellar remnants~\citep[see review by][]{ale11},
and although the rate of GW events from the GC itself is expected
to be small~\citep[but see][]{fre03}, the Galactic MBH SgrA$^{\star}$
represents the archetypal cosmic GW target. 

We adopt here a simplified model of the GC consisting of an MBH mass
of $\Mbh=4\times10^{6}\,\Mo$, surrounded by a steady-state $\alpha=7/4$
cusp of equal mass stars of either $\Ms=1\,\Mo$ or $10\,\Mo$, extending
between $a_{\mathrm{\min}}=2\times10^{-4}\,\mathrm{pc}$ ($a_{\mathrm{in}}=0.1a_{\min}$,
see Appendix~\ref{a:MCproc}) to $a_{\mathrm{out}}=a_{\max}=r_{h}=2\,\mathrm{pc}$
with a total stellar mass of $\Ms\Ns(<r_{h})=2\Mbh$ inside the radius
of influence. Test stars are injected into the nucleus with initial
sma $a_{0}=a_{\mathrm{out}}$, and with isotropic $j_{0}$. 

\begin{figure}
\noindent \begin{centering}
\begin{tabular}{c}
\includegraphics[width=1\columnwidth]{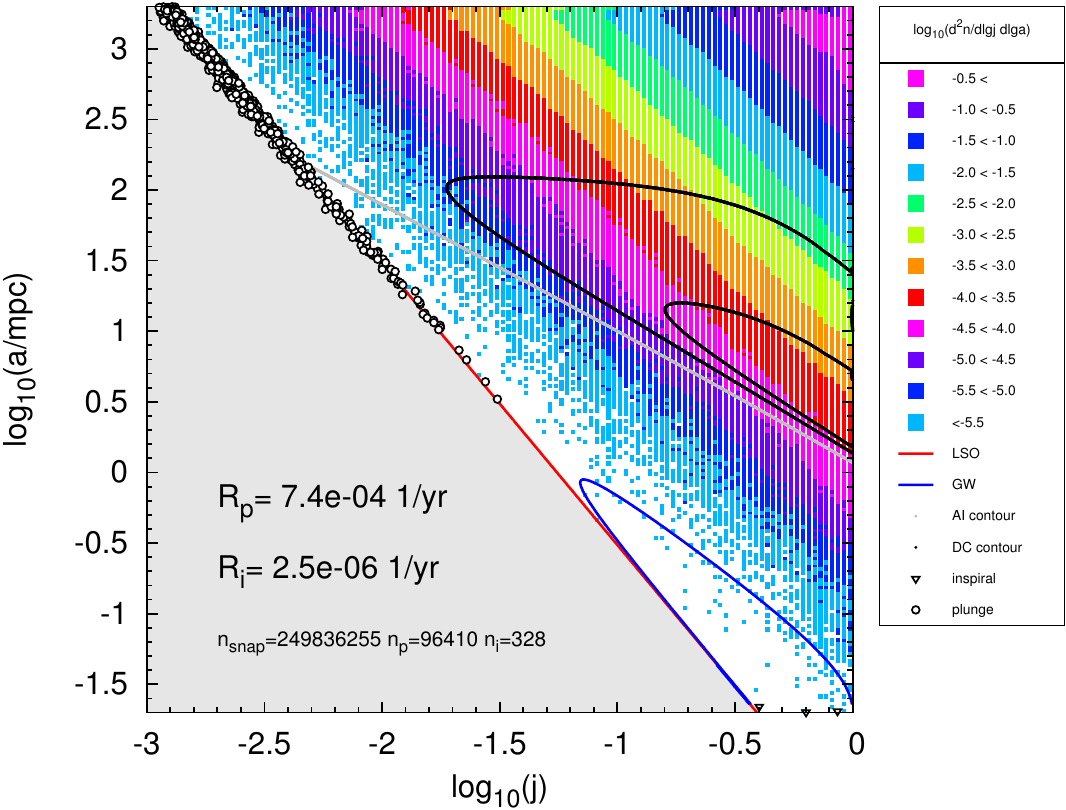}\tabularnewline
\includegraphics[width=1\columnwidth]{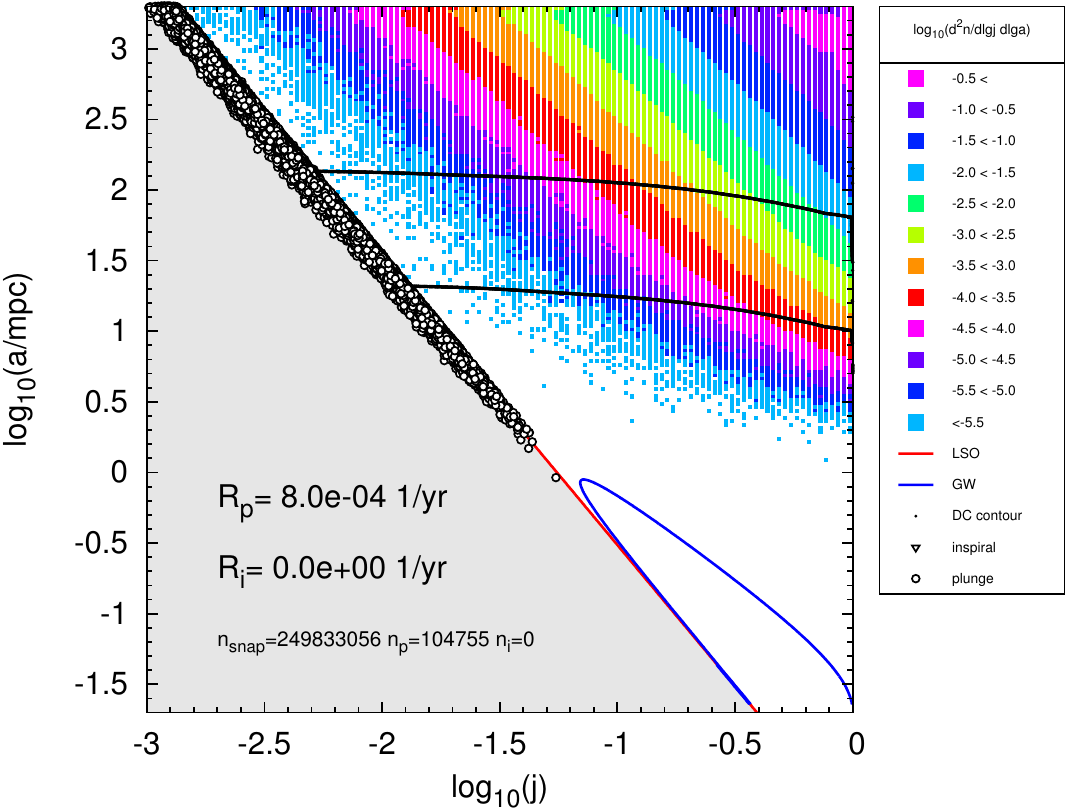}\tabularnewline
\end{tabular}
\par\end{centering}

\protect\caption{\label{f:MW}The 2D loss-cone phase-space density in a Milky Way-like
cusp model with mass-precession coherence time and Gaussian noise,
calculated by high phase-space resolution MC simulations. Stars /
stellar mass BHs of $10\,\protect\Mo$ are assumed. Top: GR precession
included. Mass precession limits the efficiency of RR beyond $\sim100\,\mathrm{mpc}$,
while GR precession limits RR below the AI locus (gray line). RR is
faster than NR only well away from the loss-cone, inside the black
contours (equally fast at the outer contour, 10 times faster at the
inner contour). Bottom: When GR precession is artificially switched
off, RR remain effective all the way down to the loss-cone and is
faster than NR below $a\sim100\,\mathrm{mpc}$. As a result, stars
are driven to plunge trajectories well before they can lose enough
energy by NR to reach the GW loss-line. A central, strongly depleted
cavity is formed, and the EMRI rate is completely suppressed.}
\end{figure}

Figure~\ref{f:MW} shows the steady-state configuration and loss-rates
for a GC model with $\Ms=10\,\Mo$ and a smooth background noise whose
coherence is set by mass precession. As expected from the fact that
the RR-dominated region in phase-space is well separated from the
loss-lines, the steady-state phase-space density is very close to
the simple NR-only solution.

Table~\ref{f:MW} explores the implications of varying some of the
assumed processes for the loss-rates: the mass of the cusp stars ($1\,\Mo$
or $10\,\Mo$), the nature of relaxation (NR only, or NR and RR) the
noise model (White, exponential, Gaussian), the coherence mechanism
(mass precession, self-quenching), or the GW dissipation approximation. 

\begin{table}
\protect\caption{\label{t:MW}The plunge and inspiral rates in Milky Way-like cusp
models}

\noindent \centering{}\begin{tabular}{llllll}
\hline 
$\Ms$~$^{1}$ & Processes$^{2}$ & $T_{c}$~$^{3}$ & Noise$^{4}$ & Plunge$^{5}$ & Inspiral$^{5}$\tabularnewline
\hline 
\hline 
$1$ & No RR & --- & --- & $730$ & $3.1$\tabularnewline
$1$ & GW1 & SQ & W & $16000$ & $0.0$\tabularnewline
$1$ & GW1 & SQ & E & $860$ & $3.3$\tabularnewline
$1$ & GW1 & SQ & G & $880$ & $2.3$\tabularnewline
$1$ & GW1 & M & W & $930$ & $0.0$\tabularnewline
$1$ & GW1 & M & E & $840$ & $3.2$\tabularnewline
$1$ & GW1 & M & G & $840$ & $3.2$\tabularnewline
\hline 
$10$  & No RR & --- & --- & $610$ & $2.8$\tabularnewline
$10$  & GW1 & SQ & W & $6060$ & $0.0$\tabularnewline
$10$  & GW1 & SQ & E & $760$ & $1.9$\tabularnewline
$10$  & GW1 & SQ & G & $690$ & $2.4$\tabularnewline
$10$  & GW1 & M & W & $800$ & $0.0$\tabularnewline
$10$  & GW1 & M & E & $730$ & $2.0$\tabularnewline
$10$  & GW1 & M & G & $730$ & $2.5$\tabularnewline
$10$  & GW2 & M & G & $730$ & $1.2$\tabularnewline
$10$  & GW3 & M & G & $740$ & $1.1$\tabularnewline
\hline 
\multicolumn{6}{l}{{\scriptsize{}$^{1}$~Stellar mass in $\Mo$.}}\tabularnewline
\multicolumn{6}{l}{{\scriptsize{}$^{2}$~GW approximations: GW1~\citet{gai+06}, GW2~\citet{pet64}, }}\tabularnewline
\multicolumn{6}{l}{{\scriptsize{}\ \ \ GW3~\citet{hop+06a}}}\tabularnewline
\multicolumn{6}{l}{{\scriptsize{}$^{3}$~Coherence time: M = Mass prec., SQ = Self-quenching.}}\tabularnewline
\multicolumn{6}{l}{{\scriptsize{}$^{4}$~Noise model: W = White, E = Exponential, G
= Gaussian.}}\tabularnewline
\multicolumn{6}{l}{{\scriptsize{}$^{5}$~Event rates in units of $10^{-6}\,\mathrm{yr}^{-1}$.}}\tabularnewline
\hline 
\end{tabular}
\end{table}

The uncorrelated (white) noise model for the resonant torques, which
is equivalent to the assumption that $\nu_{GR}\to0$, results in very
high plunge rates as strong RR rapidly drains the cusp, and as a result
the EMRI rate drops to zero. In contrast, for all other RR models,
irrespective of the assumptions about the nature of the noise or the
coherence mechanism, GR precession suppresses RR to the extent that
rates are very similar to those derived for the non-physical ``NR-only''
model: a plunge rate of $\Gamma_{p}\sim(6-9)\times10^{-4}\,\mathrm{yr^{-1}}$,
and an inspiral rate of $\Gamma_{i}\sim(1-3)\times10^{-6}\,\mathrm{yr}$.
We find that the more sophisticated GW dissipation estimate of~\citet{gai+06}{\scriptsize{}
}results in inspiral rates that are a factor $\sim2$ higher than
the estimates by~\citet{pet64} or~\citet{hop+06a}. We conclude that
to within a factor of $\sim2$, our rate estimates for relaxed steady-state
cusps are robust to variations of the physical assumptions.

\subsection{Scaling with the MBH mass}

\label{ss:MBHscale}

The MC simulations can be used to validate a simple analytic model
for estimating the loss-rates and their dependence on the parameters
of the galactic nucleus, which is based on identifying critical values
of the sma, $a_{c}$, below which the probability of a star to cross
the loss-line is ${\cal O}(1)$~\citep{lig+77,hop+05}. The loss-rate
is then $\Gamma\propto N(<a_{c})/T_{NR}(a_{c})$, where the proportionality
factor includes the suppression of the density near the loss-line.
For plunge events $a_{c}\sim{\cal O}(r_{h})$~\citep{lig+77}, while
for GW inspiral $a_{c}\sim a_{GW}\ll r_{h}$, the maximum of the GW
line (Section~\ref{s:analytic}). Figure~\ref{f:EJ} shows that the
region of phase-space where RR dominates the dynamics is well separated
from the loss-lines, is well below $r_{h}$ and well above $a_{GW}$.
The timescale relevant for estimating is therefore that of NR and
not RR\@.

In order to estimate the integrated cosmic rates of EMRIs or tidal
disruption flares, it is necessary to scale the loss-rates by the
parameters of the host galaxy, in particular the MBH mass. Here we
adopt a simplified one-parameter sequence of galactic nuclei, where
the free parameter is $\Mbh$, which together with several additional
fixed parameters define the sequence. The $\Mbh$-scaling is based
on the empirical $\Mbh/\sigma$ relation $\Mbh=M_{0}(\sigma/\sigma_{0})^{\beta}$
where $\sigma$ is the stellar velocity dispersion outside the MBH
radius of influence $r_{h}=\eta_{h}G\Mbh/\sigma^{2}$, which encloses
a stellar mass of order the mass of the MBH $\Ns(r_{h})=\mu_{h}Q$.
The power law parameter $\beta\sim4-5$ and the normalization $M_{0}/\sigma_{0}^{\beta}$
are determined empirically~\citep{fer+00,geb+00}. It then follows
that $r_{h}=\eta_{h}\left(M_{\bullet}/M_{0}\right)^{1-2/\beta}GM_{0}/\sigma_{0}^{2}$~\citep{ale11}. 

Using this parameterization, and the approximation that the steady-state
distribution is given by a BW76 cusp, the total plunge and inspiral
rates can be estimated from Eqs.~\eqref{eq:Rp_SS} and~\eqref{eq:Ri_SS},
\begin{eqnarray}
R_{p}^{\mathrm{tot}} & \approx & \frac{10\mu_{h}^{2}}{\eta_{h}^{3/2}}\frac{\sigma_{0}^{3}}{2\pi GM_{0}}\left(\frac{M_{\bullet}}{M_{0}}\right)^{3/\beta-1}\nonumber \\
 &  & \times\frac{\log Q}{\log\left[\sqrt{\eta_{h}}\left(M_{0}/M_{\bullet}\right)^{1/\beta}\left(c/4\sigma_{0}\right)\right]}\,,\nonumber \\
\label{eq:Rp}
\end{eqnarray}
and
\begin{eqnarray}
R_{i}^{\mathrm{tot}} & \approx & A_{GW}\frac{10\mu_{h}^{6/5}}{\eta_{h}^{3/2}}\frac{\sigma_{0}^{3}}{2\pi GM_{0}}\left(Q_{0}\right)^{3/\beta-1}\nonumber \\
 &  & \times\frac{\left(\log Q\right)^{1/5}}{\log\left[\sqrt{\eta_{h}A_{GW}}\left(\mu_{h}\log Q\right){}^{-2/5}Q_{0}^{-1/\beta}c/\left(4\sigma_{0}\right)\right]}\,,\nonumber \\
\label{eq:Ri}
\end{eqnarray}
where $Q_{0}=M_{\bullet}/M_{0}$, $a_{GW}=A_{GW}\left(\mu_{h}\log Q\right){}^{-4/5}$
and $A_{GW}$ is a numerical factor which depend on the GW dissipation
approximation (Appendix~\ref{a:GWline}).

\begin{figure}
\centering{}\includegraphics[width=0.9\columnwidth]{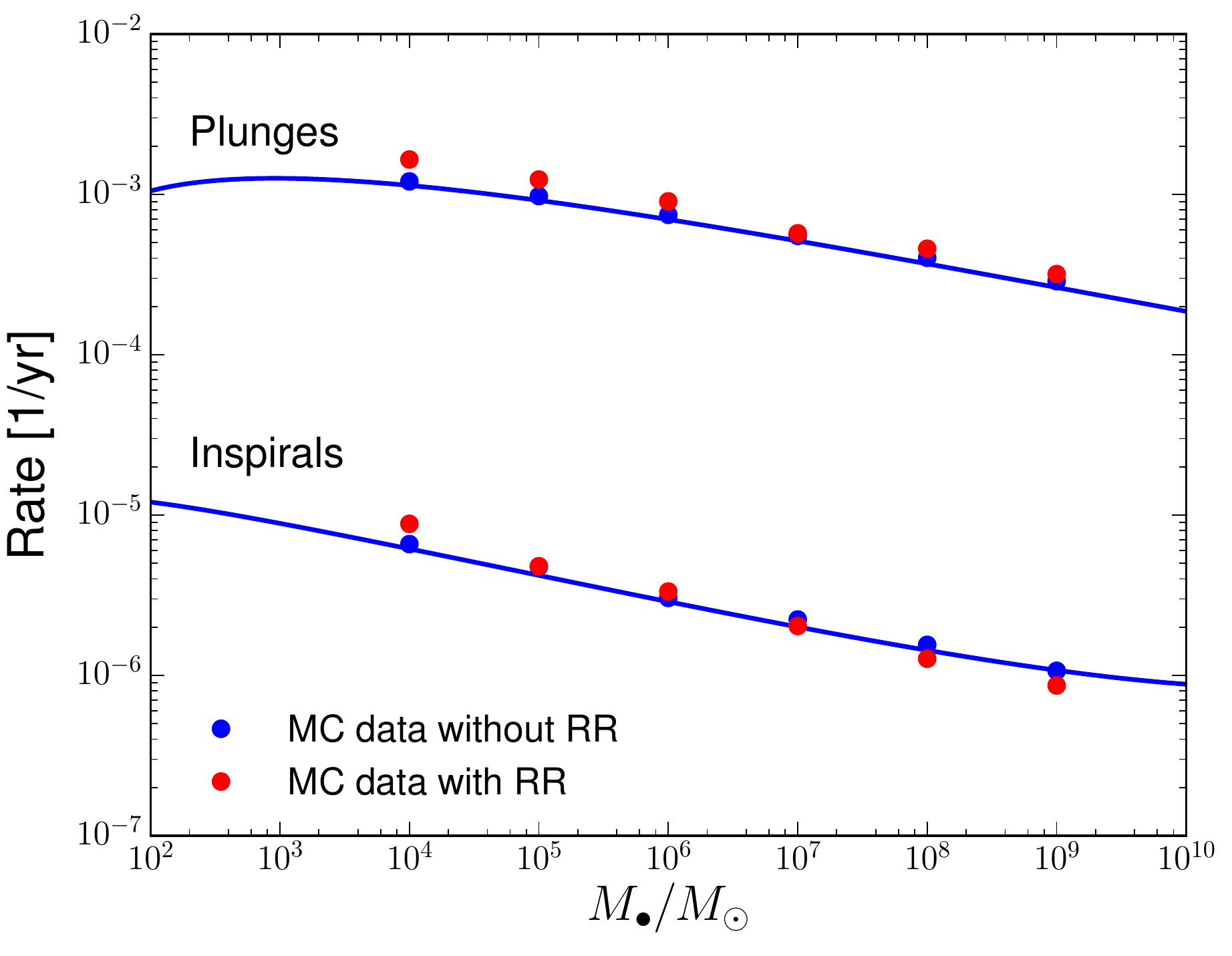}\protect\caption{\label{fig:MC-Rp-MBH}The total Plunge and inspiral rates as a function
of MBH mass. The MC simulations (circles) agree with the analytic
approximations for dynamics without RR (solid lines), Eqs.~\eqref{eq:RpMW},~\eqref{eq:RiMW}. Simulations with RR show that the contribution of
RR is small: the discrepancy between the rates with and without RR
does not exceed $\sim30\%$ over 5 orders of magnitude in $\protect\Mbh$. }
\end{figure}

In our MC simulations, we adopted for simplicity $\beta=4$, $\mu_{h}=2$,
$\eta_{h}=1$, $M_{0}=5.4\times10^{6}\,\Mo$ and $\sigma_{0}=100\,\mathrm{km\,s^{-1}}$.
Thus 
\begin{equation}
r_{h}=2\left(M_{\bullet,MW}\right)^{1/2}\,\mathrm{pc},
\end{equation}
where $M_{\bullet,MW}=M_{\bullet}/4\times10^{6}M_{\odot}$ is the
MBH mass scaled to the mass of the Galactic MBH\@. The rates as function
of the MBH mass $\Mbh$ and the mass ratio $Q=\Mbh/\Ms$ are then
\begin{eqnarray}
R_{p}^{\mathrm{tot}} & = & 3\times10^{-4}M_{\bullet,MW}^{-1/4}\nonumber \\
 &  & \times\frac{\log Q}{6.70-0.25\log\left(M_{\bullet,MW}\right)}\,\mathrm{yr^{-1}}\,,\nonumber \\
\label{eq:RpMW}
\end{eqnarray}
and
\begin{eqnarray}
R_{i}^{\mathrm{tot}} & \approx & 5\times10^{-6}M_{\bullet,MW}^{-1/4}\nonumber \\
 &  & \times\frac{\left(\log Q\right)^{1/5}}{4.65-0.25\log\left(M_{\bullet,MW}\right)-2\log\left(\log Q\right)/5}\,,\nonumber \\
\label{eq:RiMW}
\end{eqnarray}
where we used the value $A_{GW}\approx0.029$, corresponding to the
GW dissipation approximation of~\citet{gai+06} (Appendix~\ref{a:GWline}).
As shown in Figure~\ref{fig:MC-Rp-MBH}, these analytic approximations
are in agreement with the results of the MC simulations over several
orders of magnitude of $M_{\bullet}$.

\makeatletter{}

\global\long\def\Mo{M_{\odot}}
\global\long\def\Ro{R_{\odot}}
\global\long\def\Lo{L_{\odot}}
\global\long\def\Mbh{M_{\bullet}}
\global\long\def\Ms{M_{\star}}
\global\long\def\Rs{R_{\star}}
\global\long\def\Ts{T_{\star}}
\global\long\def\vs{v_{\star}}
\global\long\def\s{\sigma_{\star}}
\global\long\def\n{n_{\star}}
\global\long\def\tr{t_{\mathrm{rlx}}}
\global\long\def\Ns{N_{\star}}
\global\long\def\ns{n_{\star}}
\global\long\def\jlc{j_{lc}}

\section{Discussion and summary}

\label{s:discussion}

The determination of the steady-state of galactic nuclei is a fundamental
open question in stellar dynamics, with many implications and ramifications,
and has been the focus of numerous numerical and analytical studies.
In particular, current estimates of loss-rates vary over several orders
of magnitude due to theoretical and empirical uncertainties. Previous
studies either used post-Newtonian $N$-body simulations, which are
limited to small-$N$, or did not include the relevant relativistic
physics (Section~\ref{s:intro}). Building on recent progress in the
formal description of RR as a correlated diffusion process (the $\boldsymbol{\eta}$-formalism,~\citealt{bar+14}), we obtain here a MC procedure and analytic expressions
for the steady-state distribution and loss-rates in galactic nuclei,
taking into account two-body relaxation, RR, mass precession and the
GR effects of in-plane precession and GW emission. By cross-validating
the analytic estimates and the MC results with a high degree of accuracy,
and without the introduction of any free fit parameters, we are able
to confirm our analysis and interpretation of the dynamics of the
loss-cone in the context of our underlying assumptions.

\subsection{Discussion of main results}

\label{ss:mainresults}

The advantage of modeling RR by the $\boldsymbol{\eta}$-formalism,
over previous attempts by other approaches~\citep{rau+96,hop+06a,gur+07,mad+11,mer+11,ant+13,ham+14,mer15a,mer15b},
is that it allows to derive the FP equation rigorously from the stochastic
leading-order relativistic 3D Hamiltonian. The resulting effective
DCs, which are thus derived from first principles, are then guarantied
to obey the fundamental fluctuation-dissipation relation and the correct
3D maximal entropy solution~(\citealt[Section 7.4.3]{bin+08}; Appendix~\ref{a:MaxEntropy}).

These constraints on the functional form of valid DCs are critical,
since the correct steady-state is the result of a fragile near-cancellation
of two large opposing currents (the diffusion and drift); even small
deviations from this relation (e.g., due to approximations, empirical
fits, or reduction to lower dimensions), will result in large errors.
For example,~\citet{ham+14} obtained the RR DCs from numerical simulations
using an assumed functional form, $D_{jj}\propto\sqrt{1-j^{2}}$,
based on the fit of~\citet{gur+07}\footnote{We obtain a more accurate expression for $D_{jj}$ (Appendix~\ref{a:RRtorque}),
which fits torques measured in static wires simulations very well,
over the entire range $j\in[0,1]$. } and on the ad hoc expression $D_{j}=j^{-1}D_{jj}$, which is inconsistent
with the fluctuation-dissipation relation and therefore leads to invalid
steady-sate solution. This was then partially remedied by~\citet{mer15a}
who treated separately the Newtonian ($j\to1$) and relativistic ($j\to0$)
regimes. In the Newtonian regime, the~\citet{ham+14} data was re-fitted
to DCs that effectively satisfy the fluctuation-dissipation relation,
which means that in the absence of a loss-cone, the dynamics asymptote
to the maximal entropy limit $n(j)=2j$. However, in the relativistic
limit $j\to1$, where the simulation statistics are poorer due to
the smaller phase-space volume,~\citet{mer15a} used analytic DCs
based on the Hamiltonian model of~\citet{mer+11}, which represented
the stochastic background by an ad hoc dipole pseudo-potential and
a recipe for switching its direction every coherence time. This recipe
corresponded to the $\boldsymbol{\eta}$-formalism's ``Steps'' or
``Exponential ACF'' noise (depending on the exact switching procedure),
which both converge to the same form in the $j\to0$ limit~\citep[Figure 1]{bar+14}.
As shown by~\citet[Eq. 42]{bar+14}, in that limit $D_{jj}\approx j^{4}/\tilde{T}_{c}$
and $D_{j}\approx(5/2)j^{3}/\tilde{T}_{c}$, where $\tilde{T}_{c}=0.5T_{c}\nu_{GR}^{2}\left(j=1\right)/\nu_{j}^{2}\left(j=0\right)$.
This indeed satisfies the fluctuation-dissipation relation, as any
Hamiltonian model is guaranteed to do. These DCs are different from
the ones derived by~\citet{mer15a}, $D_{jj}\propto j^{4}/T_{c}$
and $D_{j}=2D_{jj}/j$, who implicitly forced the solution to 2D in-plane
motion by setting $\sin i=1$ in the derivation~\citep[Eqs.C.8-C.9]{mer15a}.
Therefore, these derived DCs satisfy the 2D fluctuation-dissipation
relation $2D_{j}=\partial D_{jj}/\partial j$, rather than the correct
3D one, $2jD_{j}=\partial jD_{jj}/\partial j$. These DCs therefore
imply the steady-state solution $n(j)=\mathrm{const}$ in the relativistic
regime (assuming no loss-cone); this is \emph{not} the correct solution
for 3D orbital motion (the correct one is $n(j)=2j$). We note that
the concatenation of two diffusion solutions, a 3D one for the Newtonian
regime and a 2D one for the relativistic regime, may create an artificial
discontinuity in the dynamical behavior at the interface.

We have shown that the representation of stochastic dynamics near
a MBH in terms of the streamlines of the probability flow provides
a powerful tool for analyzing the loss fluxes, and leads to the identification
of the exact separatrix between plunges and inspirals. We show that
the typical sma of this separatrix, $a_{GW}$, yields an excellent
analytical estimate for the inspiral rates found in our MC simulations
(Figure~\ref{fig:MC-Rp-MBH}). This remedies the ambiguity in the
identification of the critical sma, which was used to estimate rates
in previous studies. We also explored the effect of different GW dissipation
approximations on $a_{GW}$ and the resulting GW inspiral rate, and
found that the rates are robust to within a factor 2. Nevertheless,
it is worth noting that the more accurate method of~\citet{gai+06}
predicts EMRI rates that are more than twice higher than those of
the commonly used approximation of~\citet{pet+63}. 

We have shown that GR precession plays a critical role in the dynamics
of the loss-cone by efficiently quenching the RR torques. Conversely,
in its absence all stars would rapidly plunge into the MBH, creating
a depleted central cavity (cf Figure~\ref{f:MW}). This implies that
GR precession is important even in systems that are effectively Newtonian,
where at any given time only a very small fraction of the stars are
on relativistic orbits. In particular, $N$-body simulations of stellar
dynamics and stellar populations near MBHs should include GR precession
even if the questions of interest are in the non-relativistic regime,
so that plunges do not compete, or limit the lifetime of the stars.
It is worth noting that GR precession has not yet been tested empirically
for relativistic parameters larger than $\Upsilon=2G\Mbh/c^{2}r>8\times10^{-6}$
(the double pulsar system PSR J0737-3039,~\citealt{lyn+04}), whereas
the S-stars in the Galactic center, for example, are already observed
to reach $\Upsilon=2G\Mbh/c^{2}r>1\times10^{-3}$ at periapse (star
S14,~\citealt{gil+09}, see also~\citealt{ale06}). Moreover, near
the relativistic loss-cone ($r_{p}=4r_{g}$,~\citealt{gai+06}), where
$\Upsilon\to1/2$, there is to date no empirical confirmation of GR
precession~\citep{wil06}. The existence, dynamics and loss-rates
of stars on such relativistic orbits can therefore probe GR precession
in the strong field limit.

We have shown that the influence of RR on steady-state loss-cone dynamics
of compact objects is a small ${\cal O}(1)$ effect, since the loss-lines
for both direct plunge and GW inspiral lie well outside the region
where RR is effective (e.g., Figure~\ref{f:EJ}). RR does introduce
a correction to the steady-state distribution and the loss-rates,
which is at present small in comparison to the astrophysical uncertainties,
such as stellar density and mass function, the $\Mbh/\sigma$ relation,
and deviations from spherical symmetry. Eqs.~(\ref{eq:Rp}--\ref{eq:RiMW})
provide useful analytic estimates for the plunge and inspiral rates
per galaxy, based on the NR-only approximation. The RR correction
to the rates can be obtained by numerical integration of Eq.~\eqref{eq:RpRR}.
Using a suite of MC simulations, we have verified in the context of
our assumptions that the rate estimates are robust under different
assumptions about the properties of the stellar background noise (smoothness,
coherence time). This is in large measure a reflection of the limited
role of RR in the presence of continuous noise~\citep[Figure 1]{bar+14},
which restricts the domain where RR is effective, so that slow NR
remains the bottleneck and sets the rates. 

\begin{figure}
\begin{centering}
\begin{tabular}{c}
\includegraphics[width=0.95\columnwidth]{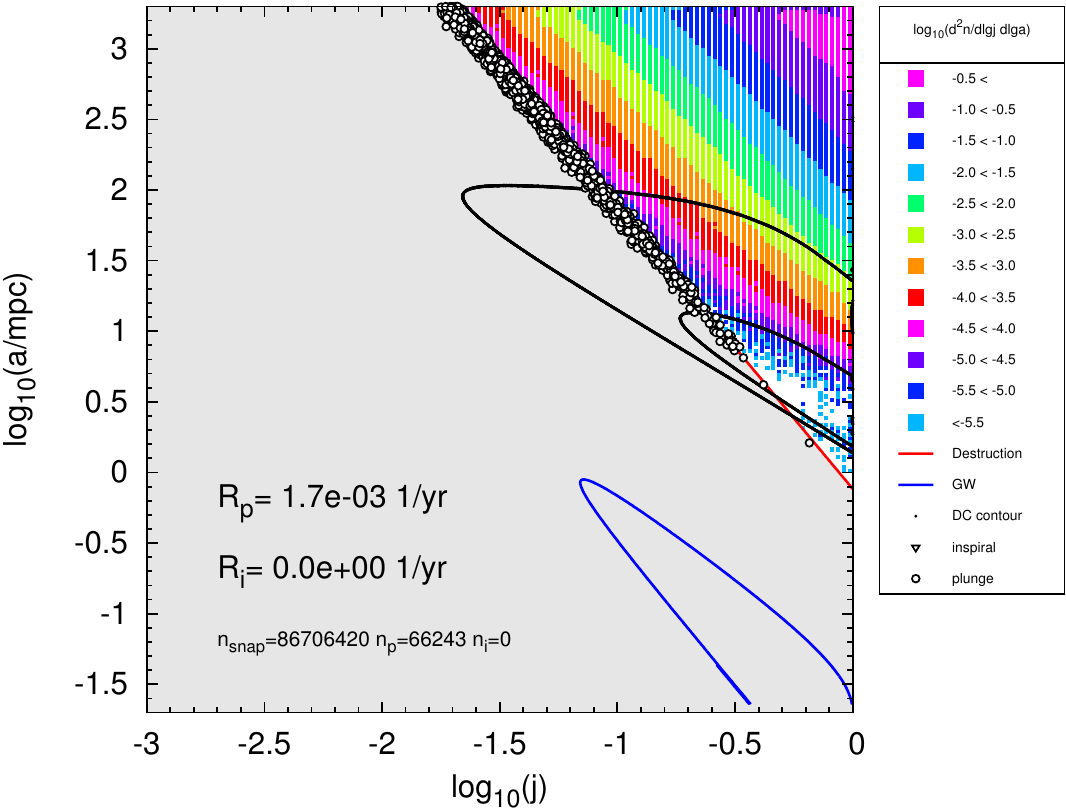}\tabularnewline
\includegraphics[width=0.95\columnwidth]{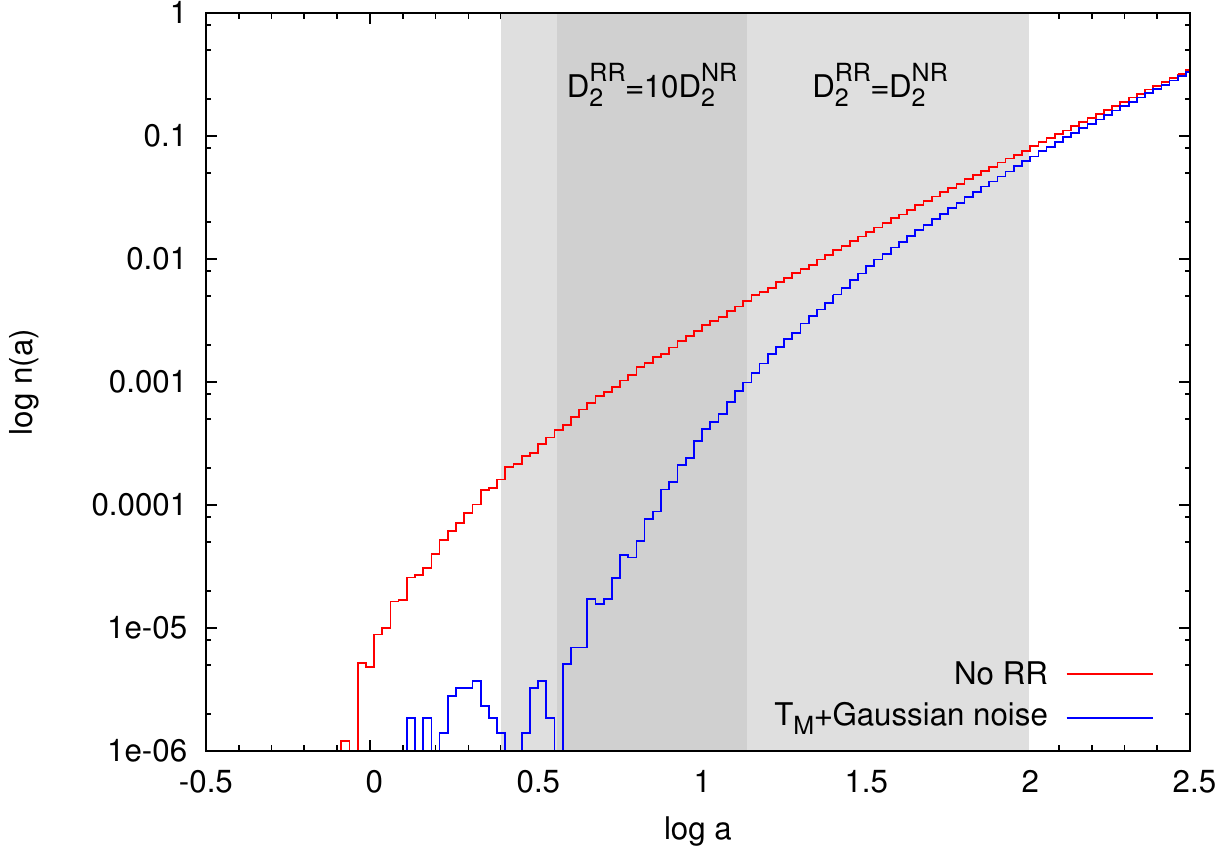}\tabularnewline
\end{tabular}
\par\end{centering}

\protect\caption{\label{f:disk}The suppression of the phase-space density of icy planetesimals
around a MBH due to RR-driven interactions with a circumnuclear accretion
disk. Top: The phase-space density and the disk interaction loss-line.
The black contours delineate the region where RR dominates ($D_{2}^{RR}/D_{2}^{NR}=1,10$).
Bottom: The resulting steady-state $a$-distribution of the planetesimals
for the NR-only case, and for the full dynamical model (top panel).
The shaded regions denote the $a$-intervals where the loss-line crosses
regions where RR dominates.}
\end{figure}

RR can substantially affect processes whose loss-lines intersect the
phase-space region where RR dominates ($D_{2}^{RR}\gg D_{1}^{RR}$
). The loss lines for tidal disruption of extended objects by the
MBH, such as red giants or binaries, do lie closer to the RR line.
However, it can be readily shown that neither class of objects is
long-lived enough for RR-driven tidal destruction to play a dominant
role in Milky Way-like galactic centers. Red giants are relatively
short-lived, and the more extended and tidally-susceptible they are,
the shorter their lifespan. Soft stellar binaries are destroyed by
3-body ionization before they are affected by RR-driven tidal separation~\citep{ale+13}. 

One class of processes where RR may have more than an ${\cal O}(1)$
effect is the hydrodynamical destruction (or removal) of objects by
interaction with a large circumnuclear accretion disk. To demonstrate
this point, we consider here as an idealized simple example the case
of a massive accretion disk of radius $R_{d}=2000r_{g}$~\citep{goo03}
and a population of long-lived icy planetesimals around it, which
are destroyed by several consecutive disk crossings (it is assumed
that the number of crossings for destruction is $N_{\mathrm{cross}}P\ll T_{NR}$).
In that case, the critical angular momentum for destruction is $j_{d}=\sqrt{2G\Mbh R_{d}}=15.8j_{lc}$,
which is large enough to intersect the RR-dominated zone (Figure~\ref{f:disk}
top). As a result, the differential sma distribution of the planetesimals
is depleted below $a\lesssim100\,\mathrm{mpc}$ and strongly so below
$a\lesssim10\,\mathrm{mpc}$ (Figure~\ref{f:disk} bottom).

Although RR is typically inefficient in driving stars all the way
to the loss-cone, it can randomize orbits in the phase-space regions
where it dominates, even when the NR timescale is longer than the
system's age or the lifespan of the stars. This may be a key element
in solving the ``paradox of youth''~\citep{ghe+03a} in the Galactic
Center, where young B stars are observed on tight orbits around SgrA$^{\star}$
(the so-called ``S-star'' cluster). The leading formation or migration
scenarios for the S-stars predict non-isotropic initial eccentricities~\citep[see reviews by][]{ale05,ale11}. However, most of the S-stars
are in the RR-dominated phase-space region (Figure~\ref{f:EJ}), and
so substantial evolution and isotropization of the initial eccentricities
is possible. However, many of the S-stars are short-lived, and some
are close in phase-space to the AI-suppressed region. A detailed analysis
using the $\boldsymbol{\eta}$-formalism, which can treat this intermediate
dynamical regime rigorously, has yet to be carried out.

\subsection{Limitations and caveats}

\label{ss:caveats}

The applicability and validity of our results are limited by several
simplifying assumptions. We assume a non-spinning MBH surrounded by
an isotropic (on average) non-rotating single Keplerian power-law
cusp of single-mass stars. We assume the dynamics are dominated by
single star interactions, that is, we neglect the possible effects
of binaries, or the contribution of non-stellar massive perturbers,
such as gas clouds, cluster or intermediate mass BHs~\citep{per+07}.

In our MC simulations we used NR and RR DCs, which are based on a
fixed isotropic, single power-law BW76 model. These DCs are therefore
not self-consistent with the steady-state solution. However, we showed
that the BW76 cusp is a good approximation (within a factor of two)
for the steady-state solution in the relevant region (down to $a_{BW}<a_{GW}$
for $\Mbh\lesssim10^{9}M_{\bullet}$; see Section~\ref{ss:SSNR}).
In addition, it can be shown that for the RR DCs the isotropic fluctuation-dissipation
relation holds even for the non-isotropic case as long as the total
angular momentum of the system is zero~\citep{bar+14}. Since the
fluctuation-dissipation relation results from the symmetries of the
Hamiltonian, it is reasonable to assume that the same will hold also
for the NR DCs. This means that small non-isotropies will only result
in small magnitude changes of the flux and have little effect on the
steady-state distribution.

Finally, the RR DCs are based on simplified (single timescale) background
noise models that can be treated analytically, and simple coherence
models that are functions of the sma only. However, we were able to
show using MC simulations that the results are largely independent
of the exact noise model as long as it is continuous (i.e., not white
noise).

\makeatletter{}

\global\long\def\Mo{M_{\odot}}
\global\long\def\Ro{R_{\odot}}
\global\long\def\Lo{L_{\odot}}
\global\long\def\Mbh{M_{\bullet}}
\global\long\def\Ms{M_{\star}}
\global\long\def\Rs{R_{\star}}
\global\long\def\Ts{T_{\star}}
\global\long\def\vs{v_{\star}}
\global\long\def\s{\sigma_{\star}}
\global\long\def\n{n_{\star}}
\global\long\def\tr{t_{\mathrm{rlx}}}
\global\long\def\Ns{N_{\star}}
\global\long\def\ns{n_{\star}}
\global\long\def\jlc{j_{lc}}

\acknowledgements{T.A. acknowledges support by the I-CORE Program of the PBC and ISF
(Center No. 1829/12). }

\appendix
\makeatletter{}

\global\long\def\Mo{M_{\odot}}
\global\long\def\Ro{R_{\odot}}
\global\long\def\Lo{L_{\odot}}
\global\long\def\Mbh{M_{\bullet}}
\global\long\def\Ms{M_{\star}}
\global\long\def\Rs{R_{\star}}
\global\long\def\Ts{T_{\star}}
\global\long\def\vs{v_{\star}}
\global\long\def\s{\sigma_{\star}}
\global\long\def\n{n_{\star}}
\global\long\def\tr{t_{\mathrm{rlx}}}
\global\long\def\Ns{N_{\star}}
\global\long\def\ns{n_{\star}}
\global\long\def\jlc{j_{lc}}

\section{The analytic gravitational wave line}

\label{a:GWline} 

The phase-space of the relativistic loss-cone is divided by a separatrix
into an outer region where streamlines end as plunges, and an inner
region where streamlines end as inspirals (Figure~\ref{f:flow2Dall}).
Since the probability current along an infinitesimal bundle of streamlines
is conserved (Section~\ref{s:EMRI-event-rates}), it can be evaluated
at any convenient point along the flow, in particular at $J\gg J_{lc}$,
where GWs are negligible, and the streamlines are identical to those
in the absence of GWs. Therefore, the inspiral rate is estimated by
locating the no-GW plunge streamline that corresponds to the separatrix,
and identifying the sma of its terminal point ($a_{p},j_{lc}$) as
the critical (maximal) sma for inspiral, $a_{GW}$. The inspiral rate
is then simply the integrated plunge rate in the absence of GW, up
to $a_{GW}$, i.e., $R_{i}^{\mathrm{tot}}=R_{p}^{no\,GW}\left(a_{GW}\right)$.

The separatrix streamline, $\mathrm{d}E/\mathrm{d}J=S_{E}/S_{J}$,
can be evaluated by noting that the flow in the $E$-direction is
mostly due to GW dissipation, 
\begin{equation}
S_{E}=n\left(E,J\right)E/t_{GW}\,,
\end{equation}
where $t_{GW}\equiv a/\left|\dot{a}_{GW}\right|$ is the GW orbital
decay timescale. In the presence of GW, the innermost plunge streamline
initially approaches $j_{lc}$ from $j\sim1$ at nearly constant $a$,
and then turns over to lower $a$ and runs nearly parallel to the
loss-line $j_{lc}(a)=4\sqrt{r_{g}/a}$ until it terminates when $j\to j_{lc}$
at some small enough terminal sma, $a_{insp}$, where the orbital
motion is effectively circular (i.e., inspiral, cf Figure~\ref{f:flow2Dall}).
For example, for parabolic orbits, the minimal possible value for
$a_{insp}$ is obtained where $j_{lc}=1$, i.e., for $a_{insp}=16r_{g}$.
As we demonstrate below (Eq.~\eqref{eq:aGWj1}), the separatrix solution
is a function of $a_{insp}$ only via $a_{insp}/a_{\max}\ll1$, and
is therefore independent of the exact choice of $a_{insp}$. 

The flow in the $J$-direction is approximately constant in $J$ with
a typical timescale $T_{J}$ (Eq.~\eqref{eq:S_J_djj}),
\begin{equation}
S_{J}\approx-\frac{n\left(E\right)}{T_{J}\log\left(J/J_{lc}\right)}\,.
\end{equation}
Since RR is negligible near $j=j_{lc}$, $T_{J}$ can be approximated
as $T_{J}\left(a\right)\approx\theta_{\gamma}Q^{2}P\left(a\right)/\left(N\left(a\right)\log Q\right)$
(see Eq.~\eqref{eq:SJ_nE}), where $\theta_{\gamma}$ is a numeric
pre-factor that depends on the cusp density profile ($\theta_{\gamma}\approx1/8$
for a $\gamma=7/4$ (BW76) cusp). 

It then follows that separatrix is the solution of the differential
equation
\begin{equation}
\frac{dE}{dJ}=\frac{S_{E}}{S_{J}}\approx-2\frac{J}{J_{c}^{2}}T_{J}\frac{E}{t_{GW}}\log\left(\frac{J}{J_{lc}}\right)\,,\label{eq:GW-diff}
\end{equation}
with the boundary condition $J(E_{insp})=J_{lc}$ (where $E_{insp}=G\Mbh/2a_{insp})$. 

An exact expression for the GW dissipation can be obtained only numerically.
Some useful analytical approximations are available (see~\citet{gai+06}
for a comparison between the different techniques). The simplest expression
was obtained by~\citet{pet+63}, who assumed a point-mass objects
moving on a Keplerian orbit. In this approximation the GW timescale
is 
\begin{equation}
t_{GW}\equiv\frac{a}{\left|\dot{a}_{GW}\right|}=\frac{1}{2\pi}\frac{5}{64}\frac{Q}{f(e)}\frac{r_{g}}{a}\left(\frac{2r_{p}}{r_{g}}\right)^{7/2}P\left(a\right)\,,\,\,\,f(e)=\left(\frac{1+e}{2}\right)^{-7/2}\left(1+\frac{73}{24}e^{2}+\frac{37}{96}e^{4}\right)\,,\label{eq:t_GW}
\end{equation}

In the limit $J\to J_{c}$, the periapse $r_{p}$ is related to the
angular momentum by $r_{p}/r_{g}=8J^{2}/J_{lc}^{2}$ and the streamline
equation (Eq.~\eqref{eq:GW-diff}) can be written in terms of $x=a/a_{\max}$
and $s=J/J_{lc}$,

\begin{equation}
\frac{dx}{ds}=A_{D}x^{\gamma-2}s^{-6}\log\left(s\right)\,,\label{eq:GW_line}
\end{equation}
where the competition between the GW dissipation and NR diffusion
is expressed by the parameter 
\begin{equation}
A_{D}=\frac{\pi}{20}f(1)\frac{\theta_{\gamma}Q}{N_{h}\log Q}\sim{\cal O}(10^{-3})\,.\label{e:A2}
\end{equation}
The boundary condition at the terminal sma is 
\begin{equation}
x_{1}\equiv x\left(s=1\right)=a_{insp}/a_{\max}\ll1,
\end{equation}
and the plunging branch of the solution is give by
\begin{eqnarray}
x\left(s\right) & = & \left[x_{1}^{3-\gamma}+\frac{A_{D}\left(3-\gamma\right)\left(s^{5}-5\log s-1\right)}{25s^{5}}\right]^{1/\left(3-\gamma\right)}\,.
\end{eqnarray}

In the phase-space region where GW is negligible, the streamline defined
by Eq.~\eqref{eq:GW-diff} is approximately constant in $E$. Therefore,
$a_{GW}$ can be identified by the sma at $s\gg1$ ($j\to1$), that
is 
\begin{equation}
a_{GW}\approx a_{\max}\lim_{s\to\infty}x=\left(x_{1}^{3-\gamma}+\frac{\left(3-\gamma\right)}{25}A_{D}\right)^{1/\left(3-\gamma\right)}a_{\max}\,.\label{eq:aGWj1}
\end{equation}
For a steady-state cusp (i.e., BW76, $\gamma=7/4$), $x_{1}^{3-\gamma}\ll(3-\gamma)A_{D}/25$,
and therefore
\begin{equation}
a_{GW}\approx A_{GW}\left(\frac{N_{h}\log Q}{Q}\right)^{-4/5}a_{\max}\label{eq:aGW_approx}
\end{equation}
where the value of $A_{GW}$ depends on the specific GW dissipation
approximation that is assumed. For the Keplerian approximation~$A_{GW}^{K}=(\pi f(1)/3200)^{4/5}\approx0.013$.

\begin{figure}
\begin{centering}
\includegraphics[width=0.5\textwidth]{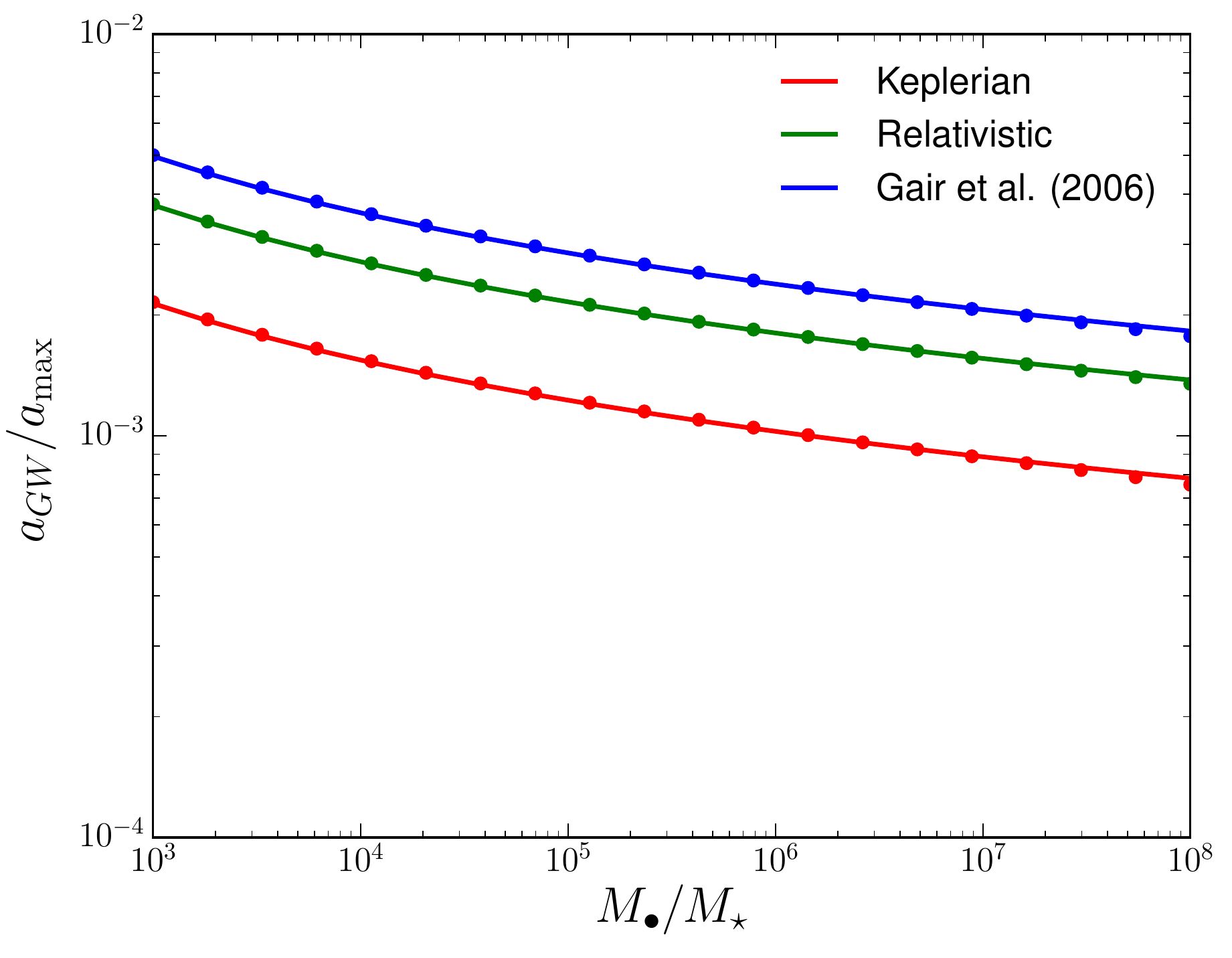}\protect\caption{\label{fig:aGW}The semi-major axis $a_{GW}$ of the outermost inspiraling
streamline for a BW76 cusp ($\gamma=7/4$) for different MBH to star
mass ratios and for different approximations of the GW dissipation.
The approximate expression (Eq.~\eqref{eq:aGW_approx}) for $a_{GW}$
(lines) is fitted to values of $a_{GW}$ found by numerically integrating
the streamline equation (Eq.~\eqref{eq:GW-diff}) (circles).}

\par\end{centering}

\end{figure}

As shown by~\citet{gai+06}, the~\citet{pet+63} estimate can be improved
by using a \textquoteleft \textquoteleft semi-relativistic\textquoteright \textquoteright{}
approximation, that is using the fully relativistic orbit in place
of the Keplerian one in~\citet{pet+63} equations. In the limit $e\to1$,
this approach, used by~\citet{hop+05}, amounts to replacing the Keplerian
$r_{p}/r_{g}$ in Eq.~\eqref{eq:t_GW} with the relativistic one~\citep{gai+06}
\begin{equation}
r_{p}/r_{g}=4s\left(s+\sqrt{s^{2}-1}\right)\,.
\end{equation}
Thus the GW separatrix (Eq.~\eqref{eq:GW_line}) is replaced by its
relativistic version, 
\begin{equation}
\frac{dx}{ds}=2^{7/2}A_{c}x^{\gamma-2}\left(1+\sqrt{1-1/s^{2}}\right)^{-7/2}s^{-6}\log\left(s\right)\,,
\end{equation}
where $a_{GW}$ can be solved numerically. As shown in Figure~\ref{fig:aGW},
$a_{GW}$ can be approximated for $\gamma=7/4$ by Eq.~\eqref{eq:aGW_approx}
with $A_{GW}^{R}=0.022$ and for the more accurate treatment of~\citet{gai+06},
$A_{GW}=0.029$, which is adopted in this study.

\makeatletter{}

\global\long\def\Mo{M_{\odot}}
\global\long\def\Ro{R_{\odot}}
\global\long\def\Lo{L_{\odot}}
\global\long\def\Mbh{M_{\bullet}}
\global\long\def\Ms{M_{\star}}
\global\long\def\Rs{R_{\star}}
\global\long\def\Ts{T_{\star}}
\global\long\def\vs{v_{\star}}
\global\long\def\s{\sigma_{\star}}
\global\long\def\n{n_{\star}}
\global\long\def\tr{t_{\mathrm{rlx}}}
\global\long\def\Ns{N_{\star}}
\global\long\def\ns{n_{\star}}
\global\long\def\jlc{j_{lc}}

\section{Monte-Carlo simulations of loss-cone dynamics}

\label{a:MCproc}

We summarize here the assumptions underlying our Monte-Carlo (MC)
simulations, the details of the implementation and the derivation
of the steady-state configuration and loss-rates.

\subsection{Assumptions and procedure}

The MC simulations in $(a,j)$ assume a fixed background model, whose
properties define the NR diffusion coefficients, an RR coherence timescale
model and a background noise model. The stellar background is approximated
as a power-law cusp with enclosed number of stars $N(a)=N_{\max}(a/a_{\max})^{3-\alpha}$,
extending between $a_{\min}$ and $a_{\max}$.  The accessible phase-space
for the test stars extends between a reflecting boundary at $a_{\mathrm{out}}=a_{\max}$
(evaporation), and an absorbing boundary at $a_{\mathrm{in}}<a_{\min}$
(destruction). The reflective boundary at $a_{\mathrm{out}}$ ensures
that the long-term distribution of test stars converges to an isotropic
distribution in $j$, $n(j)\to2j$, which is the assumed distribution
of stars far from the MBH (note that this is not guaranteed for an
absorbing outer boundary, even when the test stars are introduced
into the simulation isotropically).  The extension of phase-space
to small $a_{\mathrm{in}}$ allows inspiral trajectories to be tracked
down to a tight enough sma where their ultimate fate as EMRIs is certain
(cf Figure~\ref{f:2D}) phase-space extends in $j$ between a reflecting
boundary at $j=1$ and an absorbing boundary at the last stable orbit
(LSO), $j_{LSO}$, defined by the locus $J_{LSO}(a)=\sqrt{a/r_{g}}j_{LSO}=4r_{g}c$
(the LSO for a zero energy orbit in the Newtonian approximation; in
the limit $e\to1$ it corresponds to a critical periapse $r_{LSO}=a(1-e)=8r_{g}$). 

An MC run starts by injecting a test star in some initial phase-space
position $(a_{0},j_{0})$. This is typically chosen randomly, either
just below $a_{\max}$, with $j$ distributed isotropically above
$j_{LSO}$, to simulate a star diffusing from an isotropic galaxy
into the MBH's radius of influence, or isotropically in the cusp's
bulk, according to the power-law cusp distribution, to simulate the
initial conditions of an $N$-body simulation (see below), or the
distribution of tracer stars (e.g., red giants), which mirror the
distribution of cusp stars. The star is then evolved in small time
increments $\mathrm{d}t$, taking into account the stochastic changes
in energy and angular momentum due to NR, the deterministic changes
due to GW dissipation, and the random changes in angular momentum
due to RR\@. The star is tracked in phase-space, and its position
is recorded by snapshots taken at fixed intervals $\Delta t\gg\mathrm{d}t$.
Ultimately, after surviving for time $t_{s}$, the star leaves the
system at some terminal phase-space position $(a_{1},j_{1})$ as a
result of one of four possible outcomes. $(1)$ Evaporation. The star's
sma crosses $a_{\mathrm{out}}$ to a larger sma. This happens for
the majority of test stars. (2) Inspiral. The star's orbit decays
until it crosses $a_{insp}=\epsilon a_{\min}$ mpc ($\epsilon\ll1$).
(3) Plunge. The star crosses the LSO directly, at some $a_{p}>a_{insp}$.
(4) Finite lifespan exceeded. This is relevant for stars, which are
limited by stellar evolution or for binaries, which are also limited
by dynamical evaporation (Section~\ref{ss:mainresults}), but not
for compact remnants. Note that the branching ratio between plunges
and inspirals is independent of the exact value of the sma chosen
to distinguish between the two outcomes (parameterized by $\epsilon$),
since their respective terminal phase-space positions are clearly
separated ($\min a_{\mathrm{plunge}}\gg a_{\mathrm{inspiral}}$, cf
Figure~\ref{f:2D}; $\epsilon=0.1$ was typically used here). 

Once a star exits the system, a new star is injected (reflection at
$a_{\mathrm{out}}$ is equivalent to injection of a new star at $(a_{\mathrm{out}},j_{1})$).
This is repeated over some long accumulated time $T_{\mathrm{sim}}\gg t_{s}$
(typically $T_{\mathrm{sim}}\sim100T_{E}$, Section~\ref{ss:SSNR}).
All test stars simulated over $T_{\mathrm{sim}}$, apart for the last
one, which is omitted from the analysis, reach a definite outcome.
For each test star we record its trajectory in phase-space (in coarse
$\Delta t$ resolution); the total time it spent evolving in phase-space,
$t_{s}$; the nature of the final outcome (evaporation, inspiral,
plunge or end of lifetime); and its initial and terminal phase-space
positions. The procedure is repeated as needed (typically $10^{3}-10^{4}\times T_{\mathrm{sim}}$),
until enough test star statistics are collected. The snapshots of
the phase-space positions, the survival times and the final outcomes
are then used to estimate the steady-state configuration and loss-rates,
as detailed below.

\subsection{Representation of physical processes}

The form of the RR DCs depend on the background noise model and the
precession of the test star~\citep{bar+14}. We assume three optional
noise models: white noise (equivalent to no precession), $C^{0}$
noise with exponential ACF (an Ornstein-Uhlenbeck process), and smooth
$C^{\infty}$ noise with a Gaussian ACF\@. Prograde GR in-plane precession
is modeled by the 1PN approximation $\nu_{GR}=3\nu_{r}(a)(r_{g}/a)/j^{2}$,
where $\nu_{r}(a)$ is the Keplerian radial (orbital) frequency. Mass
precession is given exactly for an $\alpha=2$ cusp, $\nu_{M}(a,j)=-[N(a)/Q]\nu_{r}j/(1+j)$~\citep{mer+11}, and for other values of $\alpha$ by polynomial approximations
of the exact integral~\citep{ale05}. The magnitude of the RR DCs
reflects the strength of the RR torques, $\tau_{N}\simeq0.28\sqrt{1-j}\sqrt{N(2a)}G\Ms/a$,
as derived from static wire simulations (Appendix~\ref{a:RRtorque}). 

The RR DCs have explicit analytic forms that are easy to evaluate.
However, the NR DCs (Appendix~\ref{a:2bDCs}) involve multiple integrations
that are too computationally expensive to perform on the fly. We therefore
calculate them exactly beforehand on a $25\times25$ evenly-spaced
logarithmic grid extending between $a_{\min}$ and $a_{\max}$ and
$j_{\min}=10^{-3}$ and $j_{\max}=1$, and then bi-linearly interpolated
to any $(a,j)$ as needed.

Three optional analytic perturbative estimates for the rate of GW
dissipation of energy and angular momentum were studied: those of~\citet{pet64},~\citet{hop+06a}, and~\citet{gai+06}.

Each MC step, the time-step was chosen as $\Delta t=\min\left(\Delta t_{NR},\Delta t_{RR},\Delta t_{GW}\right)$,
where the time-steps for the NR and RR were chosen by criteria similar
to those used by~\citet{sha+78} for NR, and the time-step for GW
was chosen to be a small fraction of the GW dissipation times $\min(E/\dot{E}_{GW},J/\dot{J}_{GW})$.

Two optional approximate background coherence timescale models are
considered, which are assumed to be functions of $a$ only: $j$-averaged
mass precession, $T_{M}=\sqrt{\pi/2}\nu_{M}^{-1}(2a,\sqrt{1/2})$,
and self-quenching $T_{SQ}=Q\nu_{r}^{-1}(a)/\sqrt{N(2a)}$, where
the approximate numeric pre-factors~(\citealt[footnote 7]{bar+14})
were evaluated here specifically for $\alpha=7/4$, but are generally
insensitive to the exact value of $\alpha$. Note that GR precession
is not included in background coherence models of the MC, but it is
included in approximate form ($\nu_{\mathrm{prec}}=|\nu_{M}+\nu_{GR}|$)
in the analytic modeling of the coherence time (Section~\ref{s:analytic}).

\subsection{Steady state rate estimates}

\label{ss:SSrates}

The interpretation of the MC results in terms of loss-rates depends
on whether the test stars represent the underlying background cluster
that is generating the NR and RR perturbations, or whether they are
a separate trace population that is affected by the perturbations,
but does not contribute to them.

\subsubsection{Test stars as background}

In statistical steady-state, stars that exit the system ($a_{\mathrm{in}}<a<a_{\mathrm{out}}$)
are replaced at a rate that keeps their time-averaged number $N$
fixed. A star that evaporates from the system back to the infinite
reservoir at $a>a_{\mathrm{out}}$ (reflection at $a_{\mathrm{out}}$
is treated as evaporation), is replaced by another star from the reservoir,
so there is no net current through $a_{\mathrm{out}}$ due to evaporation.
The situation is different for stars that end up in the MBH, whether
by plunge or inspiral. Since they are permanently removed from the
system, a net current of stars through $a_{\mathrm{out}}$, from the
reservoir into the system, is required to compensate for their loss.
Both the stars that evaporate and those that are lost\footnote{To simplify bookkeeping, a test star that would have wandered back
and forth across $a_{\mathrm{out}}$ and is finally destroyed by the
MBH is not counted as a single star (i.e., a single trajectory). Once
it leaves the cusp ($a>a_{\mathrm{out}}$), it is considered as evaporated.
The next crossing into the cusp $a<a_{\mathrm{out}}$ is identified
as the beginning of the phase-space trajectory of a new star.} contribute to the total mean number of stars in steady-state. In
our models, this number is an input parameter, determined by the assumed
background cusp.

Designate by $P_{k}$ the probability (branching ratio) for outcome
$k=0,1,2,\ldots$, where $k=0$ denotes evaporation and $k>0$ denote
the various loss channels, so that $\Sigma_{k}P_{k}=1$. The MC statistics
provide estimates of the branching ratios, $P_{k}=n_{k}/n_{\mathrm{sim}}$,
where $n_{\mathrm{sim}}$ is the total number of test stars whose
phase-space trajectories were simulated, and $n_{k}$ the number of
times outcome $k$ has been reached. This translates to event rates
by requiring that the total number of stars be on average $N=\sum_{k}N_{k}=\sum_{k}\Gamma_{k}\bar{t}_{k}$,
where $\Gamma_{k}$ is the rate of outcome $k$, and $\bar{t}_{k}=n_{k}^{-1}\sum_{j}t_{k}^{(j)}$
is the mean survival time in the $n_{k}$ simulations that had outcome
$k$. The rate for each channel is related to the total rate of all
outcomes, $\Gamma$, by $\Gamma_{k}=\Gamma P_{k}$, where $\Gamma=N/\bar{t}_{s}$
and $\bar{t}_{s}$ is the overall mean survival time, irrespective
of outcome\footnote{This follows from summation over all channels: $N=\sum_{k}\Gamma_{k}\bar{t}_{k}$
$=\Gamma\sum_{k}P_{k}\bar{t}_{k}$ $=\Gamma\sum_{k}(n_{k}/n_{\mathrm{sim}})(n_{k}^{-1}\sum_{j}t_{k}^{(j)})$
$=\Gamma(\sum_{k,j}t_{k}^{(j)})/n_{\mathrm{sim}}$ $=\Gamma\bar{t}_{s}$.}. It then follows that the event rates are 
\begin{equation}
\Gamma_{k}=(N/\bar{t}_{s})P_{k}\,.
\end{equation}
The total replenishment rate that is required to keep the system in
steady-state is then the sum over all the loss channels,
\begin{equation}
\Gamma_{\mathrm{loss}}=\sum_{_{k>0}}\Gamma_{k}\,.
\end{equation}

Loss-rates estimated by direct $N$-body simulations~\citep{mer+11,bre+14}
can in principle be compared to the rates derived by MC simulations
of scaled-down nuclei, where the test stars are treated as representative
of the background. However, this comparison is complicated by the
fact that the $N$-body systems are not necessarily in steady-state
(the initial configuration may not be the steady-state one, and /
or stars lost in the course of the simulation are not replenished),
and do not have fixed boundaries or boundary conditions. The comparisons
discussed in Section~\ref{ss:2DMC} and Table~\ref{t:MAMW11} are
approximate. The MC loss-rates were estimated for a non-equilibrium
cusp that corresponds to the initial conditions of the $N$-body simulations,
and the MC rates were compared to the rates early in the simulations,
at times where the $N$-body configuration is still close to its initial
state.

\subsubsection{Test stars as a trace population}

It is of interest to consider how a small population of tracer stars,
which are injected into the cusp by some dynamical or evolutionary
mechanism, evolves dynamically with time, and is lost via the various
channels. Some possible injection mechanisms are capture by tidal
separation of an incoming binary~\citep{hil88}, in which case the
injection point in phase-space is a tight eccentric orbit deep inside
the cusp; the formation of a red giant when a background star evolves
off the main sequence, in which case the injection point reflects
the background distribution of progenitors; or the formation of massive
blue giant in a fragmenting gas disk, in which case the injection
point is a low-eccentricity orbit.

When the test stars in the MC simulation represent a tracer population,
the total rate of all outcomes is determined by the assumed injection
rate, and is no longer related to the total number of background stars.
In this case the MC does not predict the event rates $\Gamma_{k}$,
but rather the branching ratios $P_{k}$.

\makeatletter{}

\global\long\def\Mo{M_{\odot}}
\global\long\def\Ro{R_{\odot}}
\global\long\def\Lo{L_{\odot}}
\global\long\def\Mbh{M_{\bullet}}
\global\long\def\Ms{M_{\star}}
\global\long\def\Rs{R_{\star}}
\global\long\def\Ts{T_{\star}}
\global\long\def\vs{v_{\star}}
\global\long\def\s{\sigma_{\star}}
\global\long\def\n{n_{\star}}
\global\long\def\tr{t_{\mathrm{rlx}}}
\global\long\def\Ns{N_{\star}}
\global\long\def\ns{n_{\star}}
\global\long\def\jlc{j_{lc}}

\section{Two-body diffusion coefficients}

\label{a:2bDCs}

We derive here the relations between DCs in velocity space and the
DCs in angular momentum and energy space for stars orbiting in a spherical
potential. 

Consider a star of mass $m$ moving in a spherical potential, $\phi=\phi\left(r\right)$,
with velocity $\mathbf{v}$. The binding energy and angular momentum
are 
\begin{eqnarray}
E & = & \phi\left(r\right)-\frac{1}{2}v^{2}\,,\\
\mathbf{J} & = & \mathbf{r}\times\mathbf{v}\,.
\end{eqnarray}

Note that here $E$ is the positively defined orbital energy and $\phi\left(r\right)$
is positively defined potential. Due to gravitational encounters with
the field stars, the star changes its velocity, $\mathbf{v}$, to
$\mathbf{v}^{\prime}=\mathbf{v}+\Delta\mathbf{v}$. Consider the orthonormal
basis 
\begin{eqnarray}
\hat{\mathbf{v}} & = & \mathbf{v}/v\,,\\
\hat{\mathbf{J}} & = & \mathbf{J}/J=\mathbf{r}\times\mathbf{v}/\left|\mathbf{r}\times\mathbf{v}\right|\,,\\
\hat{\mathbf{w}} & = & \mathbf{v}\times\mathbf{J}/\left|\mathbf{v}\times\mathbf{J}\right|=\mathbf{v}\times\left(\mathbf{r}\times\mathbf{v}\right)=\left(v\hat{\mathbf{r}}-v_{r}\hat{\mathbf{v}}\right)/v_{t}\,.
\end{eqnarray}
In this basis, the change in velocity is
\begin{equation}
\Delta\mathbf{v}=\Delta v_{\parallel}\hat{\mathbf{v}}+\Delta\mathbf{v}_{\bot}\,,
\end{equation}
where
\begin{equation}
\Delta\mathbf{v}_{\bot}=\Delta v_{J}\hat{\mathbf{J}}+\Delta v_{w}\hat{\mathbf{w}}\,,
\end{equation}
and
\begin{equation}
\Delta v_{\bot}=\sqrt{\left(\Delta v_{j}\right)^{2}+\left(\Delta v_{w}\right)^{2}}\,.
\end{equation}

The change in energy is
\begin{eqnarray}
\Delta E & = & -\frac{1}{2}\left(v^{\prime2}-v^{2}\right)=-\frac{1}{2}\left(\Delta v\right)^{2}-\mathbf{v}\cdot\Delta\mathbf{v}=-\frac{1}{2}\left(\Delta v_{\parallel}\right)^{2}-\frac{1}{2}\left(\Delta v_{\bot}\right)^{2}-v\Delta v_{\parallel}\,.\nonumber \\
\label{eq:DE}
\end{eqnarray}
The position vector, $\mathbf{r}$, is 
\begin{equation}
\mathbf{r}=r\left(v_{t}/v\right)\hat{\mathbf{w}}+r\left(v_{r}/v\right)\hat{\mathbf{v}}\,,
\end{equation}
where $v_{r}$ and $v_{t}$ are the radial and transversal velocities.
Therefore, the change in the radial velocity is 
\begin{equation}
\Delta v_{r}=\frac{\Delta\mathbf{v}\cdot\mathbf{r}}{r}=\frac{v_{r}}{v}\Delta v_{\parallel}+\frac{v_{t}}{v}\Delta v_{w}\,.
\end{equation}
The change in the transverse velocity up to second order in $\Delta v/v$
is
\begin{eqnarray}
\Delta v_{t} & = & v_{t}\frac{\Delta v_{\parallel}}{v}-v_{r}\frac{\Delta v_{w}}{v}+\frac{1}{2}\frac{\Delta v_{J}^{2}}{v_{t}}\,.
\end{eqnarray}
The change in the angular momentum is 
\begin{eqnarray}
\Delta\mathbf{J} & =\mathbf{r}\times\Delta\mathbf{v}= & J\left(\frac{\Delta v_{\parallel}}{v}-\frac{v_{r}}{v_{t}^{2}}\Delta v_{w}\right)\hat{\mathbf{J}}+J\frac{\Delta v_{J}}{v}\left(\frac{v_{r}}{v_{t}}\hat{\mathbf{w}}-\hat{\mathbf{v}}\right)\,,
\end{eqnarray}
and the change in angular momentum magnitude (up to second order in
$\Delta v/v$ ) is
\begin{equation}
\Delta J=J\frac{\Delta v_{\parallel}}{v}-rv_{r}\frac{\Delta v_{w}}{v}+\frac{1}{2}\frac{r^{2}}{J}\Delta v_{J}^{2}\,.\label{eq:DJ}
\end{equation}

Using Eqs.~\eqref{eq:DE} and~\eqref{eq:DJ}, we can obtain the local
(orbital phase dependant) DCs in terms of the velocity DCs,
\begin{eqnarray}
\left\langle \Delta E\right\rangle  & = & -\frac{1}{2}\left\langle \left(\Delta v_{\parallel}\right)^{2}\right\rangle -\frac{1}{2}\left\langle \left(\Delta v_{\bot}\right)^{2}\right\rangle -v\left\langle \Delta v_{\parallel}\right\rangle \,,\\
\left\langle \left(\Delta E\right)^{2}\right\rangle  & = & v^{2}\left\langle \left(\Delta v_{\parallel}\right)^{2}\right\rangle \,,\\
\left\langle \Delta J\right\rangle  & = & \frac{J}{v}\left\langle \Delta v_{\parallel}\right\rangle +\frac{r^{2}}{4J}\left\langle \left(\Delta v_{\bot}\right)^{2}\right\rangle \,,\\
\left\langle \left(\Delta J\right)^{2}\right\rangle  & = & \frac{J^{2}}{v^{2}}\left\langle \left(\Delta v_{\parallel}\right)^{2}\right\rangle +\frac{1}{2}\left(r^{2}-\frac{J^{2}}{v^{2}}\right)\left\langle \left(\Delta v_{\bot}\right)^{2}\right\rangle \,,\\
\left\langle \Delta E\Delta J\right\rangle  & = & -J\left\langle \left(\Delta v_{\parallel}\right)^{2}\right\rangle \,,
\end{eqnarray}
were we omitted higher order terms in $\Delta v/v$, and used $\left\langle \Delta v_{w}\right\rangle =\left\langle \Delta v_{J}\right\rangle =0$
and $\left\langle \Delta v_{J}^{2}\right\rangle =\left\langle \Delta v_{w}^{2}\right\rangle =\left\langle \Delta v_{\bot}^{2}\right\rangle /2$.

The local velocity DCs are~\citep{bin+08} 
\begin{eqnarray}
\left\langle \Delta v_{\parallel}\right\rangle  & = & -\kappa\frac{m+m_{a}}{m_{a}}\int_{0}^{v}dv_{a}\frac{v_{a}^{2}}{v^{2}}f_{a}\left(v_{a}\right)\,,\\
\left\langle \Delta v_{\parallel}^{2}\right\rangle  & = & \frac{2}{3}\kappa\left[\int_{0}^{v}dv_{a}\frac{v_{a}^{4}}{v^{3}}f_{a}\left(v_{a}\right)+\int_{v}^{\infty}dv_{a}v_{a}f_{a}\left(v_{a}\right)\right]\,,\\
\left\langle \Delta v_{\bot}^{2}\right\rangle  & = & \frac{2}{3}\kappa\left[\int_{0}^{v}dv_{a}\left(\frac{3v_{a}^{2}}{v}-\frac{v_{a}^{4}}{v^{3}}\right)f_{a}\left(v_{a}\right)+2\int_{v}^{\infty}dv_{a}v_{a}f_{a}\left(v_{a}\right)\right]\,,
\end{eqnarray}
where $\kappa=\left(4\pi Gm_{a}\right)^{2}\ln\Lambda$, and where
$f_{a}\left(v_{a}\right)$ and $m_{a}$ are the velocity DF and mass
of the field stars, and where we assume that the velocity DF is isotropic.
In that case the velocity DF can be written in terms of the orbital
energies of the field stars. Using
\begin{equation}
f_{a}\left(v\right)vdv=-f\left(E\right)dE\,,
\end{equation}
we obtain
\begin{eqnarray}
\left\langle \Delta E\right\rangle  & = & \kappa\left[\frac{m}{m_{a}}\int_{E}^{\phi}\left(v_{a}/v\right)f_{a}\left(E_{a}\right)dE_{a}-\int_{-\infty}^{E}f_{a}\left(E_{a}\right)dE_{a}\right]\,,
\end{eqnarray}
\begin{equation}
\left\langle \Delta E^{2}\right\rangle =\frac{2}{3}\kappa v^{2}\left[\int_{E}^{\phi}dE_{a}\left(v_{a}/v\right)^{3}f_{a}\left(E_{a}\right)+\int_{-\infty}^{E}dE_{a}f_{a}\left(E_{a}\right)\right]\,,
\end{equation}
\begin{eqnarray}
\left\langle \Delta J\right\rangle  & = & \kappa\left\{ -\frac{J}{v^{2}}\left(\frac{m}{m_{a}}+1\right)\int_{E}^{\phi}dE_{a}\left(v_{a}/v\right)f_{a}\left(E_{a}\right)\right.\nonumber \\
 &  & +\left.\frac{r^{2}}{6J}\int_{E}^{\phi}dE^{\prime}f_{a}\left(E_{a}\right)\left[3\left(v_{a}/v\right)-\left(v_{a}/v\right)^{3}\right]+\frac{r^{2}}{3J}\int_{-\infty}^{E}dE_{a}f_{a}\left(E_{a}\right)\right\} \,,\nonumber \\
\end{eqnarray}
\begin{eqnarray}
\left\langle \Delta J^{2}\right\rangle  & = & \frac{\kappa}{v^{2}}\left\{ J^{2}\int_{E}^{\phi}\left(v_{a}/v\right)^{3}dE_{a}f_{a}\left(E_{a}\right)-J^{2}\int_{E}^{\phi}dE_{a}\left(v_{a}/v\right)f_{a}\left(E_{a}\right)\right.\nonumber \\
 &  & +\left.\frac{r^{2}v^{2}}{3}\int_{E}^{\phi}dE_{a}\left(3\left(v_{a}/v\right)-\left(v_{a}/v\right)^{3}\right)f_{a}\left(E_{a}\right)+\frac{2}{3}r^{2}v^{2}\int_{-\infty}^{E}dE_{a}f_{a}\left(E_{a}\right)\right\} \,,\nonumber \\
\end{eqnarray}
and
\begin{eqnarray}
\left\langle \Delta E\Delta J\right\rangle  & = & -J\frac{2}{3}\kappa\left[\int_{E}^{\phi}dE_{a}\left(v_{a}/v\right)^{3}f_{a}\left(E_{a}\right)+\int_{-\infty}^{E}dE_{a}f_{a}\left(E_{a}\right)\right]\,.
\end{eqnarray}
The corresponding orbit-averaged DCs are given by
\begin{align}
D_{E} & =2P^{-1}\int_{r_{a}}^{r_{p}}\left\langle \Delta E\right\rangle dr/v_{r},\\
D_{EE} & =2P^{-1}\int_{r_{a}}^{r_{p}}\left\langle \left(\Delta E\right)^{2}\right\rangle dr/v_{r},\\
D_{J} & =2P^{-1}\int_{r_{a}}^{r_{p}}\left\langle \Delta J\right\rangle dr/v_{r},\\
D_{JJ} & =2P^{-1}\int_{r_{a}}^{r_{p}}\left\langle \left(\Delta J\right)^{2}\right\rangle dr/v_{r},\\
D_{EJ} & =2P^{-1}\int_{r_{a}}^{r_{p}}\left\langle \Delta E\Delta J\right\rangle dr/v_{r}.
\end{align}
where $P$ is the orbital period. 

Assuming that the potential is Keplerian, $\phi=M_{\bullet}/r$, the
energy and angular momentum are 
\begin{eqnarray}
E & = & \frac{M_{\bullet}}{r}-\frac{1}{2}v_{r}^{2}-\frac{1}{2}\frac{J^{2}}{r^{2}}=\frac{GM_{\bullet}}{2a}\,,\\
J & = & rv_{t\,},
\end{eqnarray}
where $a$ is the sma. Using $x=\left(r/a-1\right)e^{-1}$, the orbital
average is 
\begin{equation}
\left\langle D\right\rangle _{\circlearrowright}=\frac{1}{\pi}\int_{-1}^{1}D\frac{1+xe}{\sqrt{1-x^{2}}}dx\,,
\end{equation}
where $e=\sqrt{1-J^{2}/J_{c}^{2}}$ is the eccentricity of the orbits
and $J_{c}=\sqrt{GM_{\bullet}a}$ is the maximal (circular) angular
momentum. The orbital-averaged DCs are therefore
\begin{eqnarray}
D_{E}/E & = & -\Gamma_{0}+\frac{m}{m_{a}}\Gamma_{110}\,,\\
D_{EE}/E^{2} & = & \frac{4}{3}\Gamma_{13-1}+\frac{4}{3}\Gamma_{0}\,,\\
D_{J}/J_{c} & = & \left[\frac{5-3j^{2}}{12}\Gamma_{0}-j^{2}\frac{m+m_{a}}{2m_{a}}\Gamma_{111}+\Gamma_{310}-\frac{1}{3}\Gamma_{330}\right]/j\,,\\
D_{JJ}/J_{c} & = & \frac{5-3j^{2}}{6}\Gamma_{0}+\frac{1}{2}j^{2}\Gamma_{131}-\frac{1}{2}j^{2}\Gamma_{111}+2\Gamma_{310}-\frac{2}{3}\Gamma_{330}\,,\\
D_{EJ}/\left(EJ_{c}\right) & = & -\frac{2}{3}j\left(\Gamma_{0}+\Gamma_{130}\right)\,,
\end{eqnarray}
where
\begin{eqnarray}
\Gamma_{0} & = & \kappa\int_{-\infty}^{1}dsf_{a}\left(sE\right)\,,
\end{eqnarray}
and
\begin{equation}
\Gamma_{ijk}\left(E,J\right)=2^{1+k-i}\frac{\kappa}{\pi}\int_{-1}^{1}dx\int_{1}^{2/\left(1+ex\right)}dsf_{a}\left(sE\right)\frac{1}{\sqrt{1-x^{2}}}\frac{\left(r/a\right)^{i}}{\left(v^{2}/E\right)^{k}}\left(v_{a}/v\right)^{j}\,.
\end{equation}
We can write the $\Gamma_{ijk}$ functions in terms of the~\citet{coh+78}
$F_{i}$ functions (see the definitions after Eq. (24) there)
\begin{eqnarray}
F_{0} & = & E\Gamma_{0}\,,\\
F_{1} & = & E\Gamma_{110}\,,\\
F_{2} & = & E\Gamma_{111}\,,\\
F_{3} & = & E\Gamma_{310}\,,\\
F_{4} & = & E\Gamma_{13-1}\,,\\
F_{5} & = & E\Gamma_{130}\,,\\
F_{6} & = & E\Gamma_{131}\,,\\
F_{7} & = & E\Gamma_{330}\,.
\end{eqnarray}
It is sometimes useful to consider the diffusion in the dimensionless
normalized angular momentum, $j$. For a general coordinates transformation
$\mathbf{x}^{\prime}=\mathbf{x}^{\prime}\left(\mathbf{x}\right)$
the new DCs are given by~\citep[e.g.,][]{ris89}
\begin{eqnarray}
D_{l}^{\prime} & = & \frac{\partial x_{l}^{\prime}}{\partial x_{k}}D_{k}+\frac{1}{2}\frac{\partial^{2}x_{l}^{\prime}}{\partial x_{r}\partial x_{k}}D_{rk}\,,\\
D_{lm}^{\prime} & = & \frac{\partial x_{l}^{\prime}}{\partial x_{r}}\frac{\partial x_{m}^{\prime}}{\partial x_{k}}D_{rk}\,.
\end{eqnarray}
Thus, in the $E$, $j$ coordinates the $j$-related DCs are 
\begin{eqnarray}
D_{j} & = & D_{J}/J_{c}+\frac{1}{2}jD_{E}/E+\frac{1}{2}D_{EJ}/\left(J_{c}E\right)-\frac{1}{8}jD_{EE}/E^{2}\,,\\
D_{jj} & = & D_{JJ}/J_{c}^{2}+\frac{1}{4}j^{2}D_{EE}/E^{2}+jD_{EJ}/\left(J_{c}E\right)\,,\\
D_{Ej}/E & = & D_{EJ}/\left(J_{c}E\right)+\frac{1}{2}jD_{EE}/E^{2}\,.
\end{eqnarray}
Similarly, for the $E$, $R=j^{2}$ coordinates the $R$-related DCs
are~\citep[see also][]{coh+78,coh79}
\begin{eqnarray}
D_{R} & = & 2jD_{j}+D_{jj}=2jD_{J}/J_{c}+D_{JJ}/J_{c}^{2}+j^{2}\frac{D_{E}}{E}+2\frac{D_{EJ}}{J_{c}E}j\,,\\
D_{RR} & = & 4j^{2}D_{jj}=4j^{2}D_{JJ}/J_{c}^{2}+j^{4}D_{EE}/E^{2}+4j^{3}\frac{D_{EJ}}{J_{c}E}\,,\\
D_{ER} & = & 2jD_{Ej}=2j\frac{D_{EJ}}{J_{c}E}+j^{2}D_{EE}/E^{2}\,.
\end{eqnarray}

\makeatletter{}\global\long\def\Mo{M_{\odot}}
\global\long\def\Ro{R_{\odot}}
\global\long\def\Lo{L_{\odot}}
\global\long\def\Mbh{M_{\bullet}}
\global\long\def\Ms{M_{\star}}
\global\long\def\Rs{R_{\star}}
\global\long\def\Ts{T_{\star}}
\global\long\def\vs{v_{\star}}
\global\long\def\s{\sigma_{\star}}
\global\long\def\n{n_{\star}}
\global\long\def\tr{t_{\mathrm{rlx}}}
\global\long\def\Ns{N_{\star}}
\global\long\def\ns{n_{\star}}
\global\long\def\jlc{j_{lc}}

\section{The statistical properties of the resonant torques}

\label{a:RRtorque}

We describe and measure here the residual torques acting on a test
star due to the near-Keplerian orbits of the background stars. The
torque vector $\tau$ depends on the test star's angular momentum
$\boldsymbol{J}$ and argument of pericenter $\omega$. We discuss
the symmetries of these torques and the scaling with the test star's
orbital parameters, and empirically measure the torques by static
random background simulations.

\subsection{Geometrical description}

Consider the angular momentum vector in some fixed reference frame
\[
\boldsymbol{J}=J\left(\begin{array}{c}
\ell_{x}\\
\ell_{y}\\
\ell_{z}
\end{array}\right)=\left(\begin{array}{c}
\sqrt{J^{2}-J_{z}^{2}}\cos\phi\\
\sqrt{J^{2}-J_{z}^{2}}\sin\phi\\
J_{z}
\end{array}\right),
\]
where $\hat{\boldsymbol{\ell}}=\left(\ell_{x},\ell_{y},\ell_{z}\right)$
is the unit vector in the direction of $\boldsymbol{J}$ in a fixed
Cartesian reference system $(x,y,z)$. The torque is derived from
the Hamiltonian by $\boldsymbol{\tau}=\dot{\boldsymbol{J}}=\left\{ J,H\right\} $,
where $\left\{ \dots\right\} $ denotes the Poisson brackets.  It
is more convenient to represent the torques in an orthonormal spherical
coordinate system $\left(J,\phi,u\right)$, where $u\equiv\ell_{z}=\cos\theta$,
with the associated unit vectors $\hat{e}_{i}=\left(\partial\mathbf{J}/\partial i\right)/\left|\partial\mathbf{J}/\partial i\right|$
for $i\in\left\{ J,\phi,u\right\} $. The change in the angular momentum's
magnitude is then $\tau_{J}=\dot{J}$, and the changes in its direction
are described by the angular torques, $\tau_{\phi}=J\dot{\phi}$ and
$\tau_{u}=J\dot{u}$. The transformation of the torque vector from
spherical to Cartesian coordinates is given by 
\begin{equation}
\boldsymbol{\tau}=\tau_{J}\hat{e}_{J}+\tau_{\phi}\sqrt{1-u^{2}}\hat{e}_{\phi}+\frac{\tau_{u}}{\sqrt{1-u^{2}}}\hat{e}_{u}\,.
\end{equation}

In the reference frame of the orbit we can define the orbital torques
in the direction of the semi-major axis, $\tau_{a}$, and the semi-minor
axis, $\tau_{b}$,
\begin{eqnarray}
\tau_{a} & = & \boldsymbol{\tau}\cdot\hat{a}=\sqrt{1-u^{2}}\sin\omega\tau_{\phi}-\frac{1}{\sqrt{1-u^{2}}}\cos\omega\tau_{u}\,,\\
\tau_{b} & = & \boldsymbol{\tau}\cdot\hat{b}=\sqrt{1-u^{2}}\cos\omega\tau_{\phi}+\frac{1}{\sqrt{1-u^{2}}}\sin\omega\tau_{u}\,,
\end{eqnarray}
where $\hat{a}$ and $\hat{b}$ are the direction semi-major and semi-minor
axes
\begin{eqnarray}
\hat{a} & = & \hat{e}_{\phi}\sin\omega-\hat{e}_{u}\cos\omega\,,\\
\hat{b} & = & \hat{e}_{\phi}\cos\omega+\hat{e}_{u}\sin\omega\,.
\end{eqnarray}

\subsection{Statistical properties}

Define the typical (Poisson) torque~\citep{gur+07} 
\begin{equation}
\tilde{\tau}_{N}\equiv\frac{\sqrt{N\left(2a\right)}}{a}GM_{\star}=J_{c}\nu_{r}\frac{\sqrt{N\left(2a\right)}}{Q}\,.
\end{equation}
 The mean squared values of the torques are given by 
\begin{eqnarray}
\left\langle \tau_{J}^{2}\right\rangle /\tilde{\tau}_{N}^{2} & = & T_{\parallel}\left(a,J\right)\,,\\
\left\langle \tau_{\phi}^{2}\right\rangle /\tilde{\tau}_{N}^{2} & = & \frac{1}{2}\frac{1}{1-\mu^{2}}\left[T_{+}\left(a,J\right)+T_{-}\left(a,J\right)\cos2\omega\right]\,,\\
\left\langle \tau_{u}^{2}\right\rangle /\tilde{\tau}_{N}^{2} & = & \frac{1}{2}\sqrt{1-\mu^{2}}\left[T_{+}\left(a,J\right)-T_{-}\left(a,J\right)\cos2\omega\right]\,,
\end{eqnarray}
and the cross terms are
\begin{eqnarray}
\left\langle \tau_{J}\tau_{\phi}\right\rangle  & = & \left\langle \tau_{J}\tau_{u}\right\rangle =0\,,\\
\left\langle \tau_{u}\tau_{\phi}\right\rangle /\tilde{\tau}_{N}^{2} & = & T_{-}\sin\left(2\omega\right)/2\,,
\end{eqnarray}
where we defined 
\begin{eqnarray}
T_{\parallel} & \equiv & \left\langle \tau_{J}^{2}\right\rangle /\tilde{\tau}_{N}^{2}\,,\\
T_{+} & \equiv & \left(\left\langle \tau_{a}^{2}\right\rangle +\left\langle \tau_{b}^{2}\right\rangle \right)/\tilde{\tau}_{N}^{2}\,,\\
T_{-} & \equiv & \left(\left\langle \tau_{b}^{2}\right\rangle -\left\langle \tau_{a}^{2}\right\rangle \right)/\tilde{\tau}_{N}^{2}\,.
\end{eqnarray}
Note that since the torques $\left\langle \tau_{J}^{2}\right\rangle $,
$\left\langle \tau_{a}^{2}\right\rangle $ and $\left\langle \tau_{b}^{2}\right\rangle $
are measured in the orbital plane, they have no angular dependencies,
and neither do $T_{\parallel}$, $T_{+}$ and $T_{-}$.

The symmetry of the orbits is such that for a circular orbit (i.e.,
$J=J_{c}$) there is no preferred direction in the orbital plane and
therefore $\left\langle \tau_{a}^{2}\right\rangle =\left\langle \tau_{b}^{2}\right\rangle $
and $T_{\parallel}\left(a,J_{c}\right)=T_{-}\left(a,J_{c}\right)=0$.
For a radial orbits ($J=0$), $\tau_{a}$ vanish and $\left\langle \tau_{b}^{2}\right\rangle =\left\langle \tau_{\parallel}^{2}\right\rangle $
therefore, $T_{\parallel}\left(a,0\right)=T_{+}\left(a,0\right)=T_{-}\left(a,0\right)$.

\subsection{Measuring the torques}

\begin{figure}
\begin{centering}
\includegraphics[scale=0.6]{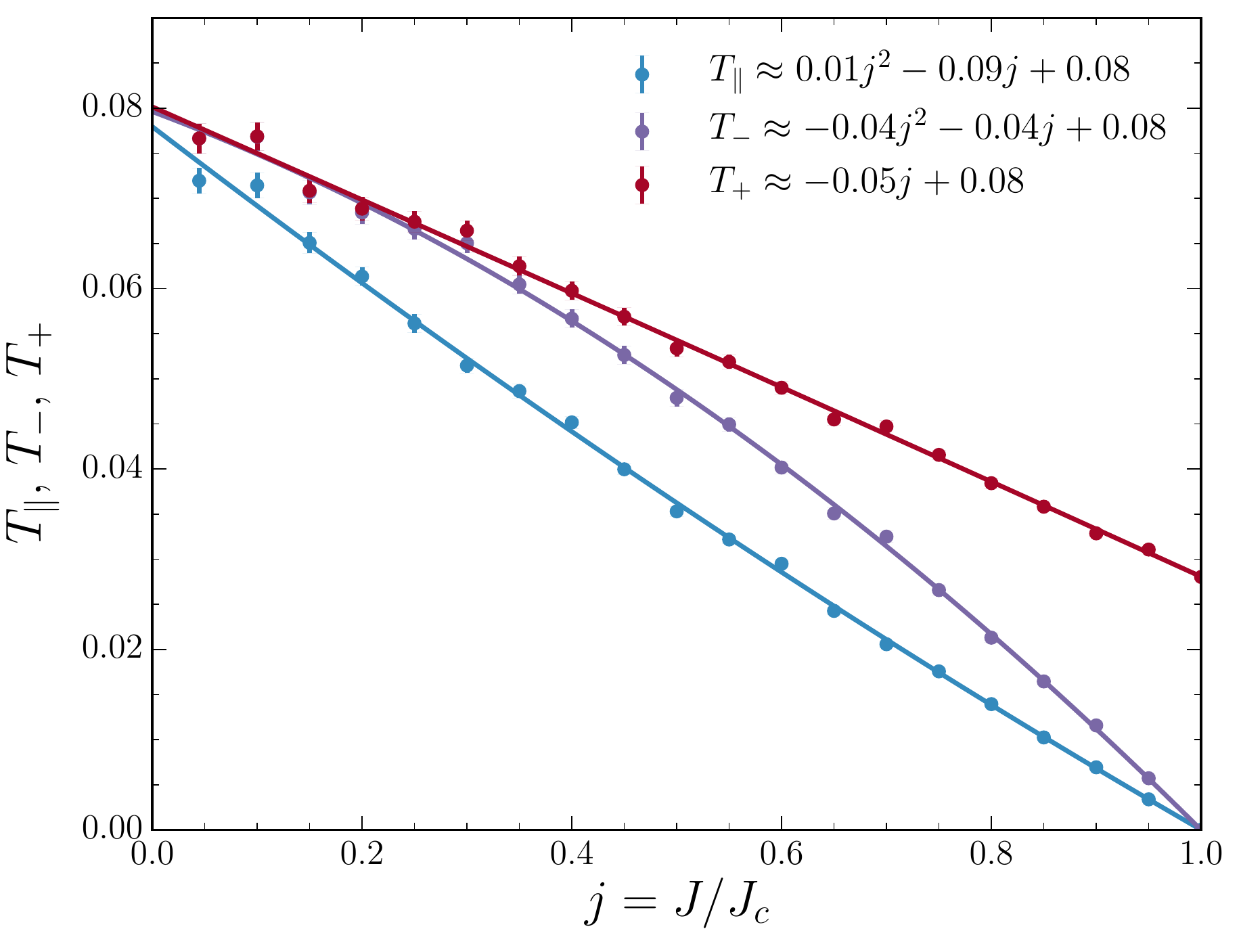}
\par\end{centering}

\centering{}\protect\caption{\label{fig:Ts_j}The mean square of the torques (normalized to the
typical (Poisson) torque $\tilde{\tau}$) as function of the angular
momentum. The torques measured from static wires simulations (circles)
are approximated by polynomial fits (solid lines). As expected from
considerations of symmetry, for a circular orbit ($J=J_{c}$), there
is no torque in the $\boldsymbol{J}$ direction and both perpendicular
torques ($\tau_{b}$,$\tau_{b}$) are equal, i.e., $T_{\parallel}=T_{-}=0$.
As $J\to0$, the orbit's geometry approaches a rod, and the torque
in the $\hat{a}$ direction vanishes because its lever arm goes to
zero, while the torque in the $\boldsymbol{J}$ and $\hat{b}$ directions
become equal, i.e., $T_{\parallel}=T_{+}=T_{-}$. }
\end{figure}

We used static wires simulations~\citep[e.g.,][]{gur+07} to measure
the $j$-dependence of the residual torque ($j=\sqrt{1-e^{2}}$) on
a test orbit with sma $a$ and eccentricity $e$. This was carried
out by simulating the background as many fixed Keplerian wire orbits,
and measuring the three components of the orbital torques $\tau_{a}$,
$\tau_{b}$ and $\tau_{J}$, for many independent random realizations
of the background, integrating over the orbit of the test star and
the orbits of the field stars with the efficient~\citet{tou+09} algorithm.
We decomposed the measured orbital torques to $T_{\parallel}$, $T_{+}$
and $T_{-}$ and fitted them to a second order polynomial in $j$.
 The best-fit results are (Figure~\ref{fig:Ts_j})
\begin{eqnarray}
T_{\parallel} & \approx & 0.08\left(1-j\right)\left(1-j/8\right)\,,\\
T_{-} & \approx & 0.08\left(1-j\right)\left(1+j/4\right)\,,\\
T_{+} & \approx & 0.08\left(1-5j/8\right)\,.
\end{eqnarray}

The residual torque in the $\boldsymbol{J}$ direction is therefore
\begin{eqnarray}
\sqrt{\left\langle \tau_{J}^{2}\right\rangle }/\tilde{\tau}_{N} & \approx & 0.28\sqrt{\left(1-j\right)\left(1-j/8\right)}\approx0.28\sqrt{1-j}\,.
\end{eqnarray}
This analytic fit is consistent with the~\citet{gur+07} results (fitted
to $\sqrt{\left\langle \tau_{J}^{2}\right\rangle }/\tilde{\tau}_{N}\approx0.25\sqrt{1-j^{2}}$
see Eq. (13) there) but is better, since it matches their data over
the entire $j$-range. For the out-of-plane torques we obtain
\begin{equation}
\sqrt{\tau_{\bot}^{2}}/\tilde{\tau}_{N}\equiv\sqrt{\tau_{a}^{2}+\tau_{b}^{2}}\approx0.28\sqrt{1-5j/8}\,.
\end{equation}
This is different from~\citet{gur+07} results (fitted to $\sqrt{\left\langle \tau_{\bot}^{2}\right\rangle }/\tilde{\tau}_{N}\approx0.28(3/2-j^{2})$
see Eq. (14) there). However, the discrepancy can be traced to an
error in their randomization procedure (Eqs. (9--11) there) which
is not truly isotropic. 

Using $N$-body simulations,~\citet{eil+09} measured the isotropic
averaged residual torques. They defined and measured the $j$-averaged
quantities 
\begin{equation}
\beta_{s}^{2}=\frac{Q^{2}}{N_{tot}}\left\langle \dot{J}^{2}P^{2}/J_{c}^{2}\right\rangle =4\pi^{2}\left\langle \tau_{\parallel}^{2}/\tilde{\tau}^{2}\right\rangle \,,
\end{equation}
and
\begin{equation}
\beta_{v}^{2}=\frac{Q^{2}}{N_{tot}}\sqrt{\left\langle \left|\dot{\mathbf{J}}\right|^{2}P^{2}/J_{c}^{2}\right\rangle }=4\pi^{2}\left\langle \left(\tau_{\parallel}^{2}+\tau_{\bot}^{2}\right)/\tilde{\tau}^{2}\right\rangle \,.
\end{equation}
Averaging the results here over $j$, we obtain $\beta_{s}\approx1.0$
and $\beta_{v}\approx1.7$ for $\gamma=7/4$, in agreement with~\citet{eil+09}\footnote{We correct here two issues in the comparison to the~\citet{gur+07}
results made by~\citet[Section 4.3]{eil+09}. First, since $\beta_{s}$
and $\beta_{v}$ are the rms values, the average values of~\citet{gur+07}
should be estimated by $\left\langle \beta_{s}\right\rangle =\sqrt{\left\langle \beta_{s}^{2}(e)\right\rangle }$
and $\left\langle \beta_{v}\right\rangle =\sqrt{\left\langle \beta_{v}^{2}\right\rangle }$.
Second, $\beta_{v}^{2}=\tau_{\parallel}^{2}+\tau_{\bot}^{2}$, defined
by~\citet{eil+09} should be compared to the sum $\tau_{\parallel}^{2}+\tau_{\bot}^{2}$
of the quantities defined in~\citet{gur+07}.}.

\makeatletter{}

\global\long\def\Mo{M_{\odot}}
\global\long\def\Ro{R_{\odot}}
\global\long\def\Lo{L_{\odot}}
\global\long\def\Mbh{M_{\bullet}}
\global\long\def\Ms{M_{\star}}
\global\long\def\Rs{R_{\star}}
\global\long\def\Ts{T_{\star}}
\global\long\def\vs{v_{\star}}
\global\long\def\s{\sigma_{\star}}
\global\long\def\n{n_{\star}}
\global\long\def\tr{t_{\mathrm{rlx}}}
\global\long\def\Ns{N_{\star}}
\global\long\def\ns{n_{\star}}
\global\long\def\jlc{j_{lc}}

\section{The Maximal Entropy Principle}

\label{a:MaxEntropy}

The maximal entropy principle (MEP) has been shown to be a powerful
tool in determining the steady-state (or quasi-steady-state) of dynamical
systems. In particular, it was studied extensively in the context
of collisionless self-gravitating systems~\citep[e.g.,][]{lyn67}.
Here we examine the more restricted problem of a near-Keplerian system.
We prove that the RR DCs used in this study comply with the MEP\@.
For NR, we prove that the $J$-only NR DCs comply with the MEP, since
for stellar systems, the MEP is relevant only when the interactions
conserve energy. 

In a system where a central object of mass $M_{\bullet}$ dominates
the potential (e.g., planetary systems, nuclear clusters), stars move
on nearly Keplerian orbits. That is, the potential is almost regular
and the orbital elements are almost constant. In particular, since
the potential varies on much longer timescales than the orbital time,
the potential can be orbit-averaged and the Keplerian energy is conserved.

The entropy of the system is given by
\begin{equation}
S=-\int d^{3}\mathbf{r}d^{3}\mathbf{v}f\left(\mathbf{r},\mathbf{v}\right)\log f\left(\mathbf{r},\mathbf{v}\right)\,,
\end{equation}
where $f$ is the stellar DF\@. The Keplerian energy distribution
relative to the MBH, $n(E)$, is conserved
\begin{equation}
n\left(E;f\right)=\int d^{3}\mathbf{r}d^{3}\mathbf{v}f\left(\mathbf{r},\mathbf{v}\right)\delta\left(E-\frac{GM_{\bullet}}{r}+\frac{1}{2}v^{2}\right)\,.
\end{equation}
This implies the conservation of the total Keplerian energy $E_{\bullet}[f]=\int En\left(E;f\right)dE$
and of the total number of stars $N_{tot}\left[f\right]=\int n\left(E;f\right)dE$.
Conservation of the total energy $E_{tot}=E_{\bullet}[f]+E_{\star}[f]$
then implies the conservation of the total potential energy due to
the interactions with the stellar background, 
\begin{eqnarray}
E_{\star}[f] & = & \frac{1}{2}\int d^{3}\mathbf{r}\int d^{3}\boldsymbol{v}f\left(\mathbf{r},\mathbf{v}\right)\psi\left(\mathbf{r};f\right),
\end{eqnarray}
where 
\begin{equation}
\psi\left(\mathbf{r};f\right)=G\Ms\int d^{3}\mathbf{r}^{\prime}\int d^{3}\boldsymbol{v}^{\prime}\frac{f\left(\mathbf{r}^{\prime},\mathbf{v}^{\prime}\right)}{\left|\mathbf{r}-\mathbf{r}^{\prime}\right|}\,,
\end{equation}
is the star-star potential. The total angular momentum is also conserved,
\begin{equation}
\mathbf{L}_{tot}[f]=\int d^{3}\mathbf{r}d^{3}\mathbf{v}f\left(\mathbf{r},\mathbf{v}\right)\mathbf{L\,},
\end{equation}
 where $\mathbf{L}=\mathbf{r}\times\mathbf{v}$. Using the Lagrange
multipliers $\beta$, $\boldsymbol{b}$ and $\lambda(E)$, we write
the target function
\begin{eqnarray}
\mathcal{S} & = & S+\beta\left(\int d^{3}\mathbf{r}d^{3}\mathbf{v}f\left(\mathbf{r}^{\prime},\mathbf{v}^{\prime}\right)\psi\left(\mathbf{r};f\right)-2E_{\star}\right)+\mathbf{b}\cdot\left(\int d^{3}\mathbf{r}d^{3}\mathbf{v}f\left(\mathbf{r},\mathbf{v}\right)\mathbf{L}-\mathbf{L}_{tot}\right)\nonumber \\
 &  & +\int d^{3}\mathbf{r}d^{3}\mathbf{v}f\left(\mathbf{r},\mathbf{v}\right)\lambda\left(E\right)\left(\int d^{3}\mathbf{r}^{\prime}d^{3}\mathbf{v}^{\prime}f\left(\mathbf{r}^{\prime},\mathbf{v}^{\prime}\right)\delta\left(E-\frac{GM_{\bullet}}{r^{\prime}}+\frac{1}{2}v^{\prime2}\right)-n_{0}\left(E\right)\right)\,,\nonumber \\
\end{eqnarray}
which is minimized by requiring
\begin{eqnarray}
\frac{\delta\mathcal{S}}{\delta f} & = & -\log f-1+\beta\psi\left(\mathbf{r};f\right)+\mathbf{b}\cdot\boldsymbol{L}+2\lambda\left(E\right)n\left(E;f\right)=0.
\end{eqnarray}
Therefore, the DF that maximizes the entropy is
\begin{equation}
f\left(\mathbf{r},\mathbf{v}\right)=A\left(E\right)e^{\beta\psi+\mathbf{b}\cdot\boldsymbol{L}}\,,
\end{equation}
where $A(E)=\exp[2\lambda(E)n(E)]$, and $\beta$ and $\boldsymbol{b}$
are constants determined by the constraints on $n(E)$, $E_{\star}$
and $\boldsymbol{L}_{tot}$. Note that an isotropic system must have
$b=0$ to ensure that the DF does not depend on $\boldsymbol{L}$,
and must also have $\beta=0$ since the star-star potential depends
on $L$ even in an isotropic system\footnote{The $L$ dependence of $\psi$ arises from the eccentricity dependence
of the enclosed mass seen by the star. This is manifested dynamically
by the retrograde evolution of the argument of periapse---mass precession. }.

\subsection{Fluctuation dissipation relation for a spherical symmetric system}

Since stars are assumed to move on Keplerian orbits, it is convenient
to work in action-angle coordinates. Choosing the $z$-coordinate
in the direction of the total angular momentum, the steady-state is
\begin{equation}
n\left(E,J,J_{z}\right)=n\left(E\right)\frac{e^{\beta\psi+bL_{z}}}{\int\int e^{\beta\psi+bL_{z}}dJ_{z}dJ}\,.
\end{equation}
For a spherical symmetric system with $\boldsymbol{L}_{tot}=0$, the
steady-state DF is given by an implicit integral equation 
\begin{equation}
n\left(E,J\right)=n\left(E\right)\frac{2Je^{\beta\psi(E,J,n)}}{\int e^{\beta\psi(E,J,n)}dJ^{2}}\,.\label{eq:nEJMEP}
\end{equation}

MEP considerations do not require the additional assumptions that
go into the FP equation (e.g., diffusion described by a Markovian
process), and the MEP solution is independent of the path that the
system took to reach it from its initial conditions. Therefore, the
MEP solution must also be satisfied by the FP equation in steady-state,
which enforces a connection between the DCs, known as the fluctuation-dissipation
(F-D) relation. For the symmetries and conserved quantities of the
system studied here (described by Eq.~\eqref{eq:nEJMEP}), the functional
form of the F-D relation is 
\begin{equation}
2JD_{J}e^{\beta\psi}=\frac{\partial}{\partial J}\left[JD_{JJ}e^{\beta\psi}\right].
\end{equation}
From this point on we will restrict ourselves to isotropic systems
which reach a MEP solution with $\beta=0$. This is a solution of
the form $n\left(E,J\right)=2n\left(E\right)J/J_{c}^{2}$. This means
that DCs derived under the assumption of an isotropic background (and
fixed Keplerian energy) must satisfy the F-D relation $2JD_{J}=\partial\left(JD_{JJ}\right)/\partial J$.
In particular, the RR DCs that were used in this study obey this relation,
as required~\citep{bar+14}.

\subsection{Fluctuation-dissipation relation for $J$-only two-body relaxation}

For NR, the coherence time is shorter than the orbital period (Section~\ref{s:MBHrlx}), and therefore orbital energy is not conserved. However,
we argue in Section~\ref{s:analytic} that the flow pattern in phase-space
justifies the approximate treatment of the fast $J$-diffusion as
separate from the slower $E$-diffusion. We now use this property
to show that in that case, the NR $J$-only DCs also satisfy the fluctuation
dissipation relation, which is a partial test of the validity of the
general NR DCs and is used in Section~\ref{s:analytic}.

Using Eq.~\eqref{eq:DE} and setting $\Delta E=0$, we obtain
\begin{equation}
2v\Delta v_{\parallel}+\Delta v^{2}=0\,.
\end{equation}
Therefore, to first order in $\Delta v^{2}/v^{2}$, the local diffusion
coefficients are
\begin{eqnarray}
\left\langle \Delta J\right\rangle  & = & \frac{1}{2J}\left(\frac{r^{2}}{2}-\frac{J^{2}}{v^{2}}\right)\left\langle \Delta v_{\bot}^{2}\right\rangle \,,\\
\left\langle \left(\Delta J\right)^{2}\right\rangle  & = & \frac{1}{2}\left(r^{2}-\frac{J^{2}}{v^{2}}\right)\left\langle \left(\Delta v_{\bot}\right)^{2}\right\rangle \,.
\end{eqnarray}
The orbit-averaged diffusion coefficients are
\begin{eqnarray}
D_{xy} & = & \left\langle \left\langle \Delta x\Delta y\right\rangle \right\rangle _{\circlearrowright}=2\int_{r_{p}}^{r_{a}}\left\langle \Delta x\Delta y\right\rangle \frac{dr}{v_{r}}\,,
\end{eqnarray}
where $v_{r}$ is the radial velocity. Assuming a spherical potential
$\Phi\left(r\right)$, the energy $E$ and angular momentum $J$ are
\begin{equation}
E=\frac{1}{2}v_{r}^{2}+\frac{1}{2}\left(J^{2}/r^{2}\right)+\Phi\left(r\right),
\end{equation}
and
\begin{equation}
J=rv_{t},
\end{equation}
where $v_{t}$ is the transverse velocity. Therefore,
\begin{equation}
v_{r}=\sqrt{2E-2\Phi\left(r\right)-J^{2}/r^{2}},
\end{equation}
and
\begin{equation}
\frac{\partial\left(1/v_{r}\right)}{\partial J}=\frac{1}{v_{r}^{2}}\frac{1}{v_{r}}\frac{J}{r^{2}}=\frac{1}{\left(r^{2}-J^{2}/v^{2}\right)}\frac{1}{v_{r}}\frac{J}{v^{2}}\,.
\end{equation}
It then follows that 
\begin{equation}
\frac{\partial}{\partial J}\left\langle JX\right\rangle _{\circlearrowright}=\left\langle \frac{\partial}{\partial J}JX\right\rangle _{\circlearrowright}+J^{2}\left\langle \frac{X}{\left(r^{2}v^{2}-J^{2}\right)}\right\rangle _{\circlearrowright}\,,
\end{equation}
and in particular 
\begin{eqnarray}
\frac{1}{2J}\frac{\partial}{\partial J}JD_{JJ} & = & \frac{1}{2J}\left\langle \left(\frac{1}{2}r^{2}-\frac{J^{2}}{v^{2}}\right)\left\langle \left(\Delta v_{\bot}\right)^{2}\right\rangle \right\rangle _{\circlearrowright}=D_{J}\,,
\end{eqnarray}
which therefore proves that the $J$-only NR DCs indeed satisfy the
F-D relation.

\bibliographystyle{apj}
\bibliography{LC_EJ}

\end{document}